%% file: main.tex
\documentclass[onecolumn,prd,floatfix,showpacs,preprintnumbers,nofootinbib,%
superscriptaddress]{revtex4}
\usepackage[T1]{fontenc}
\usepackage{lmodern} 
\usepackage[utf8]{inputenc} 
\usepackage{mathrsfs}
\usepackage{amssymb}	
\usepackage{amsmath}
\usepackage{graphicx}
\usepackage{xcolor}
\usepackage{placeins}
\usepackage[normalem]{ulem}
\usepackage{stmaryrd}

\definecolor{lcolor}{rgb}{0.5,0,0}
\definecolor{citcolor}{rgb}{0,0.3,0.0}
\definecolor{coloryksi}{rgb}{0.5,0.0,0.0}
\definecolor{colorkaksi}{rgb}{0.0,0.0,0.5}
\definecolor{colorkolme}{rgb}{0.0,0.3,0.1}

\usepackage{mathtools}
\usepackage{physics}
\usepackage{tensor}
\usepackage{cancel}
\usepackage{mathrsfs}

\newcommand{\beq}{\begin{equation}}
\newcommand{\eeq}{\end{equation}}

\newcommand{\qqb}{{q\Bar{q}}}
\newcommand{\qqbg}{{q\Bar{q}g}}
\newcommand{\ggb}{{g \tilde{g}}}
\newcommand{\xpom}{{x_\mathbb{P}}}
\newcommand{\xbj}{{x_\mathrm{Bj}}}

\usepackage[breaklinks,colorlinks,urlcolor=blue,citecolor=citcolor,linkcolor=lcolor]{hyperref}

\usepackage{tikz}
\usetikzlibrary{snakes}

\input{defs.tex}

\begin{document}

\author{G. Beuf}
\affiliation{National Centre for Nuclear Research, 02-093 Warsaw, Poland}
\author{H. H\"anninen}
\author{T. Lappi}
\affiliation{
Department of Physics, %
 P.O. Box 35, 40014 University of Jyv\"askyl\"a, Finland}
\affiliation{
Helsinki Institute of Physics, P.O. Box 64, 00014 University of Helsinki,
Finland}
\author{Y. Mulian}
\affiliation{Instituto Galego de Física de Altas Enerxías IGFAE,
Universidade de Santiago de Compostela, 15782 Santiago de Compostela, Galicia-Spain}
\author{H. M\"antysaari}
\affiliation{
Department of Physics, %
 P.O. Box 35, 40014 University of Jyv\"askyl\"a, Finland}
\affiliation{
Helsinki Institute of Physics, P.O. Box 64, 00014 University of Helsinki,
Finland}

\title{Diffractive deep inelastic scattering at NLO in the dipole picture: the $q\Bar q g$ contribution}


\begin{abstract}
We calculate the contribution from the $q\Bar q g$ state production to the diffractive cross sections in deep inelastic scattering at high energy. The obtained cross section is finite by itself and a part of the full next-to-leading order result for the diffractive structure functions. We perform the calculation in exact kinematics in the eikonal limit, and show that the previously known high-$Q^2$ and large $M_X^2$ results for the structure functions can be extracted from our results in the appropriate limits. We furthermore discuss the steps required to obtain the full next-to-leading order results for the structure functions.
\end{abstract}

\maketitle

\tableofcontents

\newpage 

\section{Introduction}

In collisions with high scattering energy, one is measuring degrees of freedom of hadronic and nuclear states that only have a small fraction of the full momentum of the state, small-$x$ degrees of freedom. The large amount of phase space available at high collision energies leads to an exponentially cascading emission of gluons. At some point, however, this  cascade must be limited by unitarity requirements on scattering amplitudes. Thus gluon mergings eventually start to be equally important, even at transverse resolution scales where a weak coupling description is appropriate. The kinematical region where these two effects balance each other is referred to as the gluon saturation regime, and understanding it is the topic of many theoretical and experimental efforts. Experimentally the saturation regime is relevant for understanding hadronic collision processes and the formation of quark gluon plasma at RHIC and the LHC. A particularly precise and clean way to access the small-$x$ degrees of freedom is, however, provided by high energy Deep Inelastic Scattering (DIS), both in the HERA experiments, and at the future Electron-Ion Collider (EIC)~\cite{Accardi:2012qut,Aschenauer:2017jsk,AbdulKhalek:2021gbh} and LHeC~\cite{LHeCStudyGroup:2012zhm,LHeC:2020van}. Theoretically, a convenient way to discuss the physics of gluon saturation is provided by the Color Glass Condensate (CGC)~\cite{Iancu:2003xm,Weigert:2005us,Gelis:2010nm} effective field theory, where the nonlinear gluon system is described as a classical color field. For the DIS 
process, the CGC framework naturally leads to the dipole picture~\cite{Nikolaev:1990ja,Nikolaev:1991et,Mueller:1993rr,Mueller:1994jq,Mueller:1994gb}. 

In the dipole picture one factorizes the DIS
process of a virtual photon off a hadronic target  into two ingredients. Firstly the perturbative part of the process is the development of the photon into a partonic state, to leading order a color neutral quark-antiquark dipole. The second ingredient is the scattering of this partonic state with the gluonic target, which in the high collision energy limit can be treated as an eikonal interaction with the classical color field. 
With the prospect of higher luminosities and the availability of nuclear targets in future DIS experiments, there has been a systematic push in the field to improve the perturbative accuracy of the dipole picture by going to higher orders in perturbation theory.
In recent years the dipole picture has been extended to NLO accuracy for the  high energy BK/JIMWLK evolution~\cite{Balitsky:1995ub,Kovchegov:1999ua,Kovchegov:1999yj,Balitsky:2008zza,Balitsky:2013fea,Kovner:2013ona,Balitsky:2014mca,Beuf:2014uia,Lappi:2015fma,Iancu:2015vea,Iancu:2015joa,Albacete:2015xza,
Lappi:2016fmu,Lublinsky:2016meo} and the inclusive DIS cross section~\cite{Balitsky:2010ze,Balitsky:2012bs,Beuf:2011xd,Beuf:2016wdz,Beuf:2017bpd,Hanninen:2017ddy,Beuf:2020dxl,Lappi:2020srm,Caucal:2021ent,Taels:2022tza,Beuf:2021qqa,Beuf:2021srj,Beuf:2022ndu}. 

Exclusive or diffractive DIS is  expected to be even more sensitive to gluon saturation than inclusive cross sections~\cite{Golec-Biernat:1999qor,Kowalski:2007rw,Armesto:2019gxy,AbdulKhalek:2021gbh}. One way to understand this is to note that, due to the optical theorem, the total cross section is proportional to the elastic dipole-target amplitude, proportional to the gluon distribution in the target. Exclusive cross sections, on the other hand, are calculated as the square of the amplitude, and are thus much more sensitive to the large amplitudes, a signature of the saturation regime. Correspondingly, the recent work on inclusive scattering has been accompanied by several calculations of exclusive vector meson and diffractive dijet production at NLO in the dipole picture~\cite{Boussarie:2014lxa,Boussarie:2016ogo,Boussarie:2016bkq,Escobedo:2019bxn,Lappi:2020ufv,Mantysaari:2021ryb,Mantysaari:2022bsp,Mantysaari:2022kdm,Iancu:2021rup,Hatta:2022lzj}. While these processes are an extremely important part of the coming experimental program, they both have some drawbacks for the purpose of understanding gluon saturation. Exclusive vector meson production requires some knowledge or modeling of the bound state physics of the meson. While there are systematical ways to do this  perturbatively, e.g. by a nonrelativistic QCD approach as in Ref.~\cite{Lappi:2020ufv} or by using universal Parton Distribution Amplitudes that can independently be measured in other processes~\cite{Lepage:1980fj,Boussarie:2016bkq,Mantysaari:2022bsp}, this unavoidably adds an additional source of uncertainty. Dijets, on the other hand, are well defined perturbative objects in a given jet algorithm, but only if the jet transverse momenta are sufficiently large. At realistic collider energies this has a tendency to push jet measurements to larger $x$ and thus outside the saturation regime. 

In this paper we will focus on a process that has gathered somewhat less attention in the work to push the dipole picture to NLO accuracy, namely inclusive diffractive DIS. Here the experimental signature is a large rapidity gap between the diffractive system ($X$) consisting of the photon remnants, and the target or its remnants. In the dipole picture the photon fluctuates into a variety of partonic states, which scatter off the target without exchanging color. In this sense the rapidity gap makes the process fundamentally an exclusive one, with a cross section given by the square of an amplitude. On the other hand, the measurement is inclusive in the sense that one sums over all of the different final states of the diffractive system, measuring the cross section differentially only in its total invariant mass $M_X$. This latter inclusive aspect makes it possible to extend the perturbative description to much lower invariant masses and to lower $x$ than for diffractive dijets, even if the parton level final states are the same.
The cross sections for such processes are expressed in terms of the diffractive structure functions $F_2^{\textrm{D}(3)}(\beta,Q^2,\xpom)$ and $ F_L^\textrm{D(3)}(\beta,Q^2,\xpom)$ or, equivalently, the diffractive virtual photon cross sections 
$\ud \sigma^{\gamma^* + A \to A+n}/\ud [\mathcal{PS}]_n$, which are the quantities that we will calculate here. 

At leading order in $\as$, the diffractive final state only consists of a quark-antiquark dipole. This already provides a good description of the general features of the experimental measurements at small $M_X^2\sim Q^2$~\cite{Golec-Biernat:1999qor}. However, a strict leading order picture fails to describe the rise of the cross section towards larger $M_X$ where, in order to make a high invariant mass partonic state, additional gluon radiation is required. The phenomenologically most successful approach has been to use the ``Wüsthoff result''~\cite{Wusthoff:1997fz}, which includes the radiation of one extra gluon into the final state~\cite{Golec-Biernat:1999qd,GolecBiernat:1999qd,Marquet:2007nf,Kowalski:2008sa,Kugeratski:2005ck,Bendova:2020hkp} in a large $Q^2$ kinematical approximation. In our terminology, this tree-level gluon emission is already a part of the NLO result, being explicitly proportional to $\as$. In this paper we will calculate the same contributions as in the Wüsthoff result at what we call the exact kinematics in the eikonal limit. This means that the kinematics within diffractive system is treated exactly without a large $Q^2$ approximation, while the interaction with the target is eikonal.

Our result presented in this paper corresponds to a subset of the NLO result that is finite by itself, and suited for explicit numerical evaluation.
The completion of the full NLO calculation requires the inclusion of virtual corrections with gluons that are not produced in the final state. We plan to return to these contributions in future work. Here we will merely outline steps that are needed to calculate these virtual contributions in our formalism. We have also here opted to calculate only the contributions where the gluon is emitted before the shockwave, not combining them with emissions after.
This avoids issues with collinear and soft divergences in final state emissions, 
which eventually cancel against virtual corrections.

 This paper is structured in the following way. We will start by introducing the experimental observable, the diffractive structure function, in Sec.~\ref{sec:observables} and the dipole picture formulation in terms of LCPT in Sec.~\ref{sec:nlodiffraction}. Before moving to specific diagrams we will then discuss in Sec.~\ref{sec:phasesp} the general strategy to calculate phase space integrals for 2- and 3-particle final states of a fixed invariant mass $M_X$ in the context of an eikonal scattering picture where the interaction with the target happens at a fixed transverse coordinate. We will rederive the known result for the leading $q\Bar{q}$ component of the wavefunction in our notations in Sec.~\ref{sec:lof2d}. We then move to the main new result of this paper, the calculation of the $q\Bar{q}g$ component of the cross section in full kinematics in Sec.~\ref{sec:gluonradiation}. A more detailed exposition of intermediate stages of the calculation has been presented earlier in Ref.~\cite{Hanninen:2021byo}. We  check in Sec.~\ref{sec:limits} that our calculation  reduces to known results in the in the kinematical limits of large $Q^2$ (Wüsthoff~\cite{Wusthoff:1997fz})   and large $M_X$ (e.g. in Ref.~\cite{Munier:2003zb}), before concluding in Sec.~\ref{sec:conc}. The results in this paper cover a finite, self-contained subset of the NLO corrections to the diffractive results, generalising earlier calculations to the full kinematics. We plan to return to the calculation of the remaining parts in a future publication, as outlined in Sec.~\ref{sec:nlodiffraction}.

\section{Diffractive structure functions}
\label{sec:observables}

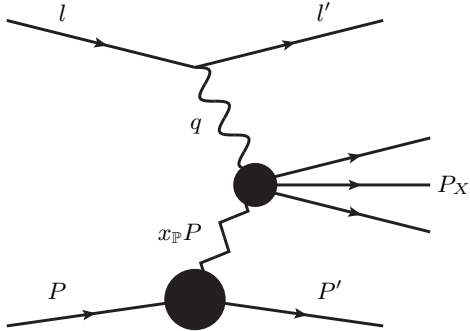
\begin{figure}
    \centering
    \includegraphics[width=6.00cm]{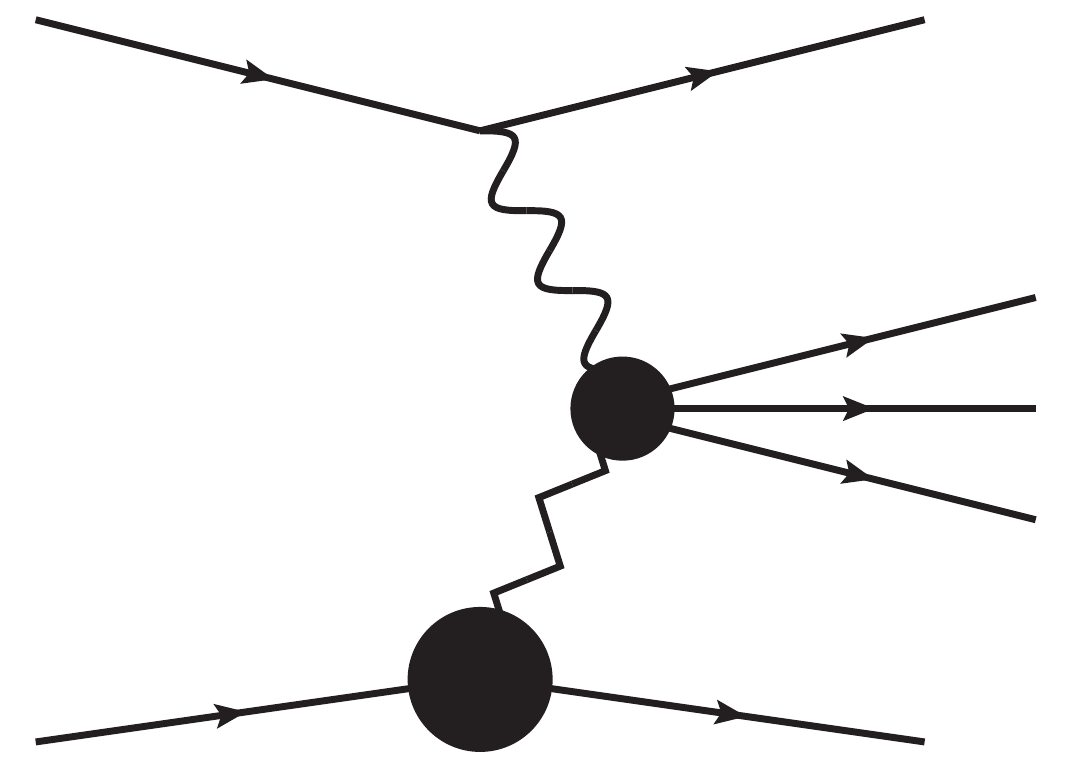}
    \begin{tikzpicture}[overlay]
         \node[anchor=south east] at (-5cm,4.1cm) {$l$};
         \node[anchor=south west] at (-1.9cm,4.1cm) {$l'$};
         \node[anchor=south east] at (-3.2cm,2.6cm) {$q$};
         \node[anchor=south east] at (-3.2cm,1.15cm) {$\xpom P$};
         \node[anchor=south west] at (-0.3cm,1.8cm) {$P_X$};
         \node[anchor=south east] at (-5cm,0.4cm) {$P$};
         \node[anchor=south west] at (-1.9cm,0.4cm) {$P'$};
    \end{tikzpicture}
    \caption{Kinematics of inclusive diffractive DIS.
    }
    \label{fig:ddis_kin}
\end{figure}

The diffractive cross section $\sigma^\textrm{D}_{e + A \to M_X + p}$ in electron-nucleus (or electron-proton) DIS integrated over the squared momentum transfer $t$ is usually expressed in terms of the diffractive structure functions $F_2^\textrm{D}$ and $F_L^\textrm{D}$ defined as
\begin{equation}
\label{eq:diffxs_F2D}
    \frac{ \ud \sigma^\textrm{D}_{e + A \to M_X + p}}{\ud \beta \ud Q^2 \ud \xpom} = \frac{2\pi \alpha_\mathrm{em}^2}{\beta Q^4} \qty[1+(1-y)^2] \left[ F_2^{\textrm{D}(3)}(\beta,Q^2,\xpom) - \frac{y^2}{1+(1-y)^2} F_L^\textrm{D(3)}(\beta,Q^2,\xpom) \right].
\end{equation}
The superscript $(3)$ refers to the structure functions that depend on three variables, in this case $\beta,Q^2$ and $\xpom$ discussed above. One can also consider structure functions differentially in the squared momentum transfer $t$, in which case one has the structure functions $F_{2,L}^{\textrm{D}(4)}(\beta,Q^2,\xpom,t)$. In this work we consider both $t$-differential and $t$-integrated cross sections. For simplicity we focus here on coherent diffraction which corresponds to the events where the target does not dissociate, but our results are straightforward to generalize to dissociative events in the Good-Walker~\cite{Good:1960ba} picture (see e.g. Refs.~\cite{Miettinen:1978jb,Caldwell:2009ke,Mantysaari:2016ykx,Mantysaari:2020axf}).

The diffractive structure functions are related to the total diffractive cross sections in $\gamma^* + A$ scattering as
\begin{align}
\label{eq:structurefun_diffxs}
    \xpom F_{T,L}^{\textrm{D}(4)}
    =
    \frac{Q^2}{4\pi^2 \aem} \frac{Q^2}{\beta} \frac{\ud \sigma^\textrm{D}_{\gamma^*_{T,L} + A \to M_X + A}}{\ud \mx^2 \ud t},
\end{align}
where $T$ and $L$ refer to transversely and longitudinally polarized photons.
The experimentally measured~\cite{Aktas:2006hy,Aktas:2006hx,Aktas:2006up,Aaron:2012hua} total diffractive cross sections are usually reported in terms of the diffractive reduced cross section
\begin{equation}
    \sigma^{\textrm{D}(3)}_r(\beta,Q^2\xpom) = F_2^{\textrm{D}(3)}(\beta,Q^2\xpom) - \frac{y^2}{1+(1-y)^2} F_L^{\textrm{D}(3)}(\beta,Q^2\xpom).    
\end{equation}

Here the Lorentz-invariant quantities describing the kinematics are the virtuality of the photon $-Q^2$ and the fraction of the target longitudinal momentum $\xpom$ carried by the pomeron (exchanged in the scattering process) in the frame where the target has a large longitudinal momentum, defined as
\begin{equation}
    \xpom \equiv \frac{(P-P')\cdot q}{P \cdot q} = \frac{M^2+Q^2-t}{W^2+Q^2-m_N^2} \approx \frac{\mx^2+Q^2}{W^2+Q^2}.
\end{equation}
The invariant mass of the diffractively produced system is denoted by $\mx^2$ and the nucleon mass by $m_N^2$. The variable $\beta = Q^2/(2q \cdot (P-P')) \approx Q^2/(\mx^2+Q^2)$ has, in the frame where the target momentum is large, an interpretation as the fraction of the pomeron momentum carried by the struck quark. The four vectors $P,P'$ are the target nucleon momenta before and after the scattering respectively, see Fig.~\ref{fig:ddis_kin}. Finally $y = (P \cdot q)/(P \cdot l)$ is the inelasticity describing the energy transfer from the lepton with initial momentum $l$, and $q$ is the photon momentum.

We are working in the dipole picture, where one develops the virtual photon state in a series of partonic Fock states.  Let us first consider the general case of an $n$-parton Fock state, for which 
we denote the phase space element as $[\mathcal{PS}]_n$, 
At leading order only the $n=q\Bar q$ state contributes, and in this work we focus on including the $n=q\Bar q g$ contribution which is actually the dominant component at high $\mx^2$ (small $\beta$) and at high $Q^2$~\cite{Kowalski:2008sa}. This tree-level contribution is also a necessary ingredient for the future full calculation of the diffractive structure functions at NLO accuracy. The diffractive cross section can be written as
\begin{equation}
    \frac{\ud \sigma^{\textrm{D}}_{\gamma^*_{T,L} + A \to M_X+A}}{\ud \mx^2}
    = \sum_{n}\int \ud[\mathcal{PS}]_n 
    \frac{\ud \sigma^{\gamma^*_{T,L} + A \to A+n}}{\ud [\mathcal{PS}]_n} \delta(\mx^2 - M^2_n),
\end{equation}
where $M_n^2$ is the invariant mass of the Fock state $n$. The cross section
$\frac{\ud \sigma^{\gamma^ * A \to A+n}}{\ud [\mathcal{PS}]_n}$ for a production of a color-singlet state $n$ in photon-nucleus or photon-proton scattering is expressed in terms of the scattering amplitudes (see \cite{Bjorken:1970ah}, except we now normalize with a $2q^+$ in a different place) as:
\begin{equation}
\label{eq:diffxs_An}
\ud \sigma_{\gamma^* + A \to A+n}^\mathrm{D} = 
2q^+ (2\pi)\delta(q^+ - q^+_n) \prod_{i \in \textrm{F.s.} n} \widetilde{\ud p_i}
\left|\mathcal{M}_{\gamma \to n}\right|^2.
\end{equation}
Here $i \in \textrm{F.s.} n$ means iterating over all the particles $i$ in the Fock state $n$ and $q_n^+$ is the total plus momentum of the partons in this Fock state.
The one particle phase space element reads
\begin{equation} 
\widetilde{\ud p_i} \equiv \frac{\ud^ 2 \pt_i \ud p_i^+}{2p_i^ + (2\pi)^3}.
\end{equation}
The scattering amplitude is obtained from the matrix elements of dressed, interacting, states by leaving out a momentum conservation delta function
\begin{equation} 
\label{eq:generic_amplitude}
{_D}\bra{\textrm{F.S.}\, n} \hat{S}-1 \ket{\gamma}_D
= 2q^+(2\pi)\delta(q^+_\gamma - q^+_n)\mathcal{M}_{\gamma \to n},
\end{equation}
where in the diffractive scattering the final state $\textrm{F.S.} \, n$ is a color singlet.

At high energies, the scattering amplitudes $\mathcal{M}_{\gamma \to n}$ can be calculated by considering the $\gamma \to n$ process, and inserting an interaction with the shockwave (target color field) in all possible ways. At high energies the transverse coordinates of the partons are fixed when they propagate in the color field of the target and as such the interaction can be straightforwardly described in terms of Wilson lines at fixed transverse coordinates as discussed in more detail in Sec.~\ref{sec:kinematics_fock_states}.

\section{Outline of NLO calculation}
\label{sec:nlodiffraction}

We will in this paper compute a part of the NLO correction to the diffractive structure functions that is finite by itself. However, let us first discuss the overall structure of the full NLO contribution in terms of the contributing diagrams. This discussion will make it more clear which parts of the NLO contributions are included in our result here, and what still needs to be done to calculate the rest. We are calculating, and drawing diagrams, for an exclusive amplitude for a virtual photon-target shockwave scattering. Thus a single diagram includes both the Fock state expansion of the incoming dressed virtual photon state $\left|\gamma\right>_D$, and that of the outgoing dressed multiparton state ${_D}\left< \textrm{F.s.}n \right|$ in the amplitude \nr{eq:generic_amplitude}. In the diagrams the shockwave is represented by a blue band, with time progressing from left to right. The state furthest to the left is the asymptotic incoming state (the dressed virtual photon), which then develops into a superposition of bare parton states, corresponding to the LFWFs $\gamma\to n$. The interaction with the shockwave  then is given in terms of the bare states~\cite{Kogut:1969xa}. On the other side of the shockwave, furthest to the right, is the asymptotic final state ${_D}\left< \textrm{F.s.}n \right|$, which develops into a superposition of bare states, going leftward in the figure. Thus the part of the figure to the right of the shockwave corresponds to the complex conjugate of the LFWF of the final state, in particular with energy denominators calculated with respect to the final state.  

\begin{figure*}[tb!]
\centerline{
\includegraphics[width=6.00cm]{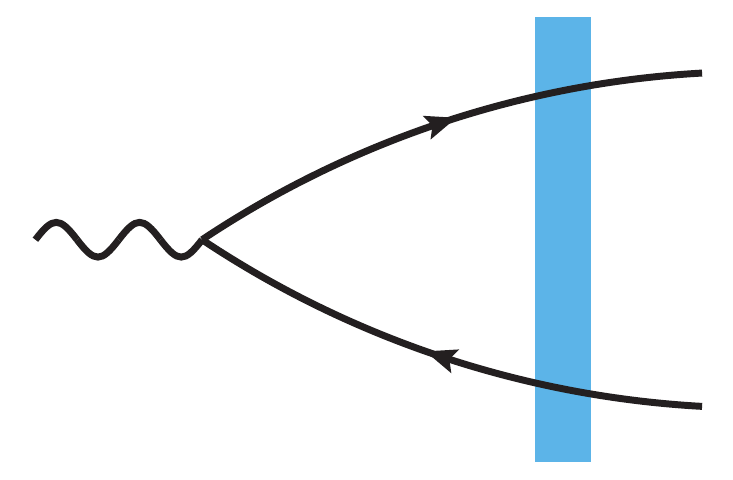}
\begin{tikzpicture}[overlay]
 \node[anchor=south east] at (-5cm,2.1cm) {$q,\lambda$};
 \node[anchor=south west] at (-1.2cm,2.6cm) {$z_0, \pt_0, \xt_0$};
  \node[anchor=south west] at (-1.2cm,0.8cm) {$z_1, \pt_1, \xt_1$};
\node[anchor=south west] at (-7cm,0cm) {\namediag{diag:loampli}};
\end{tikzpicture}
}
\caption{Leading order amplitude. The blue band represents the interaction with the target color field.}
\label{fig:ddis_loampli}
\end{figure*}
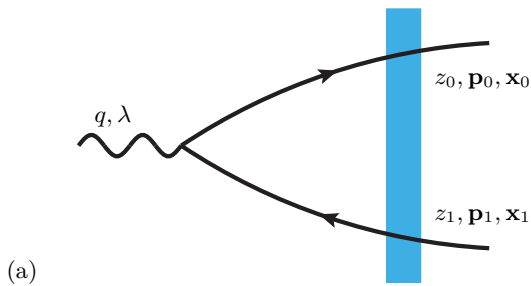

At leading order, the only diagram contributing to the scattering amplitude is diagram~\ref{diag:loampli} shown in \fig\ref{fig:ddis_loampli}, where the photon first splits to a $q\Bar q$ dipole, and then subsequently the quarks scatter off the target color field with no net color charge transfer to the target. In momentum space, we denote the quark and antiquark transverse momenta as $\pt_0$ and $\pt_1$, respectively, and in transverse coordinate space use the coordinates $\xt_0,\xt_1$. Similarly the fractions of the photon plus momentum carried by the quark and the antiquark are $z_i = p_i^+/q^+$ with $z_0+z_1=1$.

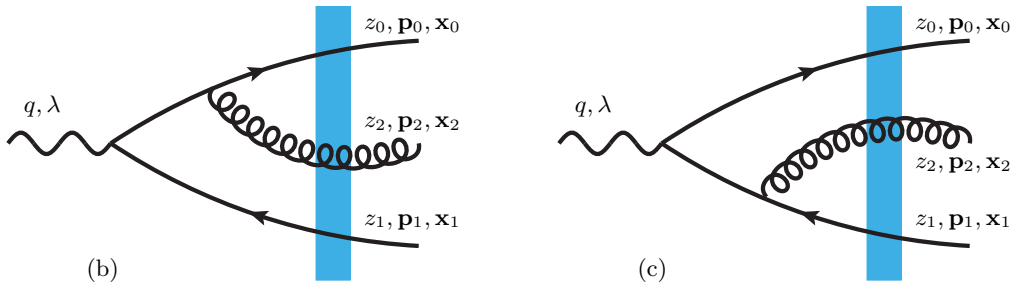
\begin{figure*}[tb!]
\centerline{
\includegraphics[width=6.00cm]{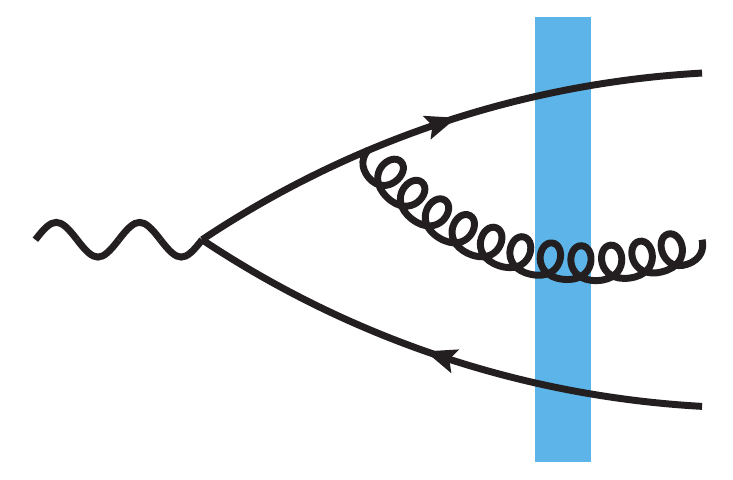}
\begin{tikzpicture}[overlay]
 \node[anchor=south east] at (-5cm,2.2cm) {$q,\lambda$};
 \node[anchor=south west] at (-1.2cm,3.3cm) {$z_0, \pt_0, \xt_0$};
 \node[anchor=south west] at (-1.2cm,2.cm) {$z_2, \pt_2, \xt_2$};
 \node[anchor=south west] at (-1.2cm,0.7cm) {$z_1, \pt_1, \xt_1$};
\node[anchor=south west] at (-5cm,0cm) {\namediag{diag:gwavef1}};
\end{tikzpicture}
\rule{1cm}{0pt}
\includegraphics[width=6.00cm]{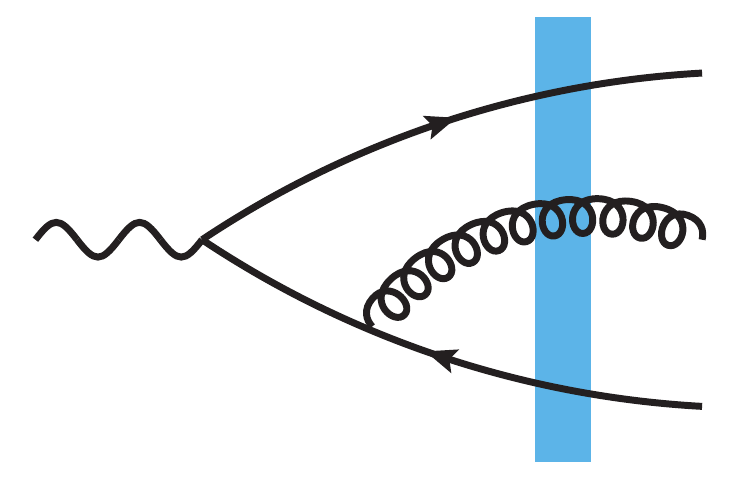}
\begin{tikzpicture}[overlay]
 \node[anchor=south east] at (-5cm,2.2cm) {$q,\lambda$};
 \node[anchor=south west] at (-1.2cm,3.3cm) {$z_0, \pt_0, \xt_0$};
 \node[anchor=south west] at (-1.2cm,1.45cm) {$z_2, \pt_2, \xt_2$};
 \node[anchor=south west] at (-1.2cm,0.7cm) {$z_1, \pt_1, \xt_1$};
\node[anchor=south west] at (-5cm,0cm) {\namediag{diag:gwavef2}};
\end{tikzpicture}
}
\caption{Gluon emission before the shock.}
\label{fig:ddis_emission_before}
\end{figure*}

\begin{figure*}[tb!]
\centerline{\includegraphics[width=6cm]{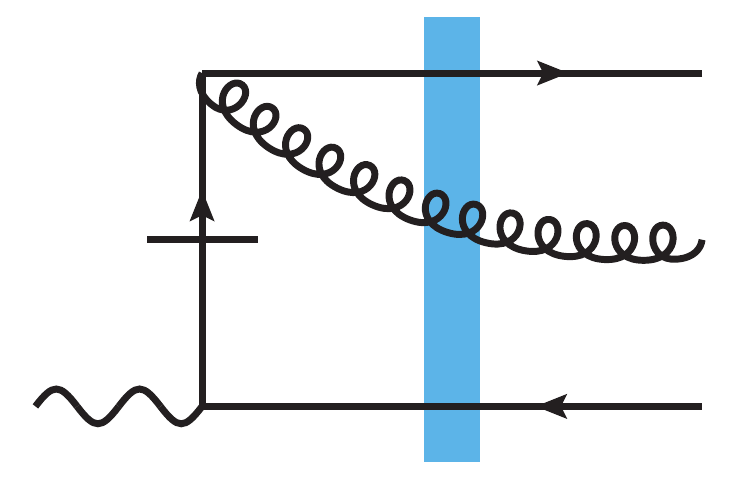}
\begin{tikzpicture}[overlay]
\node[anchor=south west] at (-6cm,3cm) {\namediag{diag:gwavefinst1}};
\end{tikzpicture}
\rule{2em}{0pt}
\includegraphics[width=6cm]{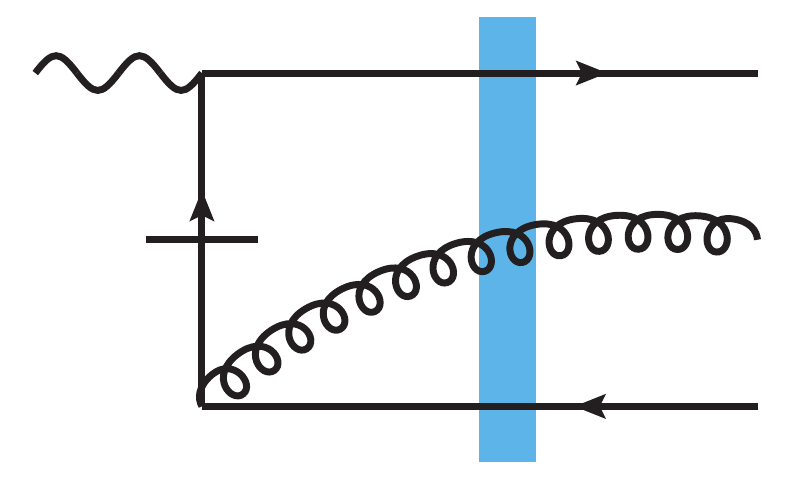}
\begin{tikzpicture}[overlay]
\node[anchor=south west] at (-6cm,0cm) {\namediag{diag:gwavefinst2}};
\end{tikzpicture}
}
\caption{
Instantaneous diagram gluon emission wavefunction with emission before the cut.
}\label{fig:ddis_instquark}
\end{figure*}
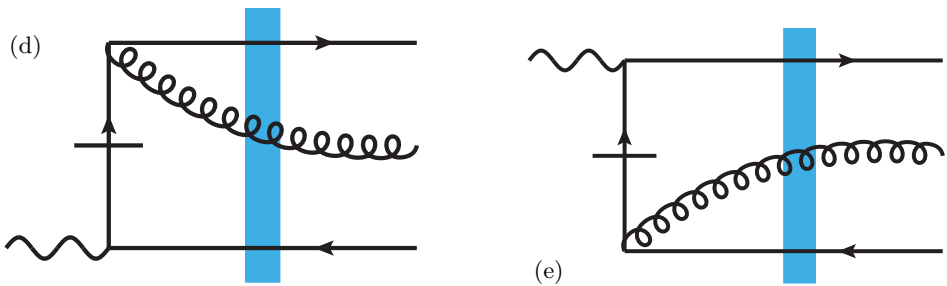

\begin{figure*}[tb!]
\centerline{
\includegraphics[width=6.00cm]{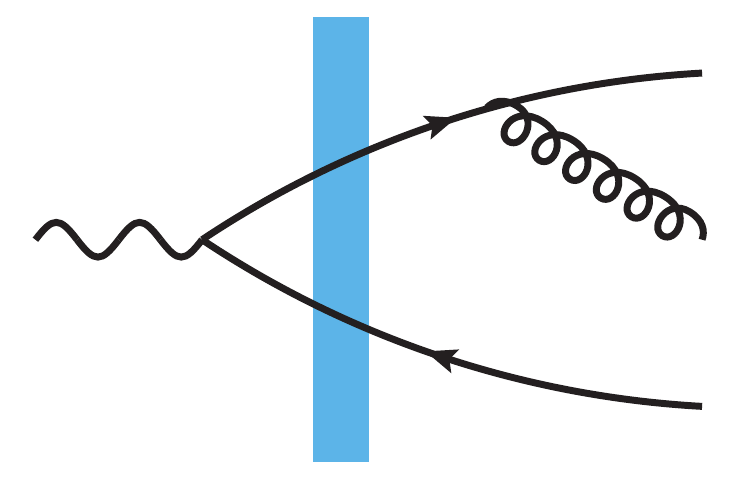}
\begin{tikzpicture}[overlay]
 \node[anchor=south east] at (-5cm,2.2cm) {$q, \lambda$};
 \node[anchor=south west] at (-3.cm,2.95cm) {$\xt_0^\prime$};
 \node[anchor=south west] at (-1.2cm,3.3cm) {$z_0, \pt_0, \xt_0$};
 \node[anchor=north west] at (-1.2cm,1.95cm) {$z_2, \pt_2, \xt_2$};
 \node[anchor=north west] at (-1.2cm,1.1cm) {$z_1, \pt_1, \xt_1$};
\node[anchor=south west] at (-5cm,0cm) {\namediag{diag:gwaveffs1}};
\end{tikzpicture}
\rule{1cm}{0pt}
\includegraphics[width=6.00cm]{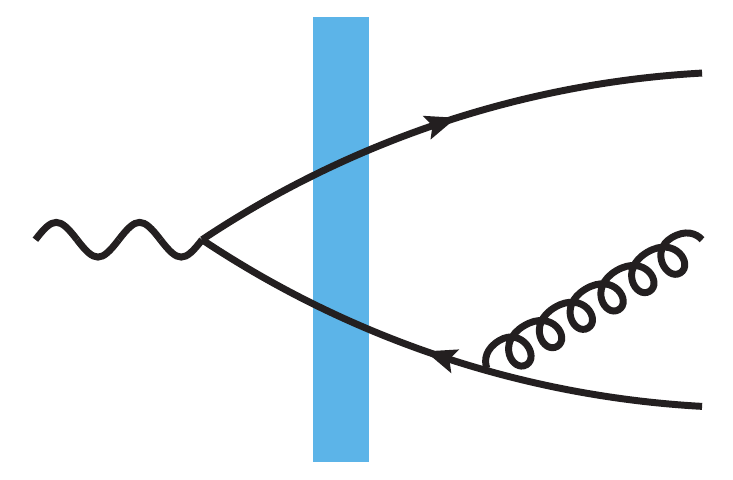}
\begin{tikzpicture}[overlay]
 \node[anchor=south east] at (-5cm,2.2cm) {$q, \lambda$};
 \node[anchor=south west] at (-3.cm,1.15cm) {$\xt_1^\prime$};
 \node[anchor=south west] at (-1.2cm,3.3cm) {$\pt_0, \xt_0$};
 \node[anchor=north west] at (-1.2cm,2.5cm) {$\pt_2, \xt_2$};
 \node[anchor=north west] at (-1.2cm,1.1cm) {$\pt_1, \xt_1$};
\node[anchor=south west] at (-5cm,0cm) {\namediag{diag:gwaveffs2}};
\end{tikzpicture}
}
\caption{Gluon emission after shock. Note that the intermediate transverse coordinate of the quark (antiquark) before the gluon emission is not the same as the coordinate at the cut. They are defined by $\xt_0^\prime \coloneqq \frac{z_0 \xt_0 + z_2 \xt_2}{z_0 + z_2}$ and $\xt_1^\prime \coloneqq \frac{z_1 \xt_1 + z_2 \xt_2}{z_1 + z_2}$.
}
\label{fig:ddis_emission_after}
\end{figure*}
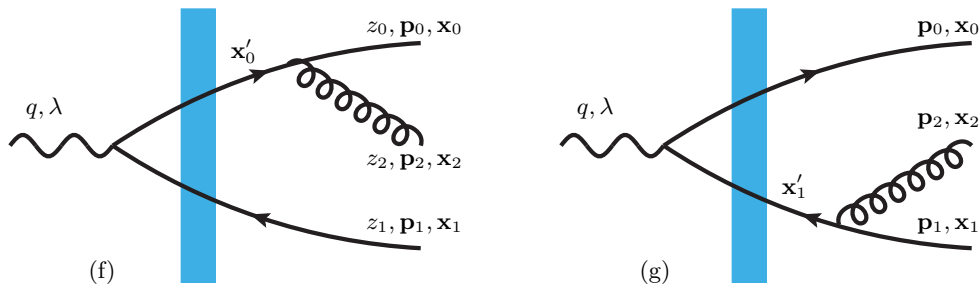

\subsection{Radiative corrections}
\label{sec:radnlo}

The purpose of this paper is to calculate the gluon emission part of the next-to-leading order contributions to the diffractive structure functions, which dominates at large $\mx^2$. 
In the CGC, if one integrates the transverse momentum of several final state particles without restriction, one can encounter spurious UV divergences in real higher order corrections, associated with the breakdown of the eikonal approximation. This occurs
when the light-cone momentum $p^-$ scales associated with the produced system become comparable or larger than that of the target.
In the case of diffractive structure functions considered here, the fixed invariant mass of the produced $q\Bar{q} g$ state ensures that the eikonal approximation stays valid for the whole integration range, by constraining the  $p^-$ scale of the diffractive system. 
For that reason, no UV divergence can arise in real NLO corrections to diffractive structure functions. 
This is to be compared to the case of  inclusive DIS~\cite{Beuf:2016wdz,Beuf:2017bpd,Hanninen:2017ddy} where one uses the optical theorem and thus the final state is completely fixed to be the same as the initial one. In dijet production~\cite{Boussarie:2014lxa,Boussarie:2016ogo,Caucal:2021ent,Taels:2022tza,Iancu:2021rup}, on the other hand, one typically fixes the momenta of some of the final state particles and integrates over the others. This can lead to a different pattern of cancellations between diagrams. The calculation of the loop corrections is left for future work, but for completeness we list all the relevant diagrams in the following subsection~\ref{sec:loop}.

In order to calculate the $n=q\Bar qg$ contribution to the diffractive scattering amplitude we include gluon emission contributions from both the quark and the antiquark.
There can be a regular gluon emission before the shockwave shown in diagrams~\ref{diag:gwavef1} and \ref{diag:gwavef2} in Fig.~\ref{fig:ddis_emission_before}.
In light cone perturbation theory there is also an instantaneous  $\gamma \to q\Bar q g$ vertex resulting in instantaneous gluon emission diagrams~\ref{diag:gwavefinst1} and \ref{diag:gwavefinst2} in Fig.~\ref{fig:ddis_instquark}. These are the contributions  that we will calculate in detail in this paper. Similarly the gluon can be emitted after the shock, see diagrams~\ref{diag:gwaveffs1} and \ref{diag:gwaveffs2} in Fig.~\ref{fig:ddis_emission_after}.  There are several relations between these diagrams.
Firstly, since the virtual photon as a whole is color neutral,  there is a destructive interference between emissions from the quark and from the antiquark. This  cancels the leading small transverse momentum (collinear) divergence and serves as a useful check of the relative sign between the contributions.  

The contributions with emissions after the shockwave, pictured in \fig\ref{fig:ddis_emission_after}, can be conveniently obtained from the corresponding ones in \fig\ref{fig:ddis_emission_before} where the emission happens before by taking a specific coordinate limit, in a procedure developed in Ref.~\cite{Iancu:2020mos}. Here one first separates out from the coordinate space $\gamma^* \to q\Bar{q}g$ wavefunction  a piece corresponding to the final gluon emission. In terms of equations this means that one writes the 
$\gamma^* \to q\Bar{q}g$ wavefunction obtained from diagrams~\ref{diag:gwavef1}, \ref{diag:gwavef2}, \ref{diag:gwavefinst1} and \ref{diag:gwavefinst2} in a factorized form as\footnote{Here $\widetilde{\psi}$ denotes a reduced coordinate space wavefunction. See \eqs\nr{eq:lfwf-qqbar}, \nr{eq:lfwf-qqbarg} for the normalization in the $\gamma^* \to \qqb$, $\gamma^* \to \qqbg$ case; the convention is trivially extended to the $1\to 2$ gluon emission case.}
\begin{equation}\label{eq:splitforsubtr}
  \widetilde{\psi}_{\gamma^{*}_\lambda \rightarrow q_0 \Bar{q}_1 g_2}
=     
 \widetilde{\psi}_{\gamma^{*}_\lambda \rightarrow q_0 \Bar{q}_1 ; q_0  \rightarrow q_0  g_2}
\widetilde{\psi}_{ q_0  \rightarrow q_0  g_2}
  +
\widetilde{\psi}_{\gamma^{*}_\lambda \rightarrow q_0 \Bar{q}_1 ; \Bar{q}_1  \rightarrow \Bar{q}_1  g_2}
 \widetilde{\psi}_{ \Bar{q}_1  \rightarrow \Bar{q}_1  g_2}. 
\end{equation}
Here $\widetilde{\psi}_{ q_0  \rightarrow q_0  g_2}$ and  $\widetilde{\psi}_{ \Bar{q}_1  \rightarrow \Bar{q}_1  g_2}$ are the $1\to 2$ particle gluon emission wavefunctions. 
Equation~\nr{eq:splitforsubtr} should be understood as the \emph{definition} of  the remaining parts $\widetilde{\psi}_{\gamma^{*}_\lambda \rightarrow q_0 \Bar{q}_1 ; q_0  \rightarrow q_0  g_2} $ and
$ \widetilde{\psi}_{\gamma^{*}_\lambda \rightarrow q_0 \Bar{q}_1 ; \Bar{q}_1  \rightarrow \Bar{q}_1  g_2}$.
 Here the notation  refers to these being the parts of the wavefunction that are associated (e.g. by the helicity structure) with the first $\gamma^* \to q\Bar{q}$ splitting ($\gamma^{*}_\lambda \rightarrow q_0 \Bar{q}_1$), but depend on the fact that the (anti)quark will later emit a gluon ($q_0  \rightarrow q_0  g_2,$
$\Bar{q}_1  \rightarrow \Bar{q}_1  g_2$). Thus  $\widetilde{\psi}_{\gamma^{*}_\lambda \rightarrow q_0 \Bar{q}_1 ; q_0  \rightarrow q_0  g_2} $ and
$ \widetilde{\psi}_{\gamma^{*}_\lambda \rightarrow q_0 \Bar{q}_1 ; \Bar{q}_1  \rightarrow \Bar{q}_1  g_2}$
depend on the coordinates of all three particles in the final state.
The calculation of the gluon emission diagrams after the shock wave requires the $q\Bar{q}$ component in the dressed ${}_D\bra{q\Bar{q}g}$ state. This, in turn, requires the $(q\Bar{q}g\to q\Bar{q})^\dag$ merging wavefunction, which is given by (minus) the hermitian conjugates of the corresponding emission wavefunctions $\widetilde{\psi}_{ q_0  \rightarrow q_0  g_2}$ and  $\widetilde{\psi}_{ \Bar{q}_1  \rightarrow \Bar{q}_1  g_2}$. In other words, one is here factoring out from the gluon emission before the shockwave the gluon emission wavefunction that appears when the gluon is emitted after.  A look at the transverse  coordinate space $\gamma\to q\Bar{q}g$ wavefunction (see e.g. \eqs(C14) and (C19) in Ref.~\cite{Hanninen:2017ddy} for explicit expressions) shows that indeed the structure of the regular emission wavefunctions naturally factorizes like this. 

Using the factorized notation~\nr{eq:splitforsubtr}, the procedure  of Ref.~\cite{Iancu:2020mos} for obtaining amplitudes for the emission-after-the-shock contributions in \fig\ref{fig:ddis_emission_after} is the following. One evaluates both the Wilson line operators and the $\gamma^*\to q\Bar{q}$ parts of the wavefunctions $\widetilde{\psi}_{\gamma^{*}_\lambda \rightarrow q_0 \Bar{q}_1 ; q_0  \rightarrow q_0  g_2}, $ $ \widetilde{\psi}_{\gamma^{*}_\lambda \rightarrow q_0 \Bar{q}_1 ; \Bar{q}_1  \rightarrow \Bar{q}_1  g_2}$
with the transverse coordinates of the gluon and its parent  (anti)quark replaced by the coordinate of the parent before the emission, and changes the sign. Thus, for the emission from the quark, diagram~\ref{diag:gwavef1}, one replaces $\xt_0\to \xt_0'$ and $\xt_2\to \xt_0'$, in both the Wilson line operator and in  $\widetilde{\psi}_{\gamma^{*}_\lambda \rightarrow q_0 \Bar{q}_1 ; q_0  \rightarrow q_0  g_2}$, with the coordinate defined as $\xt_0^\prime \coloneqq \frac{z_0 \xt_0 + z_2 \xt_2}{z_0 + z_2}$.
Correspondingly, for the emission from the antiquark, diagram~\ref{diag:gwavef2}, one replaces $\xt_1\to \xt_1'$ and $\xt_2\to \xt_1'$, with $\xt_1^\prime \coloneqq \frac{z_1 \xt_1 + z_2 \xt_2}{z_1 + z_2}$. It is clear that this corresponds to the Wilson line operator being evaluated at the correct coordinate for the diagrams~\ref{diag:gwaveffs1} and \ref{diag:gwaveffs2}. There is no ``emission after the shockwave'' contribution for the instantaneous diagrams~\ref{diag:gwavefinst1} and \ref{diag:gwavefinst2}. This is, however, built into the formalism of Ref.~\cite{Iancu:2020mos}, since it turns out that the parts of $\widetilde{\psi}_{\gamma^{*}_\lambda \rightarrow q_0 \Bar{q}_1 ; q_0  \rightarrow q_0  g_2}, $ $ \widetilde{\psi}_{\gamma^{*}_\lambda \rightarrow q_0 \Bar{q}_1 ; \Bar{q}_1  \rightarrow \Bar{q}_1  g_2}$ corresponding to the instantaneous diagrams vanish in the coordinate limits $\xt_0,\xt_2\to \xt_0'$ and $\xt_1,\xt_2\to \xt_1'$, respectively.

One can arrive at this procedure for constructing the final state emissions in multiple ways. In Ref.~\cite{Iancu:2020mos} it is derived explicitly by looking at the expressions and noticing a relation between the energy denominators of the different diagrams. More generally, using the orthogonality of the $|\gamma\rangle_D$ and $|q\Bar{q}g\rangle_D$ states one can derive the corresponding relation between the wavefunctions   for   $(q\Bar{q} g\rightarrow q\Bar{q})^\dag,$ $\gamma \to q\Bar{q}$ (diagrams~\ref{diag:gwaveffs1} and \ref{diag:gwaveffs2} in \fig\ref{fig:ddis_emission_after}), $\gamma \to q\Bar{q}g$ (diagrams~\ref{diag:gwavef1}, \ref{diag:gwavef2}, ~\ref{diag:gwavefinst1} and \ref{diag:gwavefinst2} in \figs~\ref{fig:ddis_emission_before}, \ref{fig:ddis_instquark}) and the process $(q\Bar{q} g\rightarrow \gamma)^\dag$, corresponding to the photon crossing the shockwave first and all the emissions happening after (i.e. the bare photon state in the final ${}_D\bra{\qqbg}$). Since the last one does not contribute to the amplitude because the photon is color neutral, one obtains a linear relation for the contributions of the emission diagrams of Figs.~\ref{fig:ddis_emission_before} and \ref{fig:ddis_emission_after}. As discussed above, one can also see this relation directly by looking at the coordinate space wavefunctions, using the Fourier transforms from e.g. Appendix C of~\cite{Hanninen:2017ddy}.

In an inclusive observable where one integrates over the momenta of the final state gluon and of its parent without any restrictions, it would be natural to always keep the the initial and final state gluon emissions, Figs.~\ref{fig:ddis_emission_before} and \ref{fig:ddis_emission_after}, together because they have a tendency to cancel each other in the UV region where the gluon is at the same coordinate as the emitting quark. 
Thus, they are often evaluated together such as in Refs.~\cite{Chirilli:2012jd,Iancu:2020mos}. 
However, for the case of the diffractive structure function, the restriction on the diffractive system mass $M_X$ cuts out contributions of large transverse momenta. 
Thus it is quite natural to evaluate the contributions of the emissions before and after the shockwave separately.  On the other hand, the final state gluon emissions are associated with the wavefunction renormalization constants for, and gluon exchanges between, the outgoing quarks. The relation to these contributions which  are, in our language, a part of the $q\Bar{q}$ part of the cross section is especially important for the kinematical region when the gluon becomes collinear to the quark, where corresponding IR divergences must cancel. Thus it would not be natural here to consider the diagrams with gluon emission after the shockwave, before taking into account all the loop corrections. 
In conclusion,  for a final state with a fixed $M_X$, the natural way to group diagrams together is different from some other observables. Since we are here leaving the NLO  $q\Bar{q}$  contribution overall to a future paper, we will also not calculate the final state emission contributions here. The exception to this is in Sec~\ref{sec:mslimit}, where one works in the $M_X\to \infty$ limit neglecting the restriction on final state momenta, and thus only the inclusion of the final state emissions allows one  to get a finite result.

\begin{figure*}[tb!]
\centerline{\includegraphics[width=3cm]{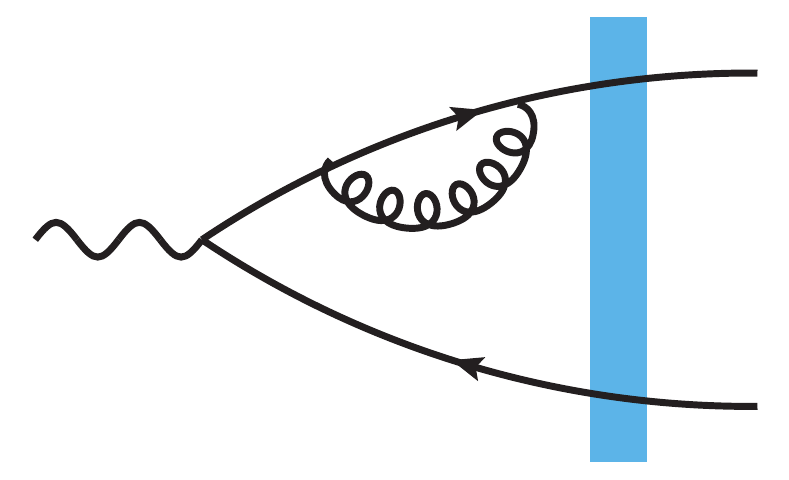}
\begin{tikzpicture}[overlay]
\node[anchor=south west] at (-3cm,0cm) {\namediag{diag:prop1}};
\end{tikzpicture}
\rule{2em}{0pt}
\includegraphics[width=3cm]{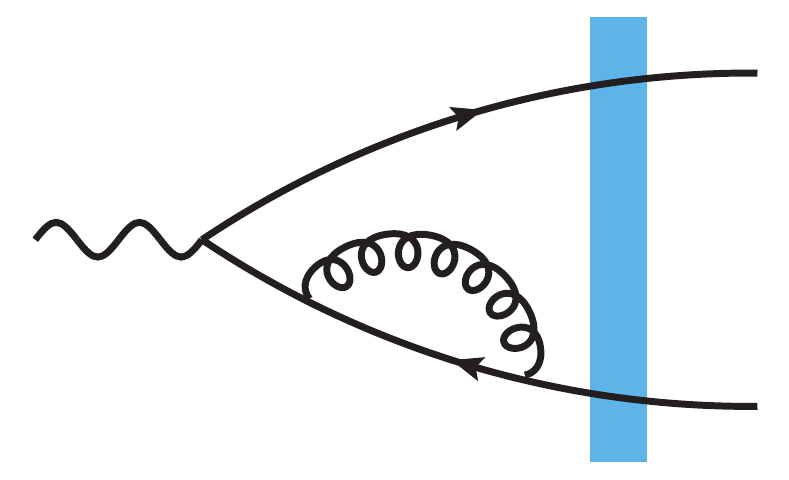}
\begin{tikzpicture}[overlay]
\node[anchor=south west] at (-3cm,0cm) {\namediag{diag:prop2}};
\end{tikzpicture}
\includegraphics[width=3cm]{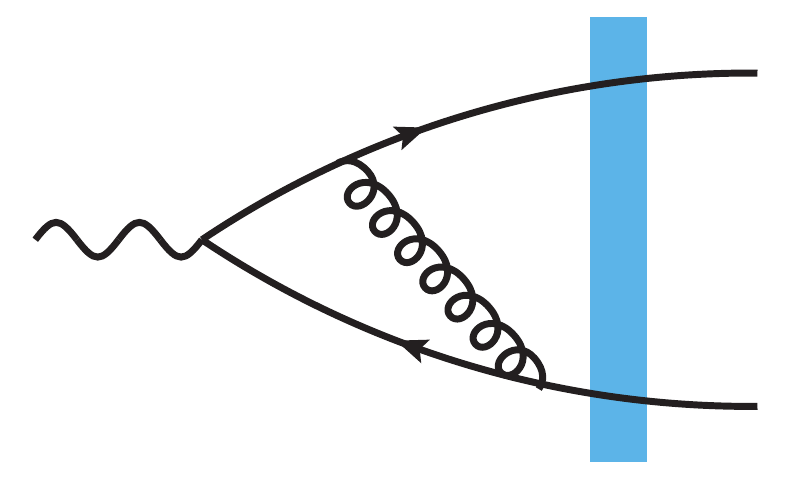}
\begin{tikzpicture}[overlay]
\node[anchor=south west] at (-3cm,0cm) {\namediag{diag:vertex1}};
\end{tikzpicture}
\includegraphics[width=3cm]{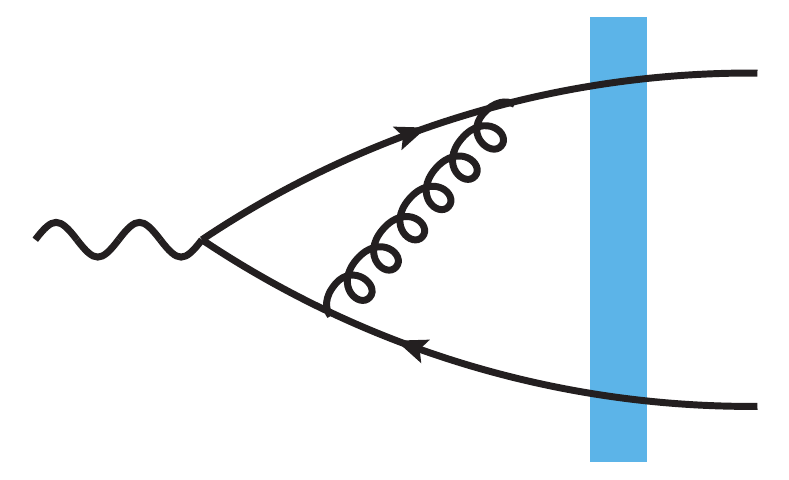}
\begin{tikzpicture}[overlay]
\node[anchor=south west] at (-3cm,0cm) {\namediag{diag:vertex2}};
\end{tikzpicture}
}
\caption{
Propagator and normal vertex correction diagrams calculated in \cite{Beuf:2017bpd,Hanninen:2017ddy}
}\label{fig:ddis_vertexprop}
\end{figure*}
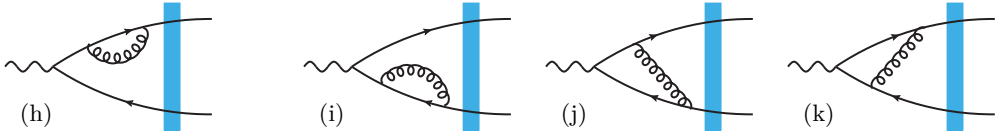

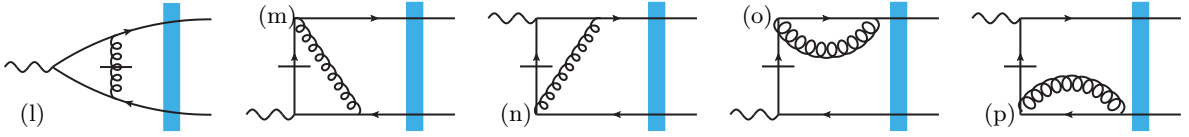
\begin{figure*}[tb!]
\centerline{
\includegraphics[width=3cm]{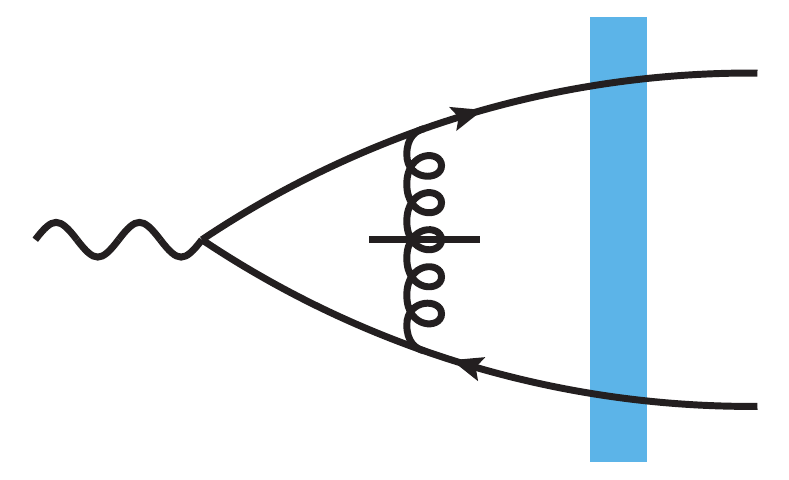}
\begin{tikzpicture}[overlay]
\node[anchor=south west] at (-3cm,0cm) {\namediag{diag:vertexi}};
\end{tikzpicture}
\includegraphics[width=3cm]{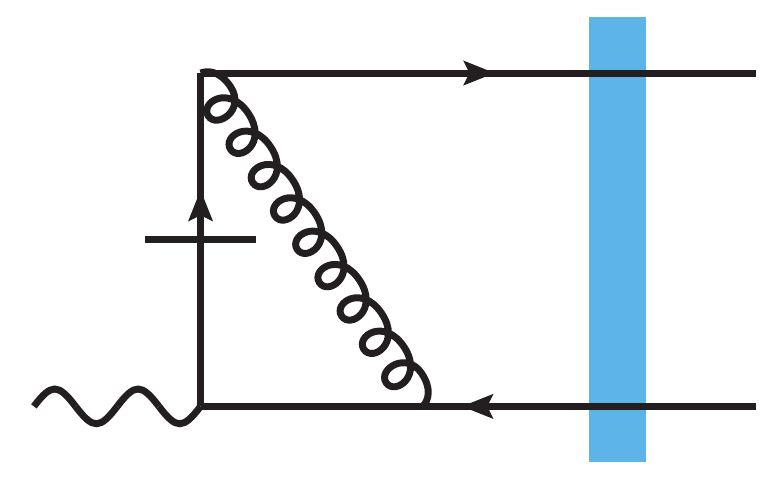}
\begin{tikzpicture}[overlay]
\node[anchor=south west] at (-3.05cm,1.3cm) {\namediag{diag:vertexiq1}};
\end{tikzpicture}
\includegraphics[width=3cm]{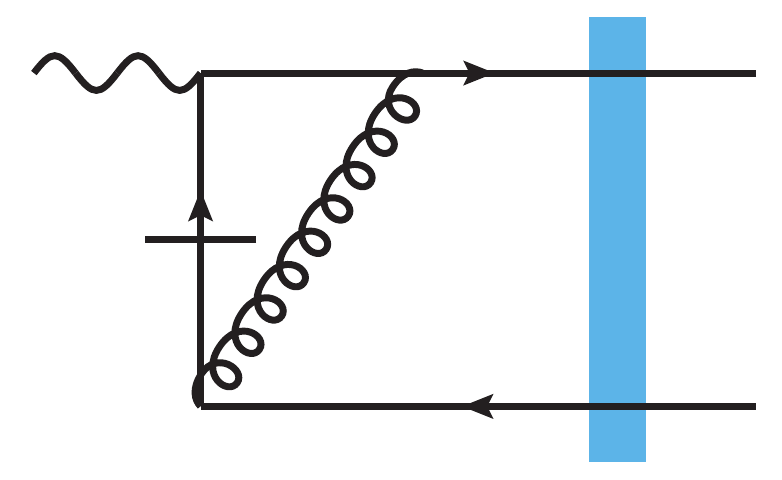}
\begin{tikzpicture}[overlay]
\node[anchor=south west] at (-3.05cm,0cm) {\namediag{diag:vertexiq2}};
\end{tikzpicture}
\includegraphics[width=3cm]{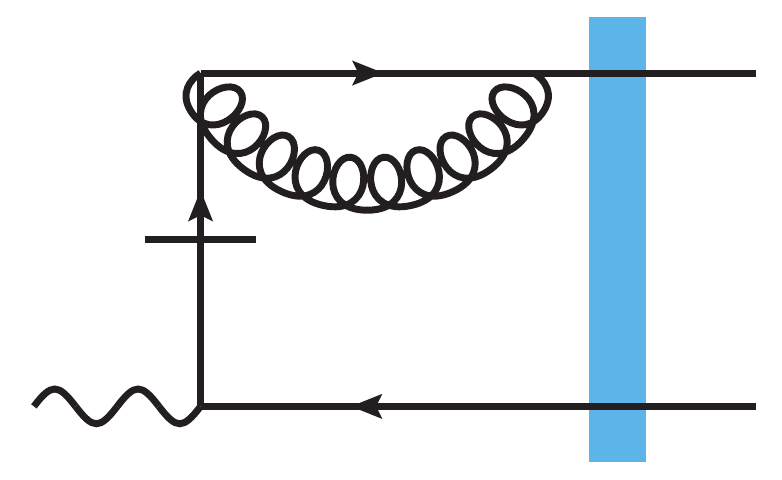}
\begin{tikzpicture}[overlay]
\node[anchor=south west] at (-3.05cm,1.3cm) {\namediag{diag:propiq1}};
\end{tikzpicture}
\includegraphics[width=3cm]{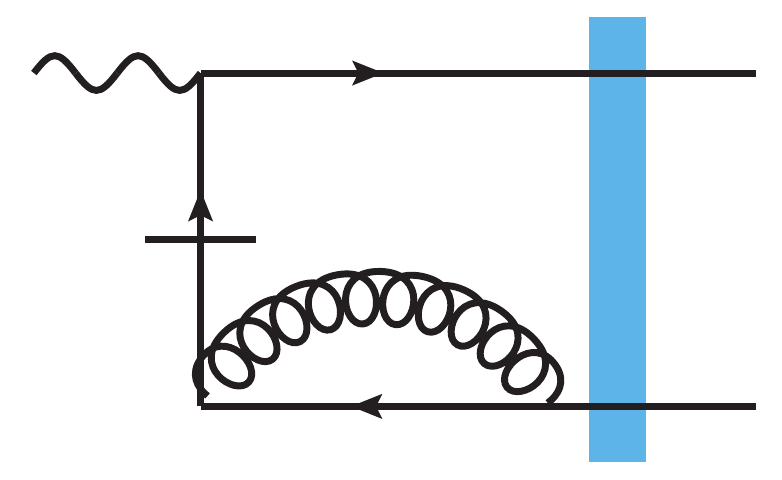}
\begin{tikzpicture}[overlay]
\node[anchor=south west] at (-3.05cm,0cm) {\namediag{diag:propiq2}};
\end{tikzpicture}
}
\caption{
Instantaneous vertex correction diagrams calculated in \cite{Beuf:2017bpd,Hanninen:2017ddy}
}\label{fig:ddis_inst}
\end{figure*}

 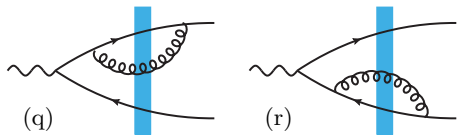
\begin{figure*}[tb!]
\centerline{\includegraphics[width=3cm]{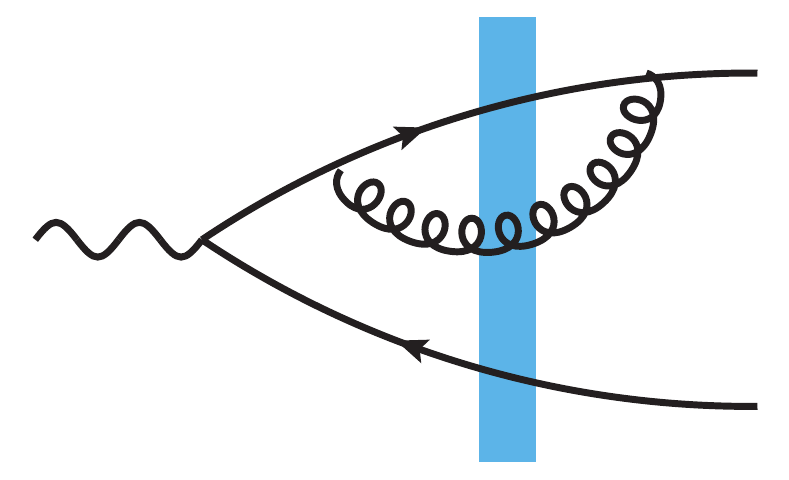}
\begin{tikzpicture}[overlay]
\node[anchor=south west] at (-3cm,0cm) {\namediag{diag:propshock1}};
\end{tikzpicture}
\includegraphics[width=3cm]{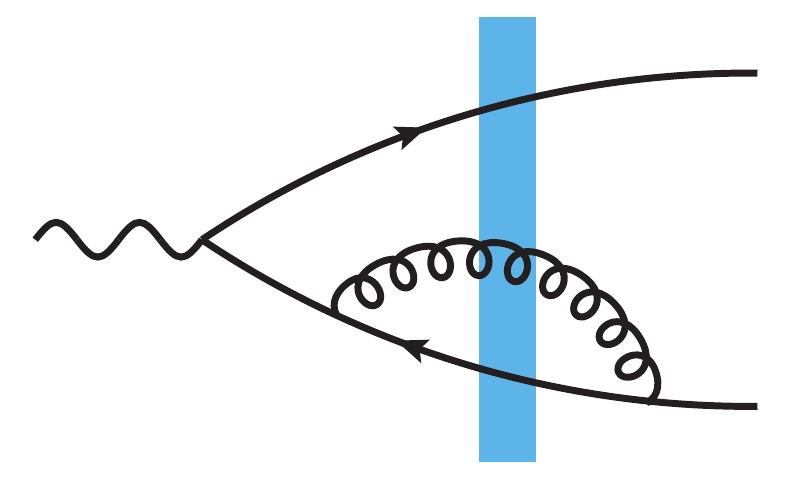}
\begin{tikzpicture}[overlay]
\node[anchor=south west] at (-3cm,0cm) {\namediag{diag:propshock2}};
\end{tikzpicture}
}
\caption{
Propagator correction diagrams where the gluon crosses the shockwave, but is not produced.
}\label{fig:ddis_propshock}
\end{figure*}
 
 \begin{figure*}[tb!]
\centerline{
\includegraphics[width=3cm]{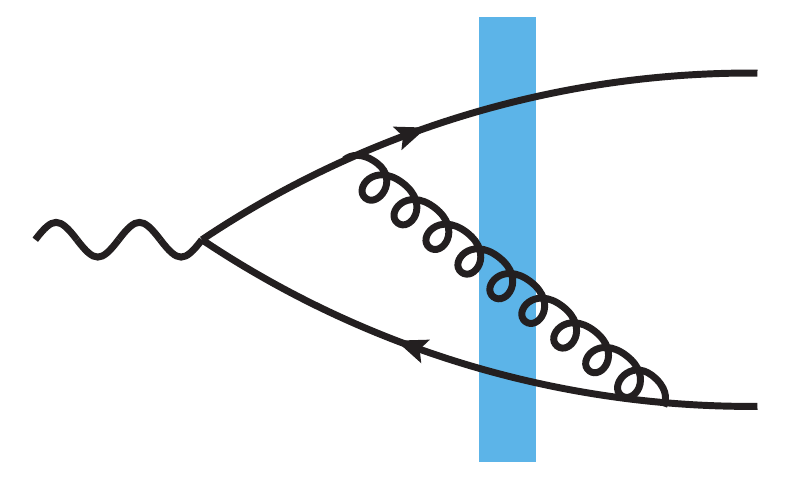}
\begin{tikzpicture}[overlay]
\node[anchor=south west] at (-3cm,0cm) {\namediag{diag:vertexshock1}};
\end{tikzpicture}
\includegraphics[width=3cm]{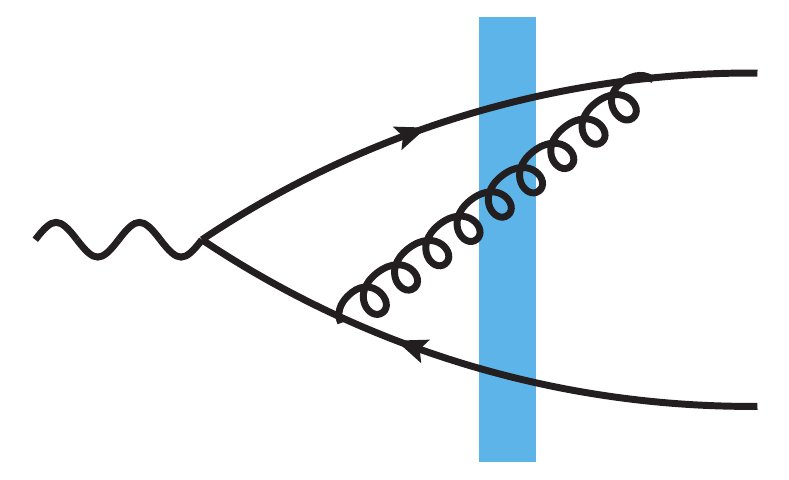}
\begin{tikzpicture}[overlay]
\node[anchor=south west] at (-3cm,0cm) {\namediag{diag:vertexshock2}};
\end{tikzpicture}
\includegraphics[width=3cm]{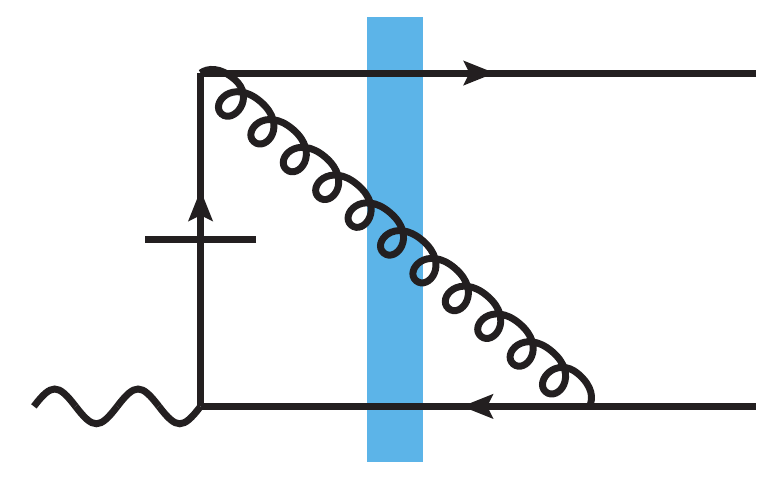}
\begin{tikzpicture}[overlay]
\node[anchor=south west] at (-3.5cm,1cm) {\namediag{diag:vertexinstshock1}};
\end{tikzpicture}
\includegraphics[width=3cm]{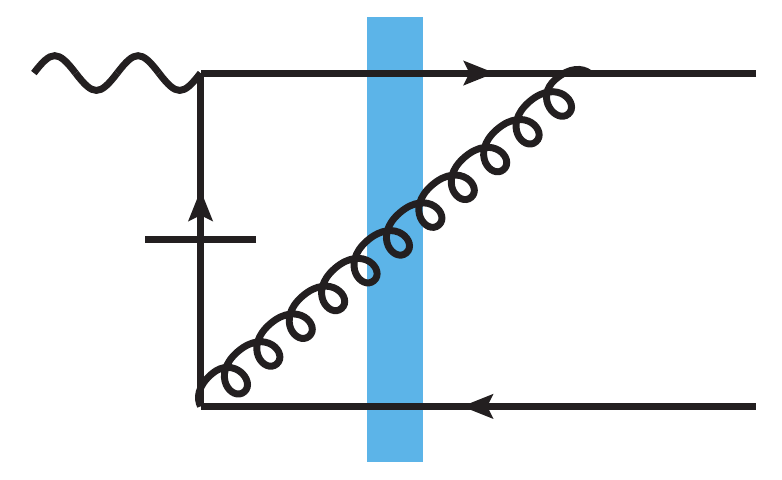}
\begin{tikzpicture}[overlay]
\node[anchor=south west] at (-3.5cm,-0.1cm) {\namediag{diag:vertexinstshock2}};
\end{tikzpicture}
}
\caption{
Gluon emission diagrams where the gluon crosses the shockwave, but is not produced.
}\label{fig:ddis_vertexshock}
\end{figure*}
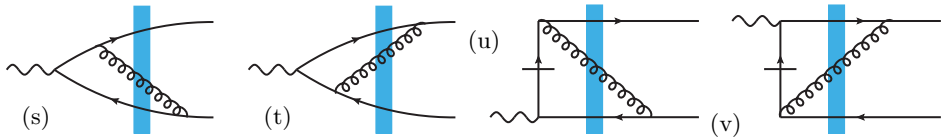

 \begin{figure*}[tb!]
\centerline{\includegraphics[width=3cm]{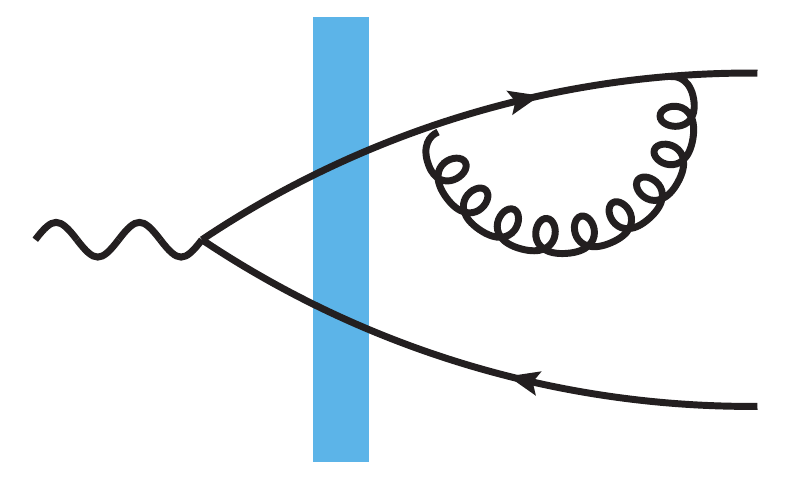}
\begin{tikzpicture}[overlay]
\node[anchor=south west] at (-3cm,0cm) {\namediag{diag:propfs1}};
\end{tikzpicture}
\includegraphics[width=3cm]{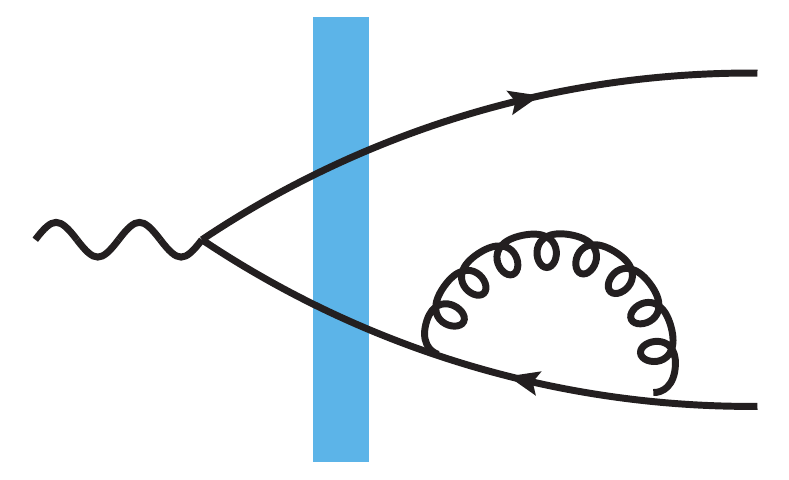}
\begin{tikzpicture}[overlay]
\node[anchor=south west] at (-3cm,0cm) {\namediag{diag:propfs2}};
\end{tikzpicture}
}
\caption{
Diagrams with propagator correction in final state. These are not included, but instead there is a wavefunction renormalization constant for the outgoing states.
}\label{fig:ddis_propfs}
\end{figure*}
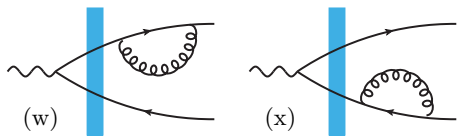

\begin{figure*}[tb!]
\centerline{\includegraphics[width=3cm]{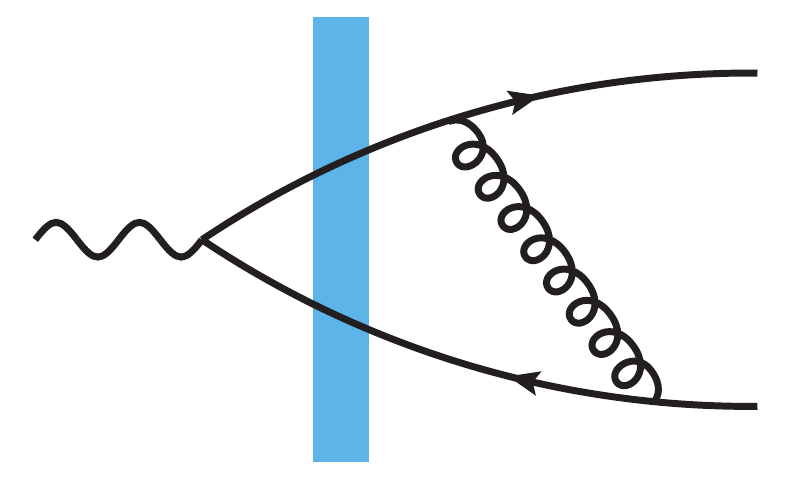}
\begin{tikzpicture}[overlay]
\node[anchor=south west] at (-3cm,0cm) {\namediag{diag:vertexfs1}};
\end{tikzpicture}
\includegraphics[width=3cm]{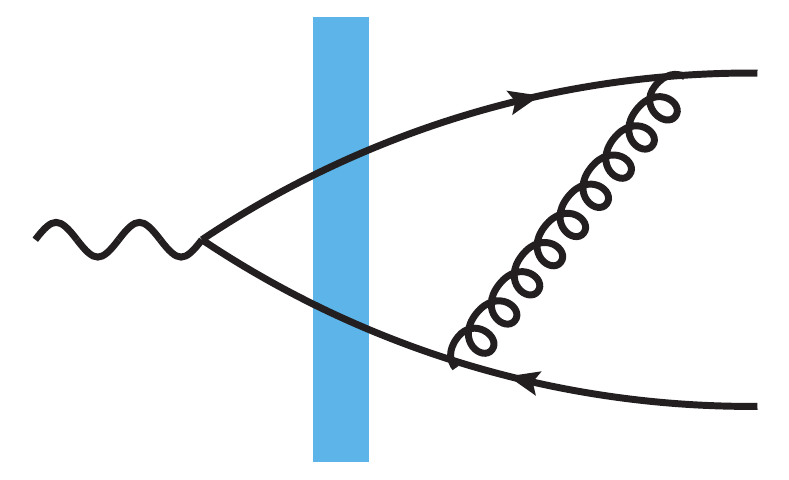}
\begin{tikzpicture}[overlay]
\node[anchor=south west] at (-3cm,0cm) {\namediag{diag:vertexfs2}};
\end{tikzpicture}
\includegraphics[width=3cm]{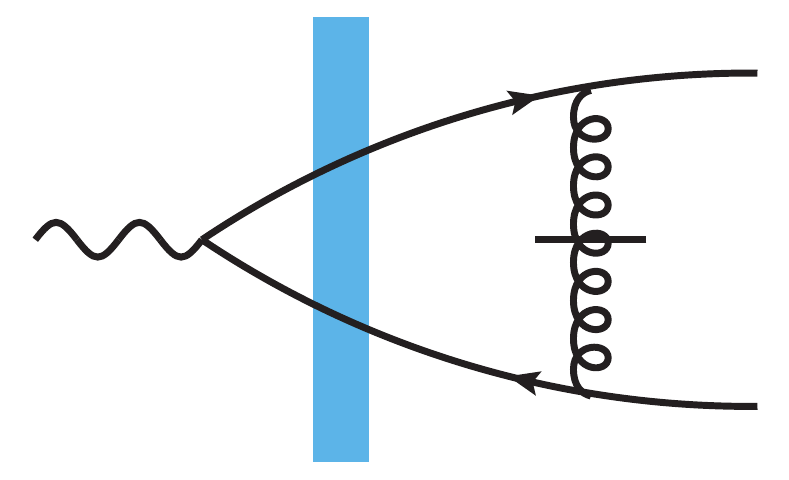}
\begin{tikzpicture}[overlay]
\node[anchor=south west] at (-3cm,0cm) {\namediag{diag:vertexfsi}};
\end{tikzpicture}
}
\caption{
Final state interaction diagrams
}\label{fig:ddis_fs}
\end{figure*}
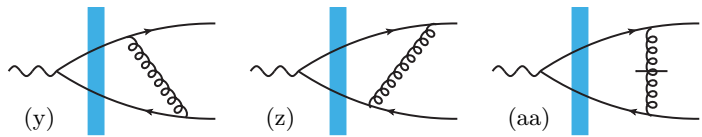

\subsection{Loop corrections}
\label{sec:loop}

In addition to the radiative contributions discussed in this paper, there are also loop corrections at NLO. We will return to a them in more detail in a future paper, but let us make a few remarks here. 
Firstly there are the one-loop corrections to the $\gamma^* \to q\Bar{q}$ wavefunction, depicted in Figs.~\ref{fig:ddis_vertexprop} and~\ref{fig:ddis_inst}.  These include a quark propagator corrections before the shockwave shown in diagrams~\ref{diag:prop1} and~\ref{diag:prop2}, and  corrections to the $\gamma^* \to q\Bar q$ vertex (including regular and instantaneous gluon or quark exchange) shown in diagrams~\ref{diag:vertex1}, \ref{diag:vertex2}, \ref{diag:vertexi}, \ref{diag:vertexiq1}, \ref{diag:vertexiq2}, \ref{diag:propiq1} and~\ref{diag:propiq2}. These one-loop wavefunctions have  already been calculated as a part of the virtual photon NLO wavefunction~\cite{Beuf:2021srj,Beuf:2022ndu,Beuf:2021qqa,Beuf:2017bpd,Hanninen:2017ddy}, and the results for the loop calculations can be directly taken from these references.

In diffractive scattering there are also additional diagrams that are not needed for total (inclusive) cross section and as such are not available in the literature. Now it will be necessary to add propagator correction diagrams where the gluon crosses the shockwave, diagrams~\ref{diag:propshock1} and~\ref{diag:propshock2}. These exhibit UV divergences in the limit when the gluon coordinate becomes equal to the emitting (anti)quark. Similarly to the calculation of the inclusive cross section these  will have to partially cancel UV divergences in the vertex correction diagrams in Figs.~\ref{fig:ddis_vertexprop}, \ref{fig:ddis_inst}. This cancellation is the reason why the one-loop $\gamma^*\to q\Bar{q}$ wavefunction alone is not sufficient to directly achieve the full NLO result. In addition to the propagator correction type diagrams, similar normal and instantaneous
 vertex correction diagrams~\ref{diag:vertexshock1},~\ref{diag:vertexshock2},~\ref{diag:vertexinstshock1} and~\ref{diag:vertexinstshock2} depicted in Fig.~\ref{fig:ddis_vertexshock} are also needed. 

In our formalism (see \cite{Lappi:2016oup,Beuf:2016wdz} for more detailed discussions, based on the seminal work of \cite{Bjorken:1970ah}) we specifically exclude 
diagrams containing self-energy corrections inserted on the external asymptotic particles.
Thus we do not explicitly have the diagrams~\ref{diag:propfs1} and~\ref{diag:propfs2} in Fig.~\ref{fig:ddis_propfs} in our calculation.
Instead, one must attach to the amplitude a wavefunction renormalization constant $\sqrt{Z_{q/\Bar{q}}}$ (again see \eq\nr{eq:fock-qqbar}), which includes the same physical contribution, and is determined by the unitarity of the evolution operator. The outgoing quark and antiquark wavefunction renormalization constants also have UV divergences, which should cancel the rest of the UV divergences from the vertex corrections. 
Finally there are also diagrams with a  gluon exchange in the final state (diagrams~\ref{diag:vertexfs1}, \ref{diag:vertexfs2} and~\ref{diag:vertexfsi}) in Fig.\ref{fig:ddis_fs}. 
Naively, one could think that these corrections correspond to a renormalization of the outgoing $q\Bar{q}$ state, but they cannot be absorbed into just a constant, since the quark and antiquark actually exchange momentum in the exchange.
A proper  discussion of how to define the dressed $q\Bar{q}$ outgoing state and treat the interactions between the outgoing particles  is a major part of the discussion of the full NLO result, which we will return to in future work.

\section{Initial and final Fock states}
\label{sec:kinematics_fock_states}

We will from now on focus on the leading order $\qqb$ part and the radiative $\qqbg$ part of the cross section. To begin, let us define explicitly our notations and normalization for the Fock states. 
The Fock expansion of the incoming virtual photon state in terms of bare partonic states is
\begin{multline}
    \label{eq:fock-gamma}
    \ket{\gamma^*_\lambda(\qplus, \qt; Q^2) \vphantom{\se}}_D
    =
    \sqrt{Z_{\gamlam}}
    \left\lbrace
        \textrm{Non-QCD Fock states}
        +
        \widetilde{\sum_{q_0 \Bar{q}_1 ~ \textrm{F. s.}}}
            \widetilde{\Psi}_{\gamma^{*}_\lambda \rightarrow q_0 \Bar{q}_1}
            \tilde{b}_0^\dagger \tilde{d}_1^\dagger \ket{0}
        \right.
        \\
        +
        \left.
        \widetilde{\sum_{q_0 \Bar{q}_1 g_2 ~ \textrm{F. s.}}}
            \widetilde{\Psi}_{\gamma^{*}_\lambda \rightarrow q_0 \Bar{q}_1 g_2}
            \tilde{b}_0^\dagger \tilde{d}_1^\dagger \tilde{a}_2^\dagger \ket{0}
        + \cdots
    \right\rbrace,
\end{multline}
where F.~s. stands for ``Fock states''.
For the outgoing partonic states the corresponding expansions are
\begin{align}
    \label{eq:fock-qqbar}
    \tensor[_D]{\bra{\Bar q_1 q_0 \vphantom{\se}}}{}
        &
        =
        \sqrt{Z_{q}}\sqrt{Z_{\Bar{q}}}
        \left\lbrace
\bra{0} \tilde d_1 \tilde b_0
+
        \widetilde{\sum_{ q_{0'} \Bar{q}_{1'} ~ \textrm{F. s.}}}'
        \bra{0} \tilde d_{1'} \tilde b_{0'} 
\widetilde\Psi_{q_0 \Bar{q}_1 \to q_{0'} \Bar{q}_{1'} }^\dagger 
+
        \widetilde{\sum_{ q_{0'} \Bar{q}_{1'} g_{2'}  ~ \textrm{F. s.}}}
        \bra{0} \tilde{a}_{2'} \tilde{d}_{1'} \tilde b_{0'} 
\widetilde\Psi_{q_0 \Bar{q}_1 \to q_{0'} \Bar{q}_{1'} g_{2'}}^\dagger + \dots
        \right\rbrace
        ,
    \\
    \label{eq:fock-qqbarg}
    \tensor[_D]{\bra{g_2 \Bar q_1 q_0 \vphantom{\se}}}{}
        &
        =
        \sqrt{Z_{g}} \sqrt{Z_{q}}\sqrt{Z_{\Bar{q}}}
        \left\lbrace
      \bra{0} \tilde a_2 \tilde d_1 \tilde b_0
        + \widetilde{\sum_{q_{0'} \Bar{q}_{1'} ~ \textrm{F. s.}}}
        \bra{0} \tilde d_{1'} \tilde b_{0'} \widetilde\Psi_{ q_0 \Bar{q}_1 g_2 \to q_{0'} \Bar{q}_{1'} }^\dagger
        + \dots
        \right\rbrace
        .
\end{align}
Here, we have only written out states that are needed for the full NLO cross section. In addition, marked with $\dots$,  there are other states needed at higher orders and non-QCD Fock states, including the photon. 
In fact, for the radiative corrections that we calculate in this paper, only the leading order terms in \eqs\nr{eq:fock-qqbar}, \nr{eq:fock-qqbarg} are needed. Strictly speaking the final state wave functions are not exactly hermitian conjugates of initial state ones, but differ by the sign of the $i\varepsilon$ in the energy denominators (see \eqs(4.2) and~(4.3) of Ref.~\cite{Bjorken:1970ah}). This difference does not play a role in the calculation of this paper, but will be crucial in the calculation of the full NLO corrections to diffractive DIS, that we leave for a future publication. 
The notation $\widetilde{\Sigma}$ denotes the sum over the quantum numbers of each parton in the Fock state and a mixed space phase-space integration for each parton~\cite{Beuf:2016wdz, Hanninen:2021byo}. 
Here the prime~$'$ in the sum denotes the fact that the original state is not included in the sum~\cite{Bjorken:1970ah}\footnote{What exactly counts as the ``same state'' requires a more detailed discussion in the case of a two-particle state than for one particle; we plan to return to this in a future paper in  the context of the full NLO calculation where this term is needed. In practice, only the diagrams including  self-energy corrections on asymptotic external legs need to be excluded, since they are already taken into account thanks to the overall renormalization constants.}.
For brevity the transverse coordinates and flavor ($f$) and helicity ($h$) indices are not written down explicitly here. We will discuss the color structure of the final state explicitly below.
The $\gamlam$-state renormalization coefficient is $Z_\gamlam = 1 + \mathcal{O}(e^2)$ and so it can be dropped in this work. The renormalization coefficients $Z_{q,\Bar{q},g}$  of the partonic states are of the order of $1 + \mathcal{O}(g^2)$, and as such do not affect the tree-level NLO corrections that we discuss in this paper.

In the mixed space $\xt_i$ are the transverse coordinates and $\kplus_i$ the longitudinal momenta of the partons, and indices $i=0,1$ refer to the quark and the antiquark, and $i=2$ to the gluon.   
The quark, antiquark and gluon creation and annihilation operators satisfy the (anti-)commutation relations
\begin{align}
    \left[
    a(\kplus_0, \xt_0, \lambda_0, a_0), a^\dagger(\kplus_1, \xt_1, \lambda_1, a_1)
    \right]
    & =
    (2 \kplus_0) (2\pi) \delta(\kplus_0 - \kplus_1) \delta^{(2)}(\xt_0 - \xt_1) \delta_{\lambda_0, \lambda_1} \delta_{a_0, a_1},
    \\
    \left\lbrace
    b(\kplus_0, \xt_0, h_0, \alpha_0), b^\dagger(\kplus_1, \xt_1, h_1, \alpha_1)
    \right\rbrace
    & =
    (2 \kplus_0) (2\pi) \delta(\kplus_0 - \kplus_1) \delta^{(2)}(\xt_0 - \xt_1) \delta_{h_0, h_1} \delta_{\alpha_0, \alpha_1},
    \\
    \left\lbrace
    d(\kplus_0, \xt_0, h_0, \alpha_0), d^\dagger(\kplus_1, \xt_1, h_1, \alpha_1)
    \right\rbrace
    & =
    (2 \kplus_0) (2\pi) \delta(\kplus_0 - \kplus_1) \delta^{(2)}(\xt_0 - \xt_1) \delta_{h_0, h_1} \delta_{\alpha_0, \alpha_1}.
    \label{eq:anticom-d}
\end{align}
Here $a_i$ and $\lambda_i$ refer to the gluon color and polarization, respectively.

In Eq.~\eqref{eq:fock-gamma} the functions $\widetilde{\Psi}_{\gamma^{*}_\lambda \rightarrow q_0 \Bar{q}_1}$ and $\widetilde{\Psi}_{\gamma^{*}_\lambda \rightarrow q_0 \Bar{q}_1 g_2}$ are the light front wavefunctions (LFWFs) describing the perturbative $\gamma^* \to q\Bar q$ and $\gamma^* \to q\Bar q g$ splittings. Furthermore it is convenient to factor out the overall color factor, momentum conservation and dependence on the photon transverse momentum $\qt$, and define the reduced wavefunctions $\widetilde{\psi}_{\gamma^{*}_\lambda \rightarrow q_0 \Bar{q}_1}$ and $\widetilde{\psi}_{\gamma^{*}_\lambda \rightarrow q_0 \Bar{q}_1 g_2}$ as
\begin{align}
    \widetilde{\Psi}_{\gamma^{*}_\lambda \rightarrow q_0 \Bar{q}_1}
    & =
    (2 \qplus) 2\pi \delta(\kplus_0 + \kplus_1 - \qplus)
    e^{i \frac{\qt}{\qplus} \cdot \qty(\kplus_0 \xt_0 + \kplus_1 \xt_1)} {\bf 1}_{\alpha_0 \alpha_1} \widetilde{\psi}_{\gamma^{*}_\lambda \rightarrow q_0 \Bar{q}_1}
    \label{eq:lfwf-qqbar}
    \\
    \widetilde{\Psi}_{\gamma^{*}_\lambda \rightarrow q_0 \Bar{q}_1 g_2}
    & =
    (2 \qplus) 2\pi \delta(\kplus_0 + \kplus_1 + \kplus_2 - \qplus)
    e^{i \frac{\qt}{\qplus} \cdot \qty(\kplus_0 \xt_0 + \kplus_1 \xt_1 + \kplus_2 \xt_2)} {t}^{a}_{\alpha_0 \alpha_1} \widetilde{\psi}_{\gamma^{*}_\lambda \rightarrow q_0 \Bar{q}_1 g_2}
    \label{eq:lfwf-qqbarg}
\end{align}
These LFWFs are currently available in the literature. The lowest order $\widetilde{\psi}_{\gamma^{*}_\lambda \rightarrow q_0 \Bar{q}_1}$ is a standard result~\cite{Kovchegov:2012mbw}. Loop corrections to it, as well as tree-level wavefunction describing the $\gamma^* \to q\Bar q g$ splitting have been recently calculated in mixed space (and $d$ dimensions) in Refs.~\cite{Beuf:2022ndu,Beuf:2021qqa,Beuf:2021srj,Beuf:2017bpd,Beuf:2011xd}; other results derived in momentum space include Refs.~\cite{Bartels:2000gt, Bartels:2001mv, Bartels:2002uz, Bartels:2004bi}, and some are compatible with BFKL evolution~\cite{Balitsky:2010ze, Balitsky:2012bs} but not the gluon saturation regime.

In this work we consider diffractive scattering, which requires that the final state partons must be in a color singlet state (see also Ref.~\cite{Altinoluk:2015dpi}). For both $q\Bar q$ and $q\Bar q g$ systems there exists exactly one such a configuration. These states are
\begin{align}
\ket{ q_0 \Bar q_1 \vphantom{\se}}_D^\text{singlet} &= 
\frac{\delta_{\beta_0,\beta_1}}{\sqrt{\nc}} \ket{q_0(\beta_0) \Bar{q}_1(\beta_1)\vphantom{\se}}_D, \quad \text{and} \label{eq:singlet_qq} \\
\ket{ q_0 \Bar q_1 g_2 \vphantom{\se}}_D^\text{singlet} &= 
\frac{t^b_{\beta_0\beta_1}}{\sqrt{\cf \nc}} \ket{ q_0(\beta_0) \Bar{q}_1(\beta_1) g_2(b)\vphantom{\se}}_D. \label{eq:singlet_qqg} 
\end{align}
Here $\beta_0,\beta_1$ are the quark and antiquark colors and $b$ is the gluon color, and $\sqrt{\nc}$ and $\sqrt{\Cf\nc}$ are normalization factors.

Using the virtual photon Fock states it now becomes possible to express the matrix elements written in Eq.~\eqref{eq:generic_amplitude} in terms of the eikonal scattering operators $\hat S_E$ describing a color rotation of a quark or a gluon in the target color field. For the $n=q\Bar q$ Fock state we have
\begin{equation}
    \label{eq:ddis-amplitude-qbarq}
    \tensor[^{\textrm{singlet}\hspace{-0.7em}}_D]{\bra{\Bar q_1 q_0 \vphantom{\se}}}{}
    \left( \se - \bf 1 \right) 
    \tensor[]{\ket{\gamma^*_\lambda(\qplus, \qt; Q^2) \vphantom{\se}}}{_D}
    =
    (2 \qplus) 2\pi \delta \qty(\pplus_0 + \pplus_1 - \qplus) i \mathcal{M}_{\gamma^* \rightarrow q \Bar q}^{\textrm{LO}} ,
\end{equation}
where the superscript LO refers to the fact that we do not include loop corrections to the $\gamma^* \to q\Bar q$ splitting in this work.
Similarly for $n=q\Bar q g$ we can write 
\begin{equation}
    \label{eq:ddis-amplitude-nlo-qbarqg}
    \tensor[^{\textrm{singlet}\hspace{-0.7em}}_D]{\bra{ g_2 \Bar q_1 q_0 \vphantom{\se}}}{}
    \left( \se - \bf 1 \right)
    \tensor[]{\ket{\gamma^*_\lambda(\qplus, \qt; Q^2) \vphantom{\se}}}{_D}
     =
    (2 \qplus) 2\pi \delta \qty(\pplus_0 + \pplus_1 + \pplus_2 - \qplus) i \mathcal{M}_{\gamma^* \rightarrow q \Bar q g}^{\textrm{NLO}} .
\end{equation}

The eikonal scattering operator $\hat{S}_E$ acts on the bare quark, antiquark and gluon creation operators as:
\begin{align}
    \hat{S}_E\; \tilde{a}^{\dagger}(k^+,\xt,\lambda,a)
    &= U_{A}(\xt)_{b a}
    \; \tilde{a}^{\dagger}(k^+,\xt,\lambda,b)\; \hat{S}_E
    \label{com_a_dag_SE}
    \\
    \hat{S}_E\; \tilde{b}^{\dagger}(k^+,\xt,h,\alpha)
    &=  U_{F}(\xt)_{\beta \alpha}\;
    \tilde{b}^{\dagger}(k^+,\xt,h,\beta)\; \hat{S}_E
    \label{com_b_dag_SE}
    \\
    \hat{S}_E\; \tilde{d}^{\dagger}(k^+,\xt,h,\alpha)
    &= \left[U_{F}^{\dag}(\xt)\right]_{\alpha \beta}\;
    \tilde{d}^{\dagger}(k^+,\xt,h,\beta)\; \hat{S}_E
    \label{com_d_dag_SE}
\end{align}
Here $\alpha$ ($a$) is the quark (gluon) color before the shock and $\beta$ ($b$) after, and $U_{F\,(A)}(\xt)$ refer to the Wilson lines at transverse coordinate $\xt$ in the fundamental (adjoint) representation, describing a color rotation of the quark (gluon) state when it propagates eikonally  through the shockwave.

Using Eq.~\eqref{eq:diffxs_An} the leading order diffractive cross section can be written in terms of the scattering amplitude $\mathcal{M}_{\gamma^* \rightarrow q \Bar q}^{\textrm{LO}}$ 
\begin{equation}
    \label{eq:diff-ddis-cs}
    \ud \sigma_{\gamma^*_\lambda \rightarrow q \Bar{q} ~ }^{\mathrm{D}, \,\textrm{LO}}
    \coloneqq
    (2 \qplus) 2\pi \delta(\pplus_0 + \pplus_1 - \qplus) 
    \widetilde{\ud p_0}    \widetilde{\ud p_1}    
    \sum_{h_0,f_0 , h_1,f_1 } 
    \left| \mathcal{M}_{\gamma^* \rightarrow q \Bar q}^{\textrm{LO}} \right|^2,
\end{equation}
where the summation is over the quantum numbers of the produced $q\Bar q $ state (for which there is exactly one color singlet color configuration as discussed above). Similarly the cross section for diffractive $q\Bar q g$ production can be written as
\begin{equation}
    \label{eq:diff-ddis-cs-nlo-qqg}
    \ud \sigma_{\gamma^*_\lambda \rightarrow q \Bar{q} g ~ \textrm{singlet}}^{\mathrm{D}, \,\textrm{NLO}}
     \coloneqq
    (2 \qplus) 2\pi \delta(\pplus_0 + \pplus_1 + \pplus_2 - \qplus) 
    \widetilde{\ud p_0}    \widetilde{\ud p_1}    \widetilde{\ud p_2}
    \sum_{h_0,f_0 , h_1,f_1 ,\lambda_2 } 
    \left| \mathcal{M}_{\gamma^* \rightarrow q \Bar q g} \right|^2,
\end{equation}
where we again sum over the final state quantum numbers with only one possible color configuration.

The Wilson line structure in the scattering amplitude \eqref{eq:ddis-amplitude-qbarq} corresponding to diffractive $q\Bar q$ production now reads 
\begin{equation}
\frac{\delta_{\beta_0 \beta_1}}{\sqrt{\nc}} \delta_{\alpha_0 \alpha_1} \left[U_F(\xt_0)_{\beta_0 \alpha_0} U_F^\dagger(\xt_1)_{\alpha_1 \beta_1} - \delta_{\beta_0 \alpha_0} \delta_{\beta_1 \alpha_1} \right] = \frac{1}{\sqrt{\nc}} \left[ \Tr \left( U_F(\xt_0)  U_F^\dagger(\xt_1) \right) - \nc\right],
\end{equation}
 where $\xt_0$ and $\xt_1$ are the quark and antiquark transverse coordinates, respectively. 
 
In this work we consider coherent diffraction in which case the target nucleus does not dissociate and the average over the target color sources is taken at the amplitude level~\cite{Good:1960ba} (see also e.g. Refs.~\cite{Miettinen:1978jb,Caldwell:2009ke,Mantysaari:2016ykx,Mantysaari:2020axf} for a discussion of the averaging procedure). 
This target average gives
\begin{equation}
\frac{1}{\sqrt{\nc} }  \left\langle \Tr \left[ U_F(\xt_0)  U_F^\dagger(\xt_1) \right] - \nc \right\rangle = \sqrt{\nc} (S_{01}-1),
\end{equation}
where $\langle \mathcal{O} \rangle$ denotes the average over the target configurations and  we have defined
\begin{equation}
\label{eq:s01}
    S_{01} \equiv \frac{1}{\nc} \left \langle \Tr \left[U_F(\xt_0) U^\dagger_F(\xt_1)\right]\right\rangle.
   \end{equation}
 From the complex conjugate amplitude one obtains exactly the same structure but with the Wilson lines evaluated at different transverse coordinates $\conj \xt_i$. The Wilson line structure at the cross section level then reads
\begin{equation}\label{eq:qqbcolor}
	\nc (S_{01}-1) (S_{\conj 0 \conj 1}^\dagger-1).
\end{equation}

The $q\Bar q g$ production case, Eq.~\eqref{eq:ddis-amplitude-nlo-qbarqg}, can be considered similarly. The Wilson line structure in the amplitude~\eqref{eq:ddis-amplitude-nlo-qbarqg} is
\begin{multline} 
\frac{(t^{b}_{\beta_0\beta_1})^*}{\sqrt{\Cf\nc}} 
t^a_{\alpha_0 \alpha_1} \left[U_A(\xt_2)^{ba} U_F(\xt_0)_{\beta_0 \alpha_0 } U_F^\dagger(\xt_1)_{\alpha_1 \beta_1} - \delta_{\alpha_0 \beta_0} \delta_{\alpha_1 \beta_1} \delta_{ab} \right] \\
= \frac{1}{\sqrt{\Cf  \nc}}  \left[ U_A^{ba}(\xt_2) \Tr \left(t^{b}\,U_F(\xt_0)\,t^{a}\,U_F^{\dagger}(\xt_1)\right)  - \Cf \nc \right]. 
\end{multline}
Note that the factor $t^a_{\alpha_0\alpha_1}$ originates from the $\gamma\to \qqbg$ wavefunction~\eqref{eq:lfwf-qqbarg}. 

Again this needs to be averaged over the target configurations at the amplitude level. Defining
\begin{equation}\label{defW}
S_{012} \equiv \frac{1}{\Cf  \nc} \left\langle U_A^{ba}(\xt_2) \Tr \left[t^{b}\,U_F(\xt_0)\,t^{a}\,U_F^{\dagger}(\xt_1)\right] \right\rangle = \frac{\nc}{2\Cf} \left( S_{02} S_{12} - \frac{1}{\nc^2} S_{01}\right)
\end{equation}
the Wilson line structure at the cross section level can be written as
\begin{equation}\label{eq:qqbgcolor}
\Cf \nc (S_{012}-1)(S_{\conj 0 \conj 1 \conj 2}^\dagger-1).
\end{equation}
Here the identity
\begin{equation}
     U_A^{ba}(\xt_2) = 2 \Tr\left[U_F(\xt_2) t^a U_F^\dagger(\xt_2) t^b\right].
\end{equation}
was used to express the adjoint Wilson line in terms of the fundamental representation Wilson lines. 
In Eq.~\eqref{defW} the mean field limit 
 $\langle \mathcal{O}_1 \mathcal{O}_2\rangle =\langle \mathcal{O}_1\rangle \langle \mathcal{O}_2\rangle$ was used to express the expectation value of the Wilson lines in terms of the dipole correlators $S_{ij}$. In particular we note that no higher multipole functions (traces of $n>2$ Wilson lines) appear even at finite $\nc$ in the mean field limit when we evaluate the diffractive cross section, in contrast to the inclusive two or three jet production case~\cite{Dominguez:2011wm}. However at finite $\nc$ the BK~\cite{Kovchegov:1999yj,Balitsky:1995ub} or JIMWLK~\cite{Jalilian-Marian:1996mkd,Jalilian-Marian:1997qno, Jalilian-Marian:1997jhx,Iancu:2001md, Ferreiro:2001qy, Iancu:2001ad, Iancu:2000hn,Mueller:2001uk} evolution would introduce an implicit dependence on such correlators.

The dipole scattering amplitude $1-S_{01}$ satisfies the small-$x$ BK or JIMWLK evolution equation. The necessary non-perturbative input for this evolution (initial condition at moderate $x$) can be determined by performing a fit to the HERA inclusive structure function data~\cite{Aaron:2009aa,Abramowicz:2015mha,H1:2018flt,H1:2012xnw} as e.g. in Refs.~\cite{Beuf:2020dxl,Lappi:2013zma,Albacete:2010sy,Ducloue:2019jmy,Mantysaari:2018zdd}. Instead of the BK/JIMWLK evolution one can also use phenomenological parametrizations such as the IPsat~\cite{Kowalski:2003hm} model where again the model parameters can similarly be constrained by HERA data~\cite{Mantysaari:2018nng,Rezaeian:2012ji}. 

Although superficially different, our formulation here is equivalent to the ``outgoing state'' formulation used e.g. in Refs.~\cite{Iancu:2018hwa,Iancu:2020mos}.  The generic amplitude \nr{eq:generic_amplitude} is given by a matrix element between two dressed states $\mathcal{M} \sim {}_D\langle\text{out}|\hat{S}-1|\text{in}\rangle_D$. The outgoing state approach consists in first expressing the incoming dressed state $|\text{in}\rangle_D$ in terms of the bare Fock states just like we do (formally expressed as a time evolution operator acting on a bare asymptotic state). One then passes through the shockwave, and obtains the state $(\hat{S}-1)|\text{in}\rangle_D$ also in terms of the bare states. In the outgoing state formulation one then, instead of taking a matrix element with the dressed state ${}_D\langle\text{out}|$, first inverts its Fock state expansion, expressing the bare states  at the shockwave in terms of the dressed states. Then inserting this inverse Fock state expression into the expression $(\hat{S}-1)|\text{in}\rangle_D$, one obtains the outgoing state $(\hat{S}-1)|\text{in}\rangle_D$ in terms of the dressed asymptotic (future) states. From here one can either read off the amplitudes, or first square them and think of the projection operator $|\text{out}\rangle_D {}_D\langle\text{out}|$ as a particle number operator counting what are in  Refs.~\cite{Iancu:2018hwa,Iancu:2020mos} called bare particles at $x^+=\infty$, which we would here call dressed particles.  In other words, in the outgoing state formalism one is acting with the time evolution operator from $x^+=0$ to $x^+\to \infty$ on the state $(\hat{S}-1)|\text{in}\rangle_D$ to express it in terms of the states $\ket{\text{out}}_D$, whereas here we use the inverse time evolution operator from $x^+\to \infty$ to $x^+=0$ to get ${}_D\langle\text{out}|$ in terms of the bare states at $x^+=0$.  The connection is easiest to see in terms of the diagrams for the amplitude, which are the same in both approaches and lead to the same expressions\footnote{Note however that Refs.~\cite{Iancu:2018hwa,Iancu:2020mos} use a different normalization for single particle states and for phase space integrals.}.

\section{Final state phase space}
\label{sec:phasesp}

The diffractive structure functions are measured at fixed invariant mass $\mx^2$. On the other hand, the diffractive 2- or 3-parton production cross sections in Eqs.~\eqref{eq:diff-ddis-cs} and~\eqref{eq:diff-ddis-cs-nlo-qqg} are written in terms of the three-momenta of the quarks and gluons. These momenta are, in turn, obtained by Fourier-transformation from coordinate space, which is how we understand the interaction with the target color field. 
The only place where the momenta of the final state particles appear is in the exponentials of this Fourier-transform and the delta function setting the invariant mass to $\mx^2$.
In order to get the final diffractive structure function one needs to integrate over the final state momenta with the restriction on $\mx$. It can be convenient to do this before integrating over the coordinates of the particles. This results in  generic ``transfer functions'' from a coordinate space squared amplitude to the final states with mass $\mx$. These functions, one for the two- and another one for the  three-particle final states, are the same for all states with the same number of partons. Thus, it makes sense to calculate them separately. Here, we will consider the two- and three-particle phase space integrals separately in subsections~\ref{sec:2part_ps} and~\ref{sec:3part_ps}. 

\subsection{2-particle phase space}
\label{sec:2part_ps}
In terms of the reduced wavefunction $\widetilde{\psi}_{\gamma^{*}_\lambda \rightarrow q_0 \Bar{q}_1} \coloneqq \widetilde{\psi}_{\gamma^{*}_\lambda \rightarrow q_0 \Bar{q}_1}(\xt_0,\xt_1,z_0,z_1)$ defined in Eq.~\eqref{eq:lfwf-qqbar} the diffractive $q\Bar q$ production cross section~\eqref{eq:diff-ddis-cs} reads 
\begin{multline}
    \frac{\ud \sigma^{\text{D}}_{\gamma^{*}_\lambda \rightarrow q \Bar{q} \, \textrm{(LO)}}}{\widetilde{\ud p_0} \widetilde{ \ud p_1}}
    = 4 \pi q^+\delta(p_0^+ + p_1^+ -q^+)
    \nc
    \int \ud^2 \xt_0 \int \ud^2 \xt_1 \int \ud^2  \Bar \xt_{0} \int \ud^2  \Bar \xt_{1}
    e^{-i \xt_{0\Bar 0} (\pt_0 - z_0 \qt)} e^{-i \xt_{1\Bar 1} (\pt_1 - z_1 \qt)} 
    \\
    \times
    \sum_{h_0,h_1,f}  
    \qty(\widetilde{\psi}_{\gamma^*_\lambda \to q_{\conz} \Bar q_{\cono}})^\dagger
    \qty(\widetilde{\psi}_{\gamma^*_\lambda \to q_{\Bar 0} \Bar q_{\Bar 1}})
    \Bigl[S_{\conz \cono}^\dagger -1 \Bigr] \Bigl[S_{01} -1 \Bigr]
\end{multline}
where we have written  $S_{ij}=S(\xt_i,\xt_j)$ and $\xt_{ij}=\xt_i-\xt_j$. The virtual photon polarization is denoted by $\lambda$, and the transverse coordinates in the complex conjugate amplitude are $\conj \xt_i$. The overall color factor $\nc$ is obtained when performing the color algebra in  the final state requiring that the outgoing state is a color singlet, see Eq.~\eqref{eq:qqbcolor}.

In order to obtain the diffractive cross section at fixed invariant mass $\mx$ and squared momentum transfer $t$, we need to integrate over the three-momenta $\vecp{0}$ and $\vecp{1}$ and introduce delta functions that enforce the required kinematics. Towards this goal we define the following transverse momentum variables
\begin{align}
    \Deltat & \equiv\pt_0 + \pt_1 - \qt,
    \\
    \lt & \equiv  z_1\pt_0 - z_0 \pt_1,
\end{align}
which should be understood as the total momentum transfer from the target to the diffractive system, and the relative momentum of the quark-antiquark pair\footnote{In fact, in the transverse plane the light cone coordinates correspond to a 2-dimensional nonrelativistic system, where $p^+$ plays the role of a mass. The definition of the relative momentum can then be thought of as $\lt  \sim  \pt_0/z_0 - \pt_1/z_1,$ which is the nonrelativistic or Galileian  \emph{velocity} of the particles in the rest frame of the 2-particle system.}. 
These satisfy:
\begin{align}
    \Deltat^2 & \approx -t ,
    \\
    \frac{\lt^2}{z_0 z_1} & \equiv \mx^2,
\end{align}
and neatly the Jacobian is unity, i.e. in \eqref{eq:diff-ddis-cs} we may replace $\ud^3 \vecp{0} \ud^3 \vecp{1} \mapsto (q^+)^2\ud^2 \Deltat \ud^2 \lt \ud z_0 \ud z_1$ with $z_i \coloneqq p_i^+/q^+$. With this change of variables, and imposing the $\mx$ and $t$ constraints, the diffractive cross section  \eqref{eq:diff-ddis-cs} becomes
\begin{align}
    \notag
    \frac{\ud \sigma^{\text{D}}_{\lambda,  \, q \Bar q}}{\ud \mx^{2} \ud \abs{t}}
    = 
    &
    \frac{\nc}{4\pi} \int_0^1 \ud z_0 \int_0^1 \ud z_1
    \delta \qty(z_0+z_1-1)
    \int \ud^2 \xt_0 \int \ud^2 \xt_1 \int \ud^2 \Bar \xt_0 \int \ud^2 \Bar \xt_1
    \\
    &
    \times
    {\cal I}_{\Deltat}^{(2)} {\cal I}_{\mx}^{(2)}
    \sum_f \sum_{h_0, h_1}
    \qty(\widetilde{\psi}_{\gamma^{*}_\lambda \rightarrow q_{\Bar 0} \Bar{q}_{\Bar 1}})^\dagger
    \qty(\widetilde{\psi}_{\gamma^{*}_\lambda \rightarrow q_0 \Bar{q}_1})
    \Bigl[S_{\conz \cono}^\dagger -1 \Bigr] \Bigl[S_{01} -1 \Bigr] ,
    \label{eq:diffqq_mx_t}
\end{align}
where we defined
\begin{equation}
    \label{eq:IDelta_2-x}
    {\cal I}_\Deltat^{(2)}
    \coloneqq
    \int \frac{\ud^2 \Deltat}{(2\pi)^2}  \delta(\Deltat^2 - \abs{t}) e^{i \Deltat \cdot (z_0 \xt_{\Bar 0 0} - z_1\xt_{\Bar 1 1} )}
\end{equation}
and
\begin{equation}
    \label{eq:IMX_2-x}
    {\cal I}_{\mx}^{(2)}
    \coloneqq
    \int \frac{\ud^2 \lt}{(2\pi)^2} \delta(\lt^2 - z_0 z_1 \mx^2) e^{i \lt \cdot (\xt_{\conz \cono}-\xt_{01})}.
\end{equation}
As the reduced photon wavefunction $\widetilde{\psi}_{\gamma^{*}_\lambda \rightarrow q_0 \Bar{q}_1}$ can only depend on the coordinate separation $\rt \coloneqq \xt_0 - \xt_1$, it is useful to make a change of variables from $\xt_0,\xt_1$ to the dipole size $\rt$ and the impact parameter $\bt=(\xt_0+\xt_1)/2$ (using again coordinates with a bar for the complex conjugate amplitude). In these coordinates we get
\begin{align}
    {\cal I}_\Deltat^{(2)}
    &=
    \int \frac{\ud^2 \Deltat}{(2\pi)^2}  \delta \qty(\Deltat^2 - \abs{t}) e^{i \Deltat \cdot \left(\Bar \bt - \bt + \frac{2z_0-1}{2}(\Bar \rt-\rt)\right)}
    \label{eq:IDelta_2}
    \\
    {\cal I}_{\mx}^{(2)} 
    &=  
    \int \frac{\ud^2 \lt}{(2\pi)^2} \delta \qty(\lt^2 - z_0 z_1 \mx^2) e^{i \lt \cdot (\Bar \rt - \rt)}
\end{align}
In the most general case (i.e. without further assumptions), it is also possible to perform the $\lt$ integral which gives
\begin{equation}
\label{eq:IMX_2}
    {\cal I}_{\mx}^{(2)} = 
    \frac{1}{4\pi} 
    \besj_0\left( \sqrt{z_0 z_1}\mx \norm{\Bar \rt-\rt} \right).
\end{equation}
This transfer function is related to the probability to form a final state with the given invariant mass $\mx$ given the dipole sizes $\rt$ and $\conj{\rt}$ in the amplitude and conjugate amplitude with fixed longitudinal momentum fractions $z_i$ for the quarks.
The integral over $\Deltat$ is of the same form, and gives
\begin{equation}
\label{eq:I2_Delta_result}
    {\cal I}_\Deltat^{(2)} = \frac{1}{4\pi} \besj_0\left( \sqrt{\abs{t}}\: \norm{\Bar \bt \!-\! \bt \!+\! \frac{(2z_0\!-\!1)}{2}(\Bar \rt\!-\!\rt)} \right).
\end{equation}

Eventually we are also interested in $t$-integrated diffractive cross sections and structure functions. Integrating over the squared momentum transfer $t$ we find 
\begin{equation}
\label{eq:I2_Delta_tint}
    \int_{-\infty}^0 \ud t ~ {\cal I}_\Deltat^{(2)} = \delta^{(2)}\left(\Bar \bt - \bt + \frac{2z_0-1}{2}(\Bar \rt-\rt) \right).
\end{equation}

The most general result for the total diffractive cross section in the case where the final state consist of two particles is then given by Eq.~\eqref{eq:diffqq_mx_t} with the phase space integrals ${\cal I}_\Deltat^{(2)}$ and ${\cal I}_{\mx}^{(2)}$ given above. 
In particular, we emphasize that the cross section \eqref{eq:diffqq_mx_t} can not be in general written in a factorized form commonly used in the literature~\cite{GolecBiernat:2001mm,Marquet:2007nf,Kowalski:2008sa} where the result is expressed as a square of an integral over the transverse coordinates in the amplitude. In Sec.~\ref{sec:lo-ddis-cs} we discuss in detail the approximations necessary to obtain a form for the diffractive cross sections where the dependence on impact parameter, amplitude coordinates and conjugate amplitude coordinates factorize and enables one to write the result in the ``squared integral'' form.

We finally note that the ``off-forward'' phase in the amplitude coupling the dipole size and the momentum transfer is $\exp\left( i \frac{2z_0-1}{2} \Deltat \cdot \rt\right)$ as shown in Eq.~\eqref{eq:IDelta_2}, and not $\exp\left(i(1-z_0)\rt \cdot \Deltat\right)$ as has been commonly used in the literature based on Refs.~\cite{Bartels:2003yj,Kowalski:2006hc}. The correct phase factor has been discussed e.g. in Refs.~\cite{Hatta:2017cte,Mantysaari:2020lhf}. We furthermore note that if one used the center-of-mass $\bt' = z_0 \xt_0 + z_1 \xt_1$ as an impact parameter (which would be a natural variable in light cone perturbation theory), no such an off forward phase would appear and the phase factor in Eq.~\eqref{eq:IDelta_2} would be just $\exp \left(i\Deltat \cdot (\Bar \bt' - \bt')\right)$.

\subsection{3-particle phase space}
\label{sec:3part_ps}

Let us next consider the phase space integral for the case where there are three particles in the final state, referring to the $q\Bar q g$ system in this work. Analogously to the 2-particle case discussed above, the starting point is the total diffractive cross section (see Eq.~\eqref{eq:diff-ddis-cs-nlo-qqg}) at fixed $\mx^2$ and $t$:
\begin{multline}
\label{eq:phasespace_3part_start}
    \frac{\ud \sigma^{\text{D}}_{\gamma^{*}_\lambda \rightarrow q \Bar{q} g}}{\ud \mx^2 \ud \abs{t}} 
    = 
    \frac{\nc \cf}{(4\pi)^2}
    \int\! \frac{\ud^2 \pt_0}{(2\pi)^2} \!\int\! \frac{\ud^2 \pt_1}{(2\pi)^2} \!\int\! \frac{\ud^2 \pt_2}{(2\pi)^2}
    \!\!\int_0^1\! \frac{\ud z_0}{z_0} \!\int_0^1\! \frac{\ud z_1}{z_1} \!\int_0^1\! \frac{\ud z_2}{z_2}
    \delta(z_0 \!+\! z_1 \!+\! z_2 -1) \delta \qty((\pt_0 \!+\! \pt_1 \!+\! \pt_2 - \qt)^2 - \abs{t})
    \\
    \times
    \delta \qty((p_0 \!+\! p_1 \!+\! p_2)^2 - \mx^2) 
    \int_{\xt_0} \int_{\xt_1} \int_{\xt_2} \int_{\Bar \xt_0} \int_{\Bar \xt_1} \int_{\Bar \xt_2} (2 \pi)^6
    e^{i \xt_{\conj{0} 0}(\pt_0 - z_0 \qt)}
    e^{i \xt_{\conj{1} 1}(\pt_1 - z_1 \qt)}
    e^{i \xt_{\conj{2} 2}(\pt_2 - z_2 \qt)} \\
    \times 
    \sum_{f,h_0, h_1, \lambda_2}
    \qty(\widetilde{\psi}_{\gamma^{*}_\lambda \rightarrow q_{\Bar 0} \Bar{q}_{\Bar 1} g_{\Bar 2}})^\dagger
    \qty(\widetilde{\psi}_{\gamma^{*}_\lambda \rightarrow q_0 \Bar{q}_1 g_2})
    \Bigl[S_{\conj{0} \conj{1} \conj{2}}^{\dagger} -1\Bigr] 
    \Bigl[S_{012} -1\Bigr] ,
\end{multline}
where the color factor $\tr(t^a t^a) = \nc \cf$ is again obtained as shown in Eq.~\eqref{eq:qqbgcolor} in Sec.~\ref{sec:kinematics_fock_states}, $p_0,p_1,p_2$ are the four-momenta of the produced partons, and their transverse coordinates are again labeled as $\xt_0,\xt_1,\xt_2$ in the amplitude, and $\Bar \xt_0,\Bar \xt_1,\Bar \xt_2$ in the conjugate amplitude. The plus momentum fractions are again denoted by $z_i$, and the transverse integral normalization is defined as $\int_{\xt} \coloneqq \int \frac{\ud^2 \xt}{2\pi}$. Note that this introduces an explicit $(2\pi)^6$ in \eq\nr{eq:phasespace_3part_start}, but will lead to nicer expressions in the end.

Next we define the following transverse momenta
\begin{align}
    \Pt_i & \coloneqq \pt_i - z_i \qt, \label{eq:Pti}
    \\
    \Kt & \coloneqq \Pt_2 - \frac{z_2}{z_0} \Pt_0,
    \\
    \Pt & \coloneqq \Pt_0 + z_0 \Deltat + \frac{z_0}{1-z_1} \Kt,
    \\
    \Deltat & \coloneqq \qt - \pt_0 - \pt_1 - \pt_2 = - \Pt_0 - \Pt_1 - \Pt_2.
\end{align}
Here $\Pt_i$ could be interpreted as the momentum of the particle $i$ with respect to the center of mass of the 3-particle system before the scattering (i.e. the momentum $\qt$). The momentum $\Deltat$ is then the total momentum transfer from the target to the scattering system, and $\Kt$ the relative momentum of the gluon with respect to the quark. The remaining $\Pt$ then is proportional  to the relative momentum of the antiquark with respect to the quark-gluon system, which becomes more obvious if one writes it as $\Pt = z_0z_1[(\Pt_0+\Pt_2)/(1-z_1) -\Pt_1/z_1]$.
Using the above variables allows us to rewrite the invariant mass of the final state particles in a simple way as the sum of two squared momenta:
\begin{equation}
    M_{q \Bar q g}^2 \coloneqq (p_0 + p_1 + p_2)^2  = \frac{\Pt_0^2}{z_0} + \frac{\Pt_1^2}{z_1} + \frac{\Pt_2^2}{z_2} - \Deltat^2 = \frac{1-z_1}{z_1 z_0^2} \Pt^2 + \frac{z_0}{z_2 (1-z_1)} \Kt^2.
\end{equation}
Now we need to apply the same changes to the exponential phases in the integral \eqref{eq:phasespace_3part_start}:
\begin{equation}
    e^{i \xt_{\Bar 0 0}(\pt_0 - z_0 \qt)}
 e^{i \xt_{\Bar 1 1}(\pt_1 - z_1 \qt)}
     e^{i \xt_{\Bar 2 2}(\pt_2 - z_2 \qt)}
     = e^{i \left( \xt_{\Bar 0 0} + \frac{z_2}{z_0} \xt_{\Bar{2} 2} - \frac{z_0 + z_2}{z_0} \xt_{\Bar{1} 1} \right) \cdot \Pt}
     e^{i \frac{z_0}{1-z_1} \left( \xt_{\Bar 2 2} - \xt_{\Bar 0 0} \right) \cdot \Kt}
     e^{- i \left( z_0 \xt_{\Bar 0 0} + z_1 \xt_{\Bar 1 1} + z_2 \xt_{\Bar 2 2} \right) \cdot \Deltat}.
\end{equation}

To obtain the cross section differentially in invariant mass and squared momentum transfer, we again integrate over all three-momenta and include delta functions that impose the required constraints. This gives
\begin{align}
    \frac{\ud \sigma^{\text{D}}_{\gamma^{*}_\lambda \rightarrow q \Bar{q} g}}{\ud M_{X}^{2} \ud \abs{t}}
    = 
    &
    4 \pi^4 \nc \cf
    \int_0^1 \!\! \frac{\ud z_0}{z_0} \int_0^1 \!\! \frac{\ud z_1}{z_1} \int_0^1 \!\! \frac{\ud z_2}{z_2} \,
     \delta(z_0+z_1+z_2-1)
        {\cal I}_\Deltat^{(3)} {\cal I}_{\mx}^{(3)}
    \notag
    \\
    &
    \times
    \int_{\xt_0} \int_{\xt_1} \int_{\xt_2} \int_{\Bar \xt_0} \int_{\Bar \xt_1} \int_{\Bar \xt_2}
    \sum_{h_0, h_1, \lambda_2}
    \qty(\widetilde{\psi}_{\gamma^{*}_\lambda \rightarrow q_{\Bar 0} \Bar q_{\Bar 1} g_{\Bar 2}})^\dagger
    \qty(\widetilde{\psi}_{\gamma^{*}_\lambda \rightarrow q_0 \Bar{q}_1 g_2})
    \Bigl[1 - S_{\conj{0} \conj{1} \conj{2}}^{\dagger}\Bigr]
    \Bigl[1 - S_{012}\Bigr],
    \label{eq:qqg_diffxs_Mx_t}
\end{align}
where we have again separated the transverse momentum integrals:
\begin{align}
    {\cal I}_\Deltat^{(3)} & =
        \int \frac{\ud^2 \Deltat}{(2 \pi)^2}
        \delta(\Deltat^2 - \abs{t})
        e^{- i \left( z_0 \xt_{\Bar 0 0} + z_1 \xt_{\Bar 1 1} + z_2 \xt_{\Bar 2 2} \right) \cdot \Deltat},
    \\
    {\cal I}_{\mx}^{(3)} & = 
        \int \frac{\ud^2 \Pt}{(2 \pi)^2} \int \frac{\ud^2 \Kt}{(2 \pi)^2}
        \delta \left(\frac{1-z_1}{z_1 z_0^2} \Pt^2 + \frac{z_0}{z_2 (1-z_1)} \Kt^2 - \mx^2 \right)
        e^{i \left( \xt_{\Bar 0 0} + \frac{z_2}{z_0} \xt_{\Bar 2 2} - \frac{z_0 + z_2}{z_0} \xt_{\Bar 1 1} \right) \cdot \Pt}
        e^{i \frac{z_0}{1-z_1} \left( \xt_{\Bar 2 2} - \xt_{\Bar 0 0} \right) \cdot \Kt}.
\end{align}
The integral in ${\cal I}_\Deltat^{(3)}$ can be evaluated using standard methods in spherical coordinates, yielding
\begin{equation}
    {\cal I}_\Deltat^{(3)} = \frac{1}{4 \pi} \mathrm{J}_0 \left( \sqrt{-t} \norm{z_0 \xt_{\Bar 0 0} + z_1 \xt_{\Bar 1 1} + z_2 \xt_{\Bar 2 2}} \right).
\end{equation}
Eventually we want to calculate $t$ integrated diffractive structure functions. The integration over the squared momentum transfer $t$ gives:
\begin{multline}
\label{eq:IDelta3}
    \int_{-\infty}^0 \ud t \, {\cal I}_\Deltat^{(3)}
 =
    \int_{-\infty}^0 \ud t \,
    \int \frac{\ud^2 \Deltat}{(2 \pi)^2}
        \delta(\Deltat^2 - \abs{t})
        e^{- i \left( z_0 \xt_{\Bar 0 0} + z_1 \xt_{\Bar 1  1} + z_2 \xt_{\Bar 2 2} \right) \cdot \Deltat}
     =
    \delta^{(2)} \left( z_0 \xt_{\Bar 0 0} + z_1 \xt_{\Bar 1 1} + z_2 \xt_{\Bar 2 2} \right)
    \equiv
    \delta^{(2)} \left( \Bar \bt - \bt \right),
\end{multline}
where $\bt \coloneqq z_0 \xt_0 + z_1 \xt_1 + z_2 \xt_2$ is the center-of-mass of the $\qqbg$-system, and $\Bar \bt$ is the respective coordinate in the conjugate amplitude\footnote{Here we directly defined the impact parameter $\bt$ as the ``true'' momentum-weighted impact parameter and not just the average of the coordinates, cf. the discussion below \eq\nr{eq:I2_Delta_tint}.}.
The calculation of ${\cal I}_{\mx}^{(3)}$ is more involved, and proceeds by Fourier transforming the $\delta$-function:
\begin{equation}
    \label{eq:delta-fun-fourier}
    \delta \left(\frac{1-z_1}{z_1 z_0^2} \Pt^2 + \frac{z_0}{z_2 (1-z_1)} \Kt^2 - \mx^2 \right)
    =
    \int_{\mathbb{R}} \frac{\ud \eta}{2 \pi} e^{i \eta \left(\frac{1-z_1}{z_1 z_0^2} \Pt^2 + \frac{z_0}{z_2 (1-z_1)} \Kt^2 - \mx^2 \right)}.
\end{equation}
To simplify the notation, we define the following transverse coordinates:
\begin{align}
    \zt & \coloneqq \xt_{\Bar 0 0} + \frac{z_2}{z_0} \xt_{\Bar 2 2} - \frac{z_0 + z_2}{z_0} \xt_{\Bar 1 1},
    \\
    \yt & \coloneqq \xt_{\Bar 2 2} - \xt_{\Bar 0 0}.
\end{align}
With these and Eq.~\eqref{eq:delta-fun-fourier} --- and completing some squares --- we can write
\begin{align}
    {\cal I}_{\mx}^{(3)} = & 
        \int_{\mathbb{R}} \frac{\ud \eta}{2 \pi} e^{- i \eta \mx^2}
        e^{-i \eta \frac{z_0^2 z_1}{1-z_1} \left( \frac{\zt}{2 \eta} \right)^2}
        e^{-i \eta \frac{z_0 z_2}{1-z_1} \left( \frac{\yt}{2 \eta} \right)^2}
        \nonumber
        \\
        & \times
        \int \frac{\ud^2 \Pt}{(2 \pi)^2} \int \frac{\ud^2 \Kt}{(2 \pi)^2}
        e^{i \eta \frac{1-z_1}{z_0^2 z_1} \left( \Pt + \frac{z_0^2 z_1}{1-z_1} \frac{1}{2 \eta} \zt\right)^2}
        e^{i \eta \frac{z_0}{(1-z_1) z_2} \left( \Kt +  \frac{z_2}{2 \eta}\yt \right)^2}.
\end{align}
Computing the --- now Gaussian --- transverse momentum integrals requires shifting $\eta \to \eta + i \epsilon$ in the complex plane. With this we have $\int_0^\infty \ud z e^{i \eta z} = i / (\eta + i \epsilon)$, and so
\begin{equation}
    \int \frac{\ud^2 \Pt}{(2 \pi)^2} \int \frac{\ud^2 \Kt}{(2 \pi)^2}
        e^{i \eta \frac{1-z_1}{z_0^2 z_1} \left( \Pt + \frac{z_0^2 z_1}{1-z_1} \frac{1}{2 \eta}\zt \right)^2}
        e^{i \eta \frac{z_0}{(1-z_1) z_2} \left( \Kt + \frac{z_2}{2 \eta} \yt \right)^2}
    =
    \frac{1}{(4 \pi)^2} \frac{i \left( \frac{z_0^2 z_1}{1-z_1} \right)}{\eta + i \epsilon} \frac{i \left( \frac{(1-z_1) z_2}{z_0} \right)}{\eta + i \epsilon}.
\end{equation}
This allows us to write the remaining $\eta$-integral as a residue in the lower half of the complex plane:
\begin{equation}
    {\cal I}_{\mx}^{(3)} =
    i \frac{z_0 z_1 z_2}{(4 \pi)^2} 
    \textrm{Res}\left( 
        \frac{  e^{- i \eta \mx^2}
                e^{-i \eta \frac{z_0^2 z_1}{1-z_1} \left( \frac{\zt}{2 \eta} \right)^2}
                e^{-i \eta \frac{z_0 z_2}{1-z_1} \left( \frac{\yt}{2 \eta} \right)^2}}
             {(\eta + i \epsilon)^2}
        ,
        \eta \to - i \epsilon \right),
\end{equation}
where we can now take $\epsilon \to 0$. The singularity at $\eta = 0$ is an essential singularity, which means that we can read the above residue as the coefficient of the $\frac{1}{\eta}$-term in the series expansion of the residue function. Defining $\Yt_{012}^2 \coloneqq \frac{z_0^2 z_1}{1-z_1} \zt^2 + \frac{z_0 z_2}{1-z_1} \yt^2$, the expansion is
\begin{align}
    \frac{1}{\eta^2} e^{- i \eta \mx^2} e^{- i \frac{\Yt_{012}^2}{4 \eta}}
    =
    \sum_{n=0}^\infty \frac{\left(-i \mx^2 \right)^n}{n!} \sum_{m=0}^\infty \frac{\left( -i \Yt_{012}^2 \right)^m}{m!} \frac{1}{\eta^{2+m-n}},
\end{align}
and so the residue is found at $2+m-n=1 ~ \implies n = m+1$. Thus we finally have
\begin{align}
\label{eq:IMX3}
    {\cal I}_{\mx}^{(3)} =
        & i \frac{z_0 z_1 z_2}{(4 \pi)^2}
        \sum_{m=0}^\infty \frac{(-i)^{2m+1}}{m! (m +1)!} \left( \mx^2 \right)^{m+1} \left( \frac{\Yt_{012}^2}{4} \right)^{m}
        \nonumber
        \\
        = &
        2 \frac{z_0 z_1 z_2}{(4 \pi)^2} \frac{\mx}{Y_{012}} \mathrm{J}_1 \left( \mx Y_{012} \right),
\end{align}
where $Y_{012} \coloneqq \norm{\Yt_{012}}$, and the series expansion of the Bessel function of the first kind was recognized:
\begin{equation}
    \besj_1 (x) = \sum_{n=0}^\infty \frac{(-1)^n}{n! (n +1)!} \left( \frac{x}{2} \right)^{2n+1}.
\end{equation}
In terms of the quark, antiquark and gluon coordinates, the transverse distance scale appearing as a conjugate to the invariant mass reads
\begin{equation}
    \Yt_{012}^2 =
        z_0 z_1 \left( \xt_{\Bar 0 0} - \xt_{\Bar 1 1} \right)^2 +
        z_1 z_2 \left( \xt_{\Bar 2 2} - \xt_{\Bar 1 1} \right)^2 +
        z_0 z_2 \left( \xt_{\Bar 2 2} - \xt_{\Bar 0 0} \right)^2.
\end{equation}
Note that $\Yt_{012}^2$ does not depend on the center-of-mass of the $q\Bar q g$ system $\bt = z_0 \xt_0 + z_1 \xt_1 + z_2 \xt_2$ (or on $\Bar \bt$) which is the Fourier conjugate to the momentum transfer $\Deltat$. Consequently if the impact parameter dependence factorizes from the Wilson lines as $S_{012}-1 = T(\bt)(S_{012}-1)$, then the $t$-integrated diffractive cross section is proportional to $\int \! \ud^2 \bt |T(\bt)|^2$.

The transverse momentum integrals ${\cal I}_{\mx}^{(3)}$ and ${\cal I}_{\Deltat}^{(3)}$ combined with the virtual photon wavefunctions and $q\Bar q g$-target scattering amplitudes can now be directly used to calculate the total diffractive cross section at fixed invariant mass $\mx^2$ using Eq.~\eqref{eq:qqg_diffxs_Mx_t}.

\section{Leading order diffractive cross section}
\label{sec:lof2d}

    In this section we present for completeness a derivation for the $q \Bar q$ contribution to the leading order diffractive cross section. The calculation is organized as follows. First in Sec.~\ref{sec:lo-ddis-wavefun} we review the leading order photon wavefunction describing the $\gamma \to q\Bar q$ dipole dipole transition, and show the squared wavefunction needed in the case of DDIS. In Sec.~\ref{sec:lo-ddis-cs} we derive the general leading order result for the diffractive cross sections, after which we discuss in detail what approximations are necessary in order to derive the form commonly used in the literature.

\subsection{Coordinate space wavefunction}
    \label{sec:lo-ddis-wavefun}

    The wavefunction describing the tree-level $\gamma^* \to q\Bar q$ splitting is required to evaluate the leading order cross section Eq.~\eqref{eq:diffqq_mx_t}. In $D=4$ dimensions the virtual photon wavefunctions in transverse coordinate space read~\cite{Kovchegov:2012mbw} (see also Refs.~\cite{Beuf:2016wdz, Beuf:2017bpd})
    \begin{equation}
        \widetilde{\psi}_{\gamma^*_L \rightarrow q_0 \Bar q_1}
        =
        - \frac{e e_f}{2 \pi} \delta_{h_1, -h_0} z_0^{3/2} z_1^{3/2} 2 Q \besk_0 \qty( x_{{0} {1}} \Bar{Q} )
    \end{equation}
    for the longitudinal photon, and
    \begin{equation}
        \widetilde{\psi}_{\gamma^*_\lambda \rightarrow q_0 \Bar q_1}
        =
        -i \frac{e e_f}{2 \pi}
        \sqrt{z_0 z_1} \left[
            (z_0 - z_1) \delta^{ij} + 2 (-h_1) i \epsilon^{ij}
        \right]
        \delta_{h_1, -h_0}
        \frac{\epsilon^i_\lambda \xt_{01}^j}{x_{{0} {1}}}
        \Bar{Q} \besk_1 \qty( x_{{0} {1}} \Bar{Q} )
    \end{equation}
    for a transversely polarized photon with polarization $\lambda$, with $x_{01} \coloneqq \norm{\xt_{01}}$ and $\Bar{Q} \coloneqq \sqrt{z_0 z_1} Q$. The quark fractional charge is denoted by $e_f$.
    Note that our  convention to pull out  a normalization factor $(2 \qplus)$ from the definition of the reduced wavefunction in Eq.~\eqref{eq:lfwf-qqbar} allows us to write the  wavefunctions in terms of the longitudinal momentum fractions $z_i$ with no explicit dependence on $q^+$.

    In order to calculate the diffractive cross sections we need the squared wavefunctions in the case where the quark transverse coordinates are different in the amplitude and in the conjugate amplitude.  Summing over the quark helicities these squares read
    \begin{equation}
    \label{eq:photon_LO_L_squared}
        \sum_{h_0, h_1}
        \qty(\widetilde{\psi}_{\gamma^{*}_L \rightarrow q_{\conj{0}} \Bar{q}_{\conj{1}}})^\dagger
        \qty(\widetilde{\psi}_{\gamma^{*}_L \rightarrow q_0 \Bar{q}_1})
         =
         2 \frac{e^2 e_f^2}{(2 \pi)^2} z_0^3 z_1^3 4 Q^2
         \besk_0 \qty(x_{{0} {1}} \Bar{Q}) \besk_0 \qty(x_{\conj{0} \conj{1}} \Bar{Q})
    \end{equation}
    and
     \begin{equation}
     \label{eq:photon_LO_T_squared}
        \frac{1}{2} \sum_{T \, \textrm{pol.} \, \lambda}
        \sum_{h_0, h_1}
        \qty(\widetilde{\psi}_{\gamma^{*}_\lambda \rightarrow q_{\conj{0}} \Bar{q}_{\conj{1}}})^\dagger
        \qty(\widetilde{\psi}_{\gamma^{*}_\lambda \rightarrow q_0 \Bar{q}_1})
         =
        \frac{e^2 e_f^2}{(2 \pi)^2} z_0^2 z_1^2
        \left((z_0 - z_1)^2 + 1 \right)
        \frac{\xt_{01} \cdot \xt_{\conj{0} \conj{1}} }{ x_{{0} {1}} x_{\conj{0} \conj{1}} }
        Q^2 \besk_1 \qty(x_{01} \Bar{Q}) \besk_1 \qty(x_{\conj{0} \conj{1}} \Bar{Q}).
    \end{equation}
A more familiar form of the momentum fraction dependence can be obtained noting that $(z_0-z_1)^2+1 \equiv 2(z_0^2+z_1^2)$, which holds under the plus-momentum conservation $z_0+z_1=1$.

\subsection{From wavefunction to diffractive cross-section}
\label{sec:lo-ddis-cs}
    
    The total diffractive cross section at leading order can be obtained by substituting the virtual photon wavefunction in Eq.~\eqref{eq:diffqq_mx_t}, and using the   phase space integrals ${\cal I}_{\Deltat}^{(2)}$ and ${\cal I}_{\mx}^{(2)}$ given in Eqs.~\eqref{eq:IDelta_2} and~\eqref{eq:IMX_2}.
    Let us first consider the case where the virtual photon is longitudinally polarized. We note that although the squared wavefunction~\eqref{eq:photon_LO_L_squared} factorizes as $\besk_0( x_{01} \Bar Q)  
    \besk_0(x_{\conj{0} \conj{1}} \Bar Q)$, the diffractive cross section can not be written simply as a square of an amplitude, since the phase space integral ${\cal I}_{\mx}^{(2)}$ mixes the transverse coordinates in the amplitude and in the conjugate amplitude even after an integral over the total momentum transfer, see Eq.~\eqref{eq:IMX_2}.
    
    In order to derive the leading order results for the $q\Bar q$ contribution to the diffractive structure functions commonly used in the literature~\cite{Marquet:2007nf}, further approximations are required. In particular, we assume that
  \begin{itemize}
    \item The invariant mass $\mx^2$ or virtuality $Q^2$ is so large that $\exp\left[ -i \frac{2z-1}{2} \Deltat \cdot (\rt-\Bar \rt)\right]\approx 1$ (note that $\rt^2,\Bar \rt^2 \lesssim 1/Q^2, 1/\mx^2$). In this case, the momentum transfer integral of ${\cal I}_{\Deltat}^{(2)}$ gives only $\delta^{(2)}\qty(\bt-\Bar \bt)$, see Eq.~\eqref{eq:I2_Delta_tint}. This is also the case if the dipole-proton scattering amplitude depends on the center-of-mass of the $q\Bar q$ system $\bt' = z_0 \xt_0 + z_1 \xt_1$ and not on the impact parameter $\bt = (\xt_0 + \xt_1)/2$ (see discussion in Sec.~\ref{sec:2part_ps}).
    \item The dipole-target interaction does not depend on the orientation of the dipole  or that of the impact parameter, i.e. $S_{01} \equiv S(\norm{\xt_0 - \xt_1},\norm{\bt}) \eqqcolon S_{r b}$. 
    The angular dependence is commonly neglected when the initial condition for the BK evolution is determined by fitting the HERA data~\cite{Beuf:2020dxl,Lappi:2013zma,Albacete:2010sy,Ducloue:2019jmy,Mantysaari:2018zdd} and in parametrizations such as IPsat~\cite{Kowalski:2003hm}. However, in general such a (probably weak) angular dependence should exist, see e.g.~\cite{Dumitru:2021tvw,Mantysaari:2020lhf,Mantysaari:2019csc,Salazar:2019ncp,Iancu:2017fzn}. 
\end{itemize}

Under these assumptions, the only dependence on the angle $\theta_{\rt,\Bar \rt}$ between $\rt \coloneqq \xt_{01}$ and $\conj \rt \coloneqq \xt_{\conj{0} \conj{1}}$ is in the phase space integral ${\cal I}_{\mx}^{(2)}$. This integral then gives
\begin{equation}
    \int \ud \theta_{\rt,\Bar \rt} \, {\cal I}_{\mx}^{(2)} 
    =
    \int \ud \theta_{\rt,\Bar \rt} \, \besj_0 \qty(\sqrt{z_0z_1}\mx \norm{\Bar \rt-\rt}) = 2\pi \besj_0 \qty(\sqrt{z_0z_1}\mx \norm{\rt}) \besj_0 \qty(\sqrt{z_0z_1} \mx \norm{\Bar \rt}),
\end{equation}
and as such it factorizes into parts that only depend on transverse coordinates in the amplitude or in the conjugate amplitude. Integrating over $t$ results in a delta function forcing $\bt = \Bar \bt$ (see Eq.~\eqref{eq:I2_Delta_tint}), and the cross section becomes
\begin{equation}
    \frac{\ud \sigma^{\text{D}}_{L, q \Bar q}}{\ud M_{X}^{2} }
    =
    \nc  \frac{e^2}{(2\pi)^2} \sum_f e_f^2
    \int_0^1 \! \ud z_0 \, z_0^3 (1-z_0)^3 \int \ud^2 \bt 
    \left\lbrace \int \ud r \, r
    \besj_0\left(\sqrt{z_0 (1-z_0)}\mx r\right) Q \besk_0 \left(r \Bar Q\right) (S_{r b}-1)  \right\rbrace^2,
\end{equation}
where we have substituted the longitudinal photon wavefunction summed over helicities shown in Eq.~\eqref{eq:photon_LO_L_squared}.
Note that if the impact parameter dependence factorizes from the dipole scattering amplitude, i.e. $(S_{r b}-1) \equiv T(\bt)[S_r - 1]$, then the impact parameter integral completely factorizes and gives  $\int \! \ud^2\bt |T(\bt)|^2$.

On the other hand, for the transversely polarized photons, the leading order reduced wavefunction squared depends on the angle between $\rt $ and $\Bar \rt$ as shown in Eq.~\eqref{eq:photon_LO_T_squared}:
\begin{equation}
    \sum_{T \, \textrm{pol.} \, \lambda}
    \sum_{h_0, h_1}
    \qty(\widetilde{\psi}_{\gamma^{*}_\lambda \rightarrow q_{\conz} \Bar{q}_{\cono}})^\dagger
    \qty(\widetilde{\psi}_{\gamma^{*}_\lambda \rightarrow q_0 \Bar{q}_1})
    \propto
    \frac{\rt \cdot \Bar \rt }{ \norm{\rt} \norm{\Bar \rt}},
\end{equation}
which means that the part that depends on the dipole sizes $\rt$ and $\Bar \rt$ --- omitting the dipole amplitudes for now --- reads:
\begin{equation}
    \int \ud^2 \rt \! \int \ud^2 \Bar \rt ~
    {\cal I}_{\mx}^{(2)}
    \frac{\rt \cdot \Bar \rt }{ \norm{\rt} \norm{\Bar \rt}}
     = 
    \int \ud^2 \rt \! \int \ud^2 \Bar \rt
    \int \frac{\ud^2 \lt}{(2\pi)^2}
    \delta \qty(\lt^2 - z_0(1-z_0)\mx^2)
    \frac{\rt \cdot \Bar \rt }{ \norm{\rt} \norm{\Bar \rt}}
    e^{i \lt \cdot (\Bar \rt - \rt)}.
\end{equation}
Parametrizing the angles as $\angle (\lt, \rt) \eqqcolon \theta$ and $\angle (\lt, \conj{\rt}) \eqqcolon \Bar{\theta}$, we have for the dot product: $\rt \cdot \Bar \rt = r \conj{r} (\cos \theta \cos  \Bar{\theta} + \sin \theta \sin  \Bar{\theta})$, where  the sine term vanishes in the integration. Thus we are left with
\begin{multline}
    \int  r \ud r \ud \theta \int \conj{r} \ud\conj{r} \ud  \Bar{\theta}
    \int \frac{\ud^2 \lt}{(2\pi)^2} \delta \qty(\lt^2 - z_0(1-z_0) \mx^2)
    \cos\theta \cos \Bar{\theta} \,
    e^{i l \Bar r \cos \Bar{\theta}} e^{-i l r \cos\theta}
    \\
    =
    \pi \int r \ud r \int \Bar r \ud \Bar r ~
    \besj_1 \qty(\sqrt{z_0(1-z_0)} \mx r) \besj_1 \qty(\sqrt{z_0(1-z_0)} \mx \Bar r).
\end{multline}
Consequently, we find that (only) under the assumptions listed at the beginning of this subsection the diffractive cross section can be written in a factorized form independently of the photon polarization. Otherwise the transverse coordinates in the amplitude and conjugate amplitude are mixed. In the future it will be interesting to study numerically the effect of these assumptions, that were used e.g. in Ref.~\cite{Kowalski:2008sa} where a good description of the HERA diffractive structure function data was obtained. Both cross sections can now under these assumptions be expressed in terms of an auxiliary function, denoting now $z_0=z$ with the integral over $z_1$ is performed using the $\delta$ function
    \begin{equation}
        \Phi_n(z,\beta,Q,\bt) = \left[ \int \!\! \ud r \, r \besj_n \qty(\sqrt{z (1-z)} \mx r) \besk_n \qty(\sqrt{z (1-z)} Q \, r) \bigl(S_{r b}-1 \bigr) \right]^2.
    \end{equation}
Using this, we may write the diffractive structure functions as
    \begin{align}
        \xpom F^{\rm D}_{L, \qqb} \qty(\beta, \xpom, Q^2)
        & =
        \frac{\nc Q^4}{2 \pi^3 \beta} \sum e_f^2
        \int \ud^2 \bt \int_0^1 \ud z \, z^3 (1-z)^3 Q^2 \Phi_0(z,\beta,Q,\bt),
        \\
        \xpom F^{\rm D}_{T, \qqb} \qty(\beta, \xpom, Q^2)
        & =
        \frac{\nc Q^4}{8 \pi^3 \beta} \sum e_f^2
        \int \ud^2 \bt \int_0^1 \ud z \, z^2 (1-z)^2 \qty(z^2 - (1-z)^2) Q^2 \Phi_1(z,\beta,Q,\bt) ,
    \end{align}
which is in agreement with Ref.~\cite{Kowalski:2008sa}, once one accounts for the different normalization of the dipole amplitude and the different integration domain of $z$.

\section{Tree-level \texorpdfstring{$q\Bar q g$}{qqg}-contribution to the diffractive structure functions}
\label{sec:gluonradiation}

In this section we present the main result of this paper: the tree-level calculation of diffractive $q \Bar q g$ production as a function of $\mx^2$ and $t$. We consider the case where the gluon is emitted before the shockwave and the $q\Bar q g$ system then interacts with the target, corresponding to the diagrams \ref{diag:gwavef1}, \ref{diag:gwavef2}, \ref{diag:gwavefinst1} and \ref{diag:gwavefinst2}. The emission-after-shock contribution could then in principle be obtained by taking the appropriate coordinate limits following the method developed in Refs.~\cite{Iancu:2018hwa,Iancu:2020mos}. As discussed in Sec.~\ref{sec:nlodiffraction}, we will however leave it to a future publication. For simplicity we only consider the massless quark limit in this work. 

\label{sec:before_shock}

The diffractive $q\Bar q g$ production cross section was written in Sec.~\ref{sec:3part_ps} (see Eq.~\eqref{eq:qqg_diffxs_Mx_t}) in terms of the phase space integrals ${\cal I}_{\mx}^{(3)}$ and ${\cal I}_{\Deltat}^{(3)}$ given in Eqs.~\eqref{eq:IMX3} and~\eqref{eq:IDelta3} as 
\begin{align}
\label{eq:qqg_diffxs_Mx_t_copied}
    \frac{\ud \sigma^{\text{D}}_{\lambda, \, q \Bar q g}}{\ud \mx^{2} \ud \abs{t}}
    = 
    &
4 \pi^4    
\nc \cf
    \int_0^1 \frac{\ud z_0}{z_0}
	\int_0^1 \frac{\ud z_1}{z_1}
	\int_0^1 \frac{\ud z_2}{z_2}
     \delta(z_0+z_1+z_2-1)
     \int_{\xt_0} \int_{\xt_1} \int_{\xt_2} \int_{\Bar \xt_0} \int_{\Bar \xt_1} \int_{\Bar \xt_2}
        {\cal I}_\Deltat^{(3)} {\cal I}_{\mx}^{(3)}
    \\
    \notag
    &
    \times
    \sum_{h_0, h_1, \lambda_2}
    \qty(\widetilde{\psi}_{\gamma^{*}_\lambda \rightarrow q_{\Bar 0} \Bar q_{\Bar 1} g_{\Bar 2}})^\dagger
    \qty(\widetilde{\psi}_{\gamma^{*}_\lambda \rightarrow q_0 \Bar{q}_1 g_2})
    \left[S_{\conj{0} \conj{1} \conj{2}}^{\dagger} -1\right] \left[S_{012} -1\right].
\end{align}

The only part missing from the diffractive $q\Bar q g$ production cross section is thus the calculation of the square of the tree-level wavefunction $\widetilde{\psi}_{\gamma^{*}_\lambda \rightarrow q_0 \Bar{q}_1 g_2}$, with different transverse coordinates in the amplitude ($\xt_i$) and in the conjugate amplitude ($\conj{\xt}_i$), summed over
helicities. The plus momentum fractions $z_i$ are external kinematical variables and therefore the same in the direct and complex conjugate amplitude. We calculate this square using the wavefunctions for the longitudinally and transversely polarized photons in $4$ dimensions from Ref.~\cite{Beuf:2017bpd} (see also Refs.~\cite{Beuf:2022ndu,Beuf:2021qqa,Beuf:2021srj,Beuf:2011xd}),
with the modification that the factor of $2\qplus$ has been taken out of the reduced wavefunctions $\widetilde{\psi}$ in the definition~\eqref{eq:lfwf-qqbarg}. The reduced LFWF for the longitudinal photon reads
		\begin{multline}
			\widetilde{\psi}_{\gamma_L^{*}\rightarrow q_0\Bar{q}_1g_2}^{\textrm{Tree}}
			=      e\, e_f\, g\; \frac{i}{(2\pi)^2} \varepsilon_{\lambda_2}^{j *}\,
			2Q\; \textrm{K}_0\!\left(Q X_{012}\right)
			\sqrt{z_0}\, \sqrt{z_1}\; \delta_{h_1,-h_0}
			\\
			\times
			\Bigg\{ z_{1} 
			\Big[(2 z_{0} \!+\! z_{2}) \delta^{jm}
			-i(2h_0)\, z_{2}\,  \epsilon^{jm}\Big]\:
			\left(\frac{\xt_{20}^m}{\xt_{20}^2}\right)
			-z_{0}
			\Big[(2 z_{1} \!+\! z_{2} ) \delta^{jm}
			+i(2h_0)\, z_{2} \,  \epsilon^{jm}\Big]\:
			\left(\frac{\xt_{21}^m}{\xt_{21}^2}\right)
			\Bigg\}
			\, ,
			\label{eq:qqbarg_WF_L_mixed_4D}
		\end{multline}
whereas for the transverse photon we have:
		\begin{align}
			\widetilde{\psi}_{\gamma_{\lambda}^{*}\rightarrow q_0\Bar{q}_1g_2}^{\textrm{Tree}}
			=
			&
			\frac{ e\, e_f\, g}{(2\pi)^2}\;
			\varepsilon_{\lambda}^{i}\, \varepsilon_{\lambda_2}^{j *}\;
			\sqrt{z_0}\, \sqrt{z_1}\; \delta_{h_1,-h_0}\;
			\frac{Q}{X_{012}}\; \textrm{K}_1\!\left(Q X_{012}\right)
			\nonumber\\
			&
			\times
			\Bigg\{ z_1
			\Big[(2 z_0 \! + \! z_{2}) \delta^{jm}
			-i\, (2h_0)\, z_{2} \,  \epsilon^{jm} \Big]\!
			\Big[(2 z_{1} \! - \! 1) \delta^{il}
			-i\, (2h_0) \, \epsilon^{il} \Big]\;
			\xt_{0+2;1}^l\, \left(\frac{\xt_{20}^m}{\xt_{20}^2}\right)
			\nonumber\\
			&
			\hphantom{\Bigg\lbrace}
			+z_0
			\Big[(2z_{1} \! + \! z_{2}) \delta^{jm}
			+i(2h_0)\, z_{2} \,  \epsilon^{jm}\Big]\! \Big[(2 z_{0} \!-\! 1) \delta^{il}
			+ i\, (2h_0) \, \epsilon^{il}\Big] \;
			\xt_{0;1+2}^l\, \left(\frac{\xt_{21}^m}{\xt_{21}^2}\right)
			\nonumber\\
			&
			\hphantom{\Bigg\lbrace}
			- \frac{z_{0} z_{1} z_{2}}{z_{0} \! + \! z_{2}}\;
			\Big[\delta^{ij}  - i\, (2h_0)\, \epsilon^{ij}\Big]
			+ \frac{z_{0} z_{1} z_{2}}{z_{1} \!+\! z_{2}}\;
			\Big[\delta^{ij} + i\, (2h_0)\, \epsilon^{ij}\Big]
			\Bigg\}
			\, ,
			\label{eq:qqbarg_WF_T_mixed_4D}
		\end{align}
		where $X_{012}$, $\xt_{0+2;1}$ and $\xt_{0;1+2}$ are defined as:
		\begin{align}
			X_{012}^2 
			& \coloneqq
			z_{0} z_{1} \xt_{01}^2 + z_{0} z_{2} \xt_{02}^2 + z_{1} z_{2} \xt_{12}^2
			\\
			\xt_{0+2;1}
			& \coloneqq
			- \frac{z_0}{z_0+z_2} \xt_{20} + \xt_{21}
			=
			\xt_{01} + \frac{z_2}{z_0 + z_2} \xt_{{2} {0}}
			\\
			\xt_{0;1+2}
			& \coloneqq
			-\xt_{{2} {0}} +\frac{z_1}{z_1+z_2} \xt_{{2} {1}}
			=
			\xt_{01} - \frac{z_2}{z_1+z_2} \xt_{{2} {1}}
			\, .
		\end{align}
The quantity $Q^2X_{012}^2$ corresponds to the $q\Bar q g$ formation time divided by the lifetime of the virtual photon that forms the $q\Bar q g$ system, as discussed in more detail in Ref.~\cite{Beuf:2011xd}. Configurations with large $Q^2 X_{012}^2$ are exponentially suppressed, which enforces the restriction that the $\qqbg$ state must develop within a formation time that is less than the lifetime of the virtual photon. 

The calculation of the squared wavefunctions $\sum_{h_0, h_1, \lambda_2}
\qty(\widetilde{\psi}_{\gamma^{*}_\lambda \rightarrow q_{\Bar 0} \Bar q_{\Bar 1}  g_{\Bar 2}})^\dagger
\qty(\widetilde{\psi}_{\gamma^{*}_\lambda \rightarrow q_0 \Bar{q}_1 g_2})$ is cumbersome but straightforward. More technical details are given in Ref.~\cite{Hanninen:2021byo}. After a lot of algebra, we obtain the diffractive structure functions
\begin{align}
	\notag
	& \xpom F_{L, \, q \Bar q  g}^{\textrm{D}(4) \, \textrm{NLO}} (\xbj, Q^2, \beta, t)
	=
	4
	\frac{\as \nc \cf Q^4 }{\beta}
	\sum_f e_f^2
	\int_0^1 \frac{\ud z_0}{z_0}
	\int_0^1 \frac{\ud z_1}{z_1}
	\int_0^1 \frac{\ud z_2}{z_2}
	\delta(z_0 \! + \! z_1 \! + \! z_2 \! - \! 1)
	\nonumber
	\\
	& \quad \times
	\int_{\xt_0} \int_{\xt_1} \int_{\xt_2} \int_{\cxt_0} \int_{\cxt_1} \int_{\cxt_2}
	{\cal I}_{\mx}^{(3)} {\cal I}_{\Deltat}^{(3)}
    \,
	4 z_0 z_1 Q^2 \besk_0 \left( Q X_{012}\right) \besk_0 \left( Q X_{\conj{0} \conj{1} \conj{2}}\right)
	\nonumber
	\\
	&\qquad
	\times \Bigg\{
	z_1^2
	\Bigg[
	\left(2 z_0 (z_0 + z_2) + z_2^2\right)
	\left(
	\frac{\xt_{20}}{\xt_{20}^2} \cdot
	\left( \frac{\xt_{\conj{2} \conj{0}}}{\xt_{\conj{2} \conj{0}}^2}
	- \half \frac{\xt_{\conj{2} \conj{1}}}{\xt_{\conj{2} \conj{1}}^2} \right)
	- \half \frac{\xt_{\conj{2} \conj{0}} \cdot \xt_{21}}{\xt_{\conj{2} \conj{0}}^2 \xt_{21}^2} \right)
	+ \frac{z_2^2}{2}
	\left(
	\frac{\xt_{\conj{2} \conj{0}} \cdot \xt_{21}}{\xt_{\conj{2} \conj{0}}^2 \xt_{21}^2}
	+
	\frac{\xt_{{2} {0}} \cdot \xt_{\conj{2} \conj{1}}}{\xt_{{2} {0}}^2 \xt_{\conj{2} \conj{1}}^2}
	\right)
	\Bigg]
	\nonumber
	\\
	&\qquad \hphantom{\Bigg\{}
	+	
	z_0^2
	\Bigg[
	\left(2 z_1 (z_1 + z_2) + z_2^2\right)
	\left(
	\frac{\xt_{21}}{\xt_{21}^2} \cdot
	\left( \frac{\xt_{\conj{2} \conj{1}}}{\xt_{\conj{2} \conj{1}}^2}
	- \half \frac{\xt_{\conj{2} \conj{0}}}{\xt_{\conj{2} \conj{0}}^2} \right)
	- \half \frac{\xt_{{2} {0}} \cdot \xt_{\conj{2} \conj{1}}}{\xt_{{2} {0}}^2 \xt_{\conj{2} \conj{1}}^2} \right)
	+ \frac{z_2^2}{2}
	\left(
	\frac{\xt_{\conj{2} \conj{0}} \cdot \xt_{21}}{\xt_{\conj{2} \conj{0}}^2 \xt_{21}^2}
	+
	\frac{\xt_{{2} {0}} \cdot \xt_{\conj{2} \conj{1}}}{\xt_{{2} {0}}^2 \xt_{\conj{2} \conj{1}}^2}
	\right)
	\Bigg]
	\Bigg\}
	\nonumber
	\\
	&\quad \times 
	\left[ 1 - S_{\contrip}^{\dagger} \right] \left[ 1 - S_{012} \right] ,
	\label{eq:ddis-FL-qqbarg-nlo}
\end{align}
for the longitudinal structure function, and
\begin{align}
	\notag
	&\xpom F_{T, \, q \Bar q  g}^{\textrm{D}(4) \, \textrm{NLO}} (\xbj, Q^2, \beta, t)
	=
	2
	\frac{\as \nc \cf Q^4}{\beta} 
	\sum_f e_f^2
	\int_0^1 \frac{\ud z_0}{z_0}
	\int_0^1 \frac{\ud z_1}{z_1}
	\int_0^1 \frac{\ud z_2}{z_2}
	\delta(z_0 \! + \! z_1 \! + \! z_2 \! - \! 1)
	\nonumber
	\\
	& \qquad \times
	\int_{\xt_0} \int_{\xt_1} \int_{\xt_2} \int_{\cxt_0} \int_{\cxt_1} \int_{\cxt_2}
	{\cal I}_{\mx}^{(3)} {\cal I}_{\Deltat}^{(3)}
	\frac{z_0 z_1 Q^2}{X_{012} X_{\conj{0}\conj{1}\conj{2} }}
	\besk_1\left(QX_{012}\right) \besk_1\left(Q X_{\conj{0}\conj{1}\conj{2} }\right)
	\nonumber
	\\
	& \qquad \times
	\Big\lbrace
	\Upsilon^{(|\ref{diag:gwavef1}|^2)}_{\textrm{reg.}} + \Upsilon^{(|\ref{diag:gwavef2}|^2)}_{\textrm{reg.}} 
	+ \Upsilon^{\ref{diag:gwavefinst1}}_{\textrm{inst.}} + \Upsilon^{\ref{diag:gwavefinst2}}_{\textrm{inst.}} + \Upsilon^{\ref{diag:gwavef1} \! \times \! \ref{diag:gwavef2}}_{\textrm{interf.}}
	\Big\rbrace
	\left[ 1 - S_{\contrip}^{\dagger} \right] \left[ 1 - S_{012} \right]
	\label{eq:ddis-FT-qqbarg-nlo}
\end{align}
for the transverse structure function.

The $\Upsilon$-terms of the squared virtual photon light-front wavefunction are:
\begingroup
\allowdisplaybreaks
\begin{align}
	\label{eq:ddis-qqbarg-nlo-upsilon-a^2}
	\Upsilon^{(|\ref{diag:gwavef1}|^2)}_{\textrm{reg.}}
	=&
	z_1^2 \Bigg[
		(2z_0(z_0 + z_2) + z_2^2) (1 - 2z_1 (1-z_1))
		\left(\xt_{\conj{0} + \conj{2}; \conj{1}} \cdot \xt_{0+2;1} \right)
		\frac{(\xt_{\conj{2} \conj{0}} \cdot \xt_{{2} {0}})}{\xt_{\conj{2} \conj{0}}^2 \xt_{{2} {0}}^2}
		\nonumber\\
		&
		\hphantom{z_1^2 \Bigg[}
		- z_2 (2z_0 + z_2)(2z_1 - 1)
		\frac{
		    \left( \xt_{\conj{0} + \conj{2}; \conj{1}} \cdot \xt_{\conj{2} \conj{0}} \right)
		    \left( \xt_{{0} + {2}; {1}} \cdot \xt_{{2} {0}} \right)
		    -
		    \left( \xt_{\conj{0} + \conj{2}; \conj{1}} \cdot \xt_{{2} {0}} \right)
		    \left( \xt_{{0} + {2}; {1}} \cdot \xt_{\conj{2} \conj{0}} \right)
		}{\xt_{\conj{2} \conj{0}}^2 \xt_{{2} {0}}^2}
	\Bigg] ,
	\\
	\label{eq:ddis-qqbarg-nlo-upsilon-b^2}
	\Upsilon^{(|\ref{diag:gwavef2}|^2)}_{\textrm{reg.}}
	=&
	z_0^2 \Bigg[
		(2z_1(z_1 + z_2) + z_2^2) (1 - 2z_0 (1-z_0))
		\left(\xt_{\conj{0}; \conj{1} + \conj{2}} \cdot \xt_{0;1+2} \right)
		\frac{(\xt_{\conj{2} \conj{1}} \cdot \xt_{{2} {1}})}{\xt_{\conj{2} \conj{1}}^2 \xt_{{2} {1}}^2}
		\nonumber\\
		&
		\hphantom{z_0^2 \Bigg[}
		- z_2 (2z_1 + z_2)(2z_0 - 1)
		\frac{
		    \left( \xt_{\conj{0}; \conj{1}  + \conj{2}} \cdot \xt_{\conj{2} \conj{1}} \right)
		    \left( \xt_{{0}; {1}  + {2}} \cdot \xt_{{2} {1}} \right)
		    -
		    \left( \xt_{\conj{0}; \conj{1}  + \conj{2}} \cdot \xt_{{2} {1}} \right)
		    \left( \xt_{{0}; {1}  + {2}} \cdot \xt_{\conj{2} \conj{1}} \right)
		}{\xt_{\conj{2} \conj{1}}^2 \xt_{{2} {1}}^2}
	\Bigg] ,
	\\
	\label{eq:ddis-qqbarg-nlo-upsilon-a'}
	\Upsilon^{\ref{diag:gwavefinst1}}_{\textrm{inst.}}
	=&
	\frac{z_0^2 z_1^2 z_2^2}{(z_0 + z_2)^2}
	- \frac{z_0^2 z_1^3 z_2}{z_0 + z_2}
	\left( 
		\frac{\xt_{{0} + {2}; {1}} \cdot \xt_{{2} {0}}}{\xt_{{2} {0}}^2}
		+
		\frac{\xt_{\conj{0} + \conj{2}; \conj{1}} \cdot \xt_{\conj{2} \conj{0}}}{\xt_{\conj{2} \conj{0}}^2}
		\right)
	\nonumber\\
	&
	+ \frac{z_0^2 z_1 (z_1 + z_2)^2 z_2}{z_0 + z_2}
	\left( 
		\frac{\xt_{{0}; {1}+{2}} \cdot \xt_{{2} {1}}}{\xt_{{2} {1}}^2}
		+
		\frac{\xt_{\conj{0}; \conj{1}+ \conj{2}} \cdot \xt_{\conj{2} \conj{1}}}{\xt_{\conj{2} \conj{1}}^2}
		\right) ,
	\\
	\label{eq:ddis-qqbarg-nlo-upsilon-b'}
	\Upsilon^{\ref{diag:gwavefinst2}}_{\textrm{inst.}}
	=&
	\frac{z_0^2 z_1^2 z_2^2}{(z_1 + z_2)^2}
	- \frac{z_0 z_1^2 (z_0 + z_2)^2 z_2}{z_1 + z_2}
	\left( 
		\frac{\xt_{{0} + {2}; {1}} \cdot \xt_{{2} {0}}}{\xt_{{2} {0}}^2}
		+
		\frac{\xt_{\conj{0} + \conj{2}; \conj{1}} \cdot \xt_{\conj{2} \conj{0}}}{\xt_{\conj{2} \conj{0}}^2}
		\right)
	\nonumber\\
	&
	+ \frac{z_0^3 z_1^2 z_2}{z_1 + z_2}
	\left( 
		\frac{\xt_{{0}; {1}+{2}} \cdot \xt_{{2} {1}}}{\xt_{{2} {1}}^2}
		+
		\frac{\xt_{\conj{0}; \conj{1}+ \conj{2}} \cdot \xt_{\conj{2} \conj{1}}}{\xt_{\conj{2} \conj{1}}^2}
		\right) ,
	\\
	\label{eq:ddis-qqbarg-nlo-upsilon-ab}
	\Upsilon^{\ref{diag:gwavef1} \! \times \! \ref{diag:gwavef2}}_{\textrm{interf.}}
	=&
	- z_0 z_1
	\left[ z_1(z_0 + z_2) + z_0(z_1 + z_2) \right]
	\left[ z_0(z_0 + z_2) + z_1(z_1 + z_2) \right]
	\nonumber\\
	&\quad \times \!
	\left[
		\left( \xt_{\conj{0} + \conj{2}; \conj{1}} \cdot \xt_{0;1+2} \right)
		\frac{(\xt_{\conj{2} \conj{0}} \cdot \xt_{{2} {1}})}{\xt_{\conj{2} \conj{0}}^2 \xt_{{2} {1}}^2}
		+
		\left( \xt_{\conj{0}; \conj{1} + \conj{2}} \cdot \xt_{0+2;1} \right)
		\frac{(\xt_{\conj{2} \conj{1}} \cdot \xt_{{2} {0}})}{\xt_{\conj{2} \conj{1}}^2 \xt_{{2} {0}}^2}
	\right]
	\nonumber\\
	&
	+ z_0 z_1 z_2 (z_0-z_1)^2
	\nonumber\\
	& \quad \times \!
	\Bigg[
		\frac{
		    \left( \xt_{\conj{0} + \conj{2}; \conj{1}} \cdot \xt_{\conj{2} \conj{0}} \right)
		    \left( \xt_{{0}; {1}  + {2}} \cdot \xt_{{2} {1}} \right)
		    -
		    \left( \xt_{\conj{0} + \conj{2}; \conj{1}} \cdot \xt_{{2} {1}} \right)
		    \left( \xt_{{0}; {1}  + {2}} \cdot \xt_{\conj{2} \conj{0}} \right)
		}{\xt_{\conj{2} \conj{0}}^2 \xt_{{2} {1}}^2}
    	\nonumber\\
    	& \quad \quad
		+
		\frac{
		    \left( \xt_{\conj{0}; \conj{1}  + \conj{2}} \cdot \xt_{\conj{2} \conj{1}} \right)
		    \left( \xt_{{0} + {2}; {1}} \cdot \xt_{{2} {0}} \right)
		    -
		    \left( \xt_{\conj{0}; \conj{1}  + \conj{2}} \cdot \xt_{{2} {0}} \right)
		    \left( \xt_{{0} + {2}; {1}} \cdot \xt_{\conj{2} \conj{1}} \right)
		}{\xt_{\conj{2} \conj{1}}^2 \xt_{{2} {0}}^2}
	\Bigg] \!.
\end{align}%
\endgroup
It might be elucidating to note that the above expressions satisfy the following symmetries in particle exchanges:
\begin{align}
    \Upsilon^{(|\ref{diag:gwavef2}|^2)}_{\textrm{reg.}}
    & \equiv
    \Upsilon^{(|\ref{diag:gwavef1}|^2)}_{\textrm{reg.}} \qty(z_0 \leftrightarrow z_1, \xt_0 \leftrightarrow \xt_1, \Bar \xt_0 \leftrightarrow \Bar \xt_1)
    \\
    \Upsilon^{\ref{diag:gwavefinst2}}_{\textrm{inst.}}
    & \equiv
    \Upsilon^{\ref{diag:gwavefinst1}}_{\textrm{inst.}} \qty(z_0 \leftrightarrow z_1, \xt_0 \leftrightarrow \xt_1, \Bar \xt_0 \leftrightarrow \Bar \xt_1),
\end{align}
making their sum symmetric under the exchange of the quark and antiquark. Meanwhile the $\ref{diag:gwavef1} \! \times \! \ref{diag:gwavef2}$-interference term is already a sum of terms that can be obtained by this exchange, and is thus symmetric by itself.

The dipole amplitude $N_{ij}=1-S_{ij}$ satisfies the BK or JIMWLK high-energy evolution equation, where the evolution is parametrized in terms of the evolution rapidity defined as fraction of the probe photon plus momentum carried by the gluon. Following Ref.~\cite{Beuf:2020dxl} where the initial condition for the BK evolution has been determined at NLO accuracy, this evolution rapidity can be taken as 
\begin{align}
    Y &= \log \left( \frac{W^2 z_2}{Q_0^2} \right),
\end{align}
where $W$ is the center-of-mass energy for the photon-nucleon scattering and $Q_0^2$ is some typical hadronic transverse momentum scale of the target, and taken as $Q_0^2 = 1\gev^2$ in Ref.~\cite{Beuf:2020dxl}.

These contributions to the diffractive structure functions are finite without requiring any additional cancellations with other diagrams. This is because the invariant mass of the final state is fixed, and thus ultraviolet divergences do not appear. This also means that the divergences in the loop contributions that we are not calculating here should cancel against each other, see discussion in Sec.~\ref{sec:nlodiffraction}. Note that without the $M_X^2$ restriction the integration over the final state momenta would set $\conj \xt_i \to \xt_i$ and an ultraviolet divergence $\sim \int \ud^2 \xt_{02}/\xt_{02}^2$ or $\sim \int \ud^2 \xt_{12}/\xt_{12}^2$ would appear. In the diffractive structure functions the corresponding structure is $\sim \int \ud^2 \xt_{i2} \xt_{i2}/\xt_{i2}^2$ (with $i=0,1$) which is UV finite. 
The integration over the momentum transfer sets the center-of-mass of the $q\Bar q g$ system $\bt$ to be the same in the amplitude and in the conjugate amplitude, $\bt = \conj \bt$, but this does not affect the behavior of these integrals in the ultraviolet region (note that  $\mathcal{I}^{(3)}_{M_X}$ does not depend on $\bt$ or on $\conj \bt$).

A potential (transverse)  infrared divergence is removed, for a gluon emitted before the shockwave, by the dipole amplitude part vanishing for $|\xt_{02}| \sim |\xt_{12}| \to \infty$. For the emissions after the shockwave that we are not calculating here, these configurations are not suppressed by the Wilson line correlator, and need to cancel against the wavefunction renormalization of the outgoing quarks. 
Similarly there is no soft gluon divergence in the limit $z_2 \to 0$, as the invariant mass, which we keep finite, gives a lower bound for the integral $z_2 \gtrsim 1/\mx^2$.
For a parametrically large $\mx$ our result would give a large logarithm $\sim \ln \mx^2$ from the lower limit of the $z_2$ integration. While such contributions could be resummed~\cite{Kovchegov:1999ji}, they are not easily accessible at EIC energies and we will not consider this resummation further here. 
The BK/JIMWLK evolution of the target, on the other hand, is associated with the  $z_2 \to 0$ limit of contributions where the gluon crosses the shockwave, but is reabsorbed and not measured in the final state, which we are leaving to future work. 

The interpretation of the cumbersome $\Upsilon$-terms is actually straightforward. First, $\Upsilon^{(|\ref{diag:gwavef1}|^2)}_{\textrm{reg.}}$ describes the contribution where the gluon is emitted by the quark in the amplitude and absorbed by the same quark in the conjugate amplitude. Similarly, $\Upsilon^{(|\ref{diag:gwavef2}|^2)}_{\textrm{reg.}}$ corresponds to the case where the antiquark emits and absorbs the gluon. Furthermore, the instantaneous gluon emission and absorption by the quark (antiquark) is described by $\Upsilon^{\ref{diag:gwavefinst1}}_{\textrm{inst.}}$ ($\Upsilon^{\ref{diag:gwavefinst2}}_{\textrm{inst.}}$). Finally, gluon emission by the quark and absorption by the antiquark (or vice versa) contributes the term $\Upsilon^{\ref{diag:gwavef1} \! \times \! \ref{diag:gwavef2}}_{\textrm{interf.}}$. Note that the instantaneous contribution only appears as a part of the transverse cross section. The interference between the regular and instantaneous gluon emissions is included in the terms $\Upsilon^{\ref{diag:gwavefinst1}}_{\textrm{inst.}}$ and $\Upsilon^{\ref{diag:gwavefinst2}}_{\textrm{inst.}}$, with the former including the interference contributions containing the $\ref{diag:gwavefinst1}$-diagram, and similarly the latter those of the $\ref{diag:gwavefinst2}$-diagram.

We emphasize that prior to this work the $q\Bar qg$ contribution to the diffractive cross section has only been known in approximate kinematics and for a transverse photon only (which dominates at high $Q^2$), see discussion in Sec.~\ref{sec:limits}. For the longitudinal polarization even approximative results have been missing from the literature. 
The cross sections~\eqref{eq:ddis-FL-qqbarg-nlo} and~\eqref{eq:ddis-FT-qqbarg-nlo}, that are main results of this work, are finite and can be straightforwardly implemented in phenomenological applications. In a future work we plan to apply these results to describe the HERA diffractive structure function data.

\section{Recovering  known limits}
\label{sec:limits}

\subsection{The large-\texorpdfstring{$M_X$}{Mx} limit}
\label{sec:mslimit}

As a verification of our calculation, here we will extract the $q \Bar q g$-contribution for $F_T^\textrm{D}$ in the limit of large $M_X$, i.e. in the limit when the emitted gluon is soft. This limit has been considered by several authors, e.g. in Refs.~\cite{Bartels:1999tn,Kovchegov:1999ji,Kopeliovich:1999am,Kovchegov:2001ni,Munier:2003zb,Golec-Biernat:2005prq}. To be specific, we will use the result as it is written in Ref.~\cite{Munier:2003zb}  and also used in the phenomenological studies~\cite{Marquet:2007nf,Kowalski:2008sa}. The derivation simplifies considerably in the $\beta \to 0$ limit, i.e. for final states with arbitrarily large invariant masses $\mx$ in comparison to the virtuality of the photon: $\mx^2 \gg Q^2$. The result of Ref.~\cite{Munier:2003zb}  for the $\qqb g$-contribution --- including gluon emissions from the quark-antiquark dipole both \textit{before} and \textit{after} the scattering off the shockwave\footnote{The authors of Ref.~\cite{Munier:2003zb} justify the contribution from the emission after the shockwave as resulting from the normalization of the photon state, which is possible in the soft gluon limit. However, this contribution is better understood as resulting from emissions after the shockwave, see discussion in Ref.~\cite{Golec-Biernat:2005prq}. The normalization condition for the photon state can only be used to obtain such contributions  in the soft gluon limit where the $\gamma\to \qqbg$ wavefunction factorizes into $\gamma\to \qqb$ and gluon emission wavefunctions, but  not in  general kinematics~\cite{Beuf:2016wdz,Beuf:2017bpd}.} --- is, converted to our notations:
\begin{multline}
\label{eq:mslimit}
    \xpom F_{T,q\Bar q g}^{\textrm{D}\ (\text{MS})}(\xpom, \beta=0, Q^2)
    = 
    \frac{\as \nc \Cf Q^2}{16\pi^5 \aem} 
    \int \ud^2 \xt_0 \int \ud^2 \xt_1 \int \ud^2 \xt_2 \int_0^1 \frac{\ud z}{z(1-z)} 
    \left |\widetilde{\psi}_{\gamma^{*}_\lambda \rightarrow q_{\Bar 0} \Bar{q}_{\Bar 1}}^{\rm LO} \right|^2 
    \\ \times
    \frac{\xt_{01}^2}{\xt_{02}^2\xt_{12}^2} 
    \bigg[ N_{02} + N_{12} - N_{01} - N_{02}N_{12} \bigg]^2 .
\end{multline}
A key feature of this result is that the $\qqb g$ LFWF has been factorized to the leading order $\qqb$ LFWF, the BK kernel describing the gluon emission, and  the scattering of the tripole state in terms of scatterings of daughter dipoles.

Our starting point to rederive the large-$\mx$ result is  Eq.~\eqref{eq:phasespace_3part_start}, with the final state momentum integrals still undone. First we note that the leading process to create high invariant mass final states is caused by the emissions of soft gluons, with $z_2 \ll 1$, and at the limit $z_2 \to 0$ we have $M_X^2 \approx \Pt_2^2/z_2$. We consider a $t$ integrated cross section, whose momentum dependence in this limit reads
\begin{multline}
       \frac{\ud \sigma^{\text{D}}_{\gamma^{*}_\lambda \rightarrow q \Bar{q} g}}{\ud \mx^2} \sim \int \frac{\ud z_2}{z_2} \int \frac{\ud^2 \pt_0}{(2\pi)^2} \int \frac{\ud^2 \pt_1}{(2\pi)^2} \int \frac{\ud^2 \pt_2}{(2\pi)^2}  e^{i\xt_{\conj 0 0} \pt_0 + i \xt_{\conj 1 1}\pt_1 + i \xt_{\conj 2 2}\pt_2} \delta\left( \frac{\pt_2^2}{z_2} - \mx^2\right) \\
       =  \delta^{(2)}(\xt_0-\xt_{\conj 0}) \delta^{(2)}(\xt_1-\xt_{\conj 1}) 
       \int \ud z_2 \int \frac{\ud^2 \pt_2}{(2\pi)^2}  e^{i \xt_{2\conj 2}\pt_2} \frac{z_2}{\pt_2^2} \delta\left(z_2 - \frac{\pt_2^2}{\mx^2}\right) = \frac{1}{\mx^2} \delta^{(2)}(\xt_0-\xt_{\conj 0}) \delta^{(2)}(\xt_1-\xt_{\conj 1}) \delta^{(2)}(\xt_2 - \xt_{\conj 2}),
\end{multline}
where we assumed that $\mx^2$ is dominated by the gluon light cone energy $\pt_2^2/z_2$. We have here integrated over $z_2$, assuming that $\mx$ is large enough so that it is always possible to find a solution to the delta function constraint  $z_2 = \pt_2^2/\mx^2$ with $0<z_2<1$. In reality this is not possible for arbitrarily large $\pt_2$. Thus our approximation has rendered the $\pt_2$ integral unbounded, resulting in a delta function setting $\xt_2 - \xt_{\conj 2}$. This approximation, as discussed in Sec.~\ref{sec:nlodiffraction} and above in Sec. \ref{sec:gluonradiation}, would make the cross section UV divergent unless one also includes the diagrams with emission after the shockwave, using the procedure of subtracting from the ``emission before'' term the appropriate coordinate limit. Thus we will include these contributions here, unlike in the rest of this paper. 

At this point we can integrate over the coordinates in the conjugate amplitude using the delta functions. Starting from the equation  \nr{eq:phasespace_3part_start} we now get
\begin{multline}\label{eq:msstartingpoint}
   \frac{\ud \sigma^{\text{D}}_{\gamma^{*}_\lambda \rightarrow q \Bar{q} g}}{\ud \mx^2 \ud \xpom}
   =\frac{\nc \cf}{(4\pi)^2}
    \int_0^1\! \frac{\ud z_0}{z_0} \!\int_0^1\! \frac{\ud z_1}{z_1}  
    \delta(z_0 + z_1 -1)
    \int \ud^2 \xt_0 \ud^2 \xt_1 \ud^2 \xt_2  
    \frac{1}{\mx^2}
\\
        \sum_{f,h_0, h_1, \lambda_2}
    \left| \widetilde{\psi}_{\gamma^{*}_\lambda \rightarrow q_{\Bar 0} \Bar{q}_{\Bar 1} g_{\Bar 2}}   \left(S_{012}^{\dagger} -1\right)- [\text{emission after}]\right|^2
\end{multline}
For calculating the subtraction terms correctly, we need to decompose the $\gamma \to \qqbg$ wavefunction into gluon emission and effective $\gamma\to\qqb$ parts according to \eq\nr{eq:splitforsubtr}. Here the limit of $z_2 \to 0$ simplifies things dramatically. In the soft gluon limit the wavefunction is given by 
(note that in our normalization conventions for the reduced wavefunctions, see \eqs\nr{eq:lfwf-qqbar} and~\nr{eq:lfwf-qqbarg}, this relation is true without additional coefficients):
\begin{equation}
  \widetilde{\psi}_{\gamma^{*}_\lambda \rightarrow q_0 \Bar{q}_1 g_2}
\underset{z_2\to 0}{\approx}     
   \widetilde{\psi}_{\gamma^{*}_\lambda \rightarrow q_0 \Bar{q}_1}^{\rm LO}
  \left[
   \widetilde{\psi}_{ q_0  \rightarrow q_0  g_2}
  +
  \widetilde{\psi}_{ \Bar{q}_1  \rightarrow \Bar{q}_1  g_2}
\right],
\end{equation}
where the effective $\gamma \to \qqb$ wavefunctions $\widetilde{\psi}_{\gamma^{*}_\lambda \rightarrow q_0 \Bar{q}_1 ; q_0  \rightarrow q_0  g_2} $ and
$ \widetilde{\psi}_{\gamma^{*}_\lambda \rightarrow q_0 \Bar{q}_1 ; \Bar{q}_1  \rightarrow \Bar{q}_1  g_2}$ 
defined by \eq\nr{eq:splitforsubtr}
have become independent of the gluon emission in the limit $z_2\to 0$. This simplification can be understood in several  equivalent ways. In coordinate space, the argument is that for a soft emission $z_2\to 0$ the coordinate of the emitting particle does not change, and therefore the original $\gamma \to \qqb$ splitting is independent of the later gluon emission. For momentum space wavefunctions the reason is that in the limit $z_2\to 0$ the gluon light cone energy blows up and thus energy denominators for gluon emission become independent of the state emitting the gluon
\begin{equation}
    k^-_\qqbg - k^-_\gamma\approx k^-_g \approx     k^-_{qg} - k^-_q \approx k^-_{\Bar{q}g} - k^-_{\Bar{q}}.
\end{equation}
This leads to a factorization of the $\gamma\to \qqbg$ wavefunction into leading order $\gamma\to \qqb$ and $q\to qg,$ $\Bar{q}\to \Bar{q}g$ wavefunctions in momentum space, which is carried over into coordinate space. We recall from Sec.~\ref{sec:radnlo} that the subtraction procedure consists in pulling out the gluon emission wavefunction  and then taking the limit of the gluon coordinate being equal to its emitter in the remaining factors. In this case we can also factor out the common $\gamma\to\qqb$ part. Thus we get, restoring the coordinates for clarity
\begin{multline}
    \Bigl[   \widetilde{\psi}_{\gamma^{*}_\lambda \rightarrow q_0 \Bar{q}_1 g_2} (\xt_0,\xt_1,\xt_2 )  
    \left(S_{012} -1\right) - [\text{emission after}] \Bigr] 
\\
= 
 \widetilde{\psi}_{\gamma^{*}_\lambda \rightarrow q_0 \Bar{q}_1}^{\rm LO} (\xt_0,\xt_1)
\Bigg\{
 \widetilde{\psi}_{ q_0  \rightarrow q_0  g_2}(\xt_0,\xt_2)
    \left[\left(S_{012} -1\right)- \left. \left(S_{012} -1\right)\right|_{\xt_2\to \xt_0} \right]
\\
+ 
 \widetilde{\psi}_{ \Bar{q}_1  \rightarrow \Bar{q}_1  g_2}(\xt_1,\xt_2)
    \left[\left(S_{012} -1\right)- \left. \left(S_{012} -1\right)\right|_{\xt_2\to \xt_1} \right]
\Biggr\}
\end{multline}
We now use the coordinate limits of the ``tripole''  (see e.g.~\cite{Beuf:2017bpd,Hanninen:2017ddy}) operators
\begin{equation}
\left.S_{012} \right|_{\xt_2\to \xt_0}
=
\left. S_{012} \right|_{\xt_2\to \xt_1}
= S_{01}
\end{equation}
and the relation (see also e.g.~\cite{Beuf:2017bpd,Hanninen:2017ddy})
\begin{equation}
    S_{012}-S_{01} = \frac{\nc^2}{\nc^2-1}\left(S_{02}S_{21}- S_{01}\right) \overset{\nc\to \infty}{\approx} 
S_{02}S_{21}- S_{01} = N_{02}N_{21}- N_{02} -N_{21} + N_{01}
\end{equation}
with $N_{ij} = 1-S_{ij}$
to get 
\begin{multline}\label{eq:subtractedMS}
    \Bigl[   \widetilde{\psi}_{\gamma^{*}_\lambda \rightarrow q_0 \Bar{q}_1 g_2} (\xt_0,\xt_1,\xt_2 )  
    \left(S_{012} -1\right) - [\text{emission after}] \Bigr] 
\\
= 
\widetilde{\psi}_{\gamma^{*}_\lambda \rightarrow q_0 \Bar{q}_1}^{\rm LO} (\xt_0,\xt_1)
\left\{
 \widetilde{\psi}_{ q_0  \rightarrow q_0  g_2}(\xt_0,\xt_2)
+ 
 \widetilde{\psi}_{ \Bar{q}_1  \rightarrow \Bar{q}_1  g_2}(\xt_1,\xt_2)
\right\}
\left[N_{02}N_{21}- N_{02} -N_{21} + N_{01}\right].
\end{multline}
We can now use the gluon emission wavefunction, which in  our conventions (see \eq~(37) of Ref.~\cite{Lappi:2016oup}   and recall the normalization of the reduced wavefunction from \nr{eq:lfwf-qqbar},  \nr{eq:lfwf-qqbarg})  reads
\begin{equation}
 \widetilde{\psi}_{ q_0  \rightarrow q_0  g_2}(\xt_0,\xt_2) \underset{z_2\to 0}\approx 
2 g   \int  \frac{\ud^2\pt_2}{(2\pi)^2}
 \frac{\pt_2\cdot \epst^*_{\lambda_2}}{\pt_2^2}
 e^{i\pt_2\cdot (\xt_2-\xt_0)}
=    \frac{2 g i }{2\pi} \frac{(\xt_2-\xt_0)\cdot \epst^*_{\lambda_2}}{(\xt_2-\xt_0)^2}
\end{equation}
and similarly for the antiquark. 
Using this relation in \eq\nr{eq:msstartingpoint} enables the  familiar calculation of the BK kernel
\begin{equation} \label{eq:bkkernel}
\sum_{\lambda_2}
\left|
 \widetilde{\psi}_{ q_0  \rightarrow q_0  g_2}(\xt_0,\xt_2)
+ 
 \widetilde{\psi}_{ \Bar{q}_1  \rightarrow \Bar{q}_1  g_2}(\xt_1,\xt_2)
\right|^2=
\frac{g^2}{\pi^2}
    \frac{\xt_{01}^2}{\xt_{02}^2 \xt_{21}^2} 
    = \frac{4 \as}{\pi}
    \frac{\xt_{01}^2}{\xt_{02}^2 \xt_{21}^2} .
\end{equation}
Inserting the squared wavefunctions from \eq\nr{eq:bkkernel} and \nr{eq:subtractedMS} into \nr{eq:msstartingpoint} one then directly recovers the  result \nr{eq:mslimit}.

\subsection{The large-\texorpdfstring{$Q^2$}{Q²} limit}

In this section we will rederive the W\"usthoff result for the $q \Bar q g$-contribution to $F_T^D$ \cite{Wusthoff:1997fz,GolecBiernat:1999qd,GolecBiernat:2001mm}, which is an approximate result at the large-$Q^2$ limit for this next-to-leading order contribution. Specifically we will verify that the Wüsthoff result emerges
from the NLO result calculated in exact kinematics when one takes the large-$Q^2$ limit.
The Wüsthoff result for the $q \Bar q g$-contribution has been extensively used in phenomenology~\cite{GolecBiernat:1999qd, Marquet:2007nf, Kowalski:2008sa, Bendova:2020hkp}, and it has some special features we seek to understand in depth.
It can be written in coordinate space in the appealing short form:\footnote{We use the explicit form from Ref.~\cite{Kowalski:2008sa}, which is written in Color Glass Condensate formalism. Its connection to the original two-gluon exchange formulation~\cite{Wusthoff:1997fz,GolecBiernat:1999qd} is discussed in Refs.~\cite{Marquet:2007nf,Hanninen:2021byo}.}
\begin{multline}\label{eq:wusthoffqqbarg}
\xpom F_{T,q\Bar{q}g}^\textrm{D (GBW)} \qty(\xpom,\beta,Q^2) = 
\frac{\as\beta}{8\pi^4}\sum_{f}e_f^2 
\int \ud^2 \bt
\int_0^{Q^2} \ud k^2 
\int_{\beta}^{1} \ud z \Bigg\{
k^4 \ln \frac{Q^2}{k^2}
\left[\left(1\!-\!\frac{\beta}z\right)^2+\left(\frac{\beta}z\right)^2\right]
\\
\times \bigg[\int_0^\infty \! \ud r \, r 
 \dsigmaadj \qty(\bt,\rt,\xpom) 
\besk_2\qty(\sqrt{z}k r) \besj_2\qty(\sqrt{1-z}k r)
\bigg]^2
\Bigg\},
\end{multline}
while a momentum space version can be found e.g. from Ref.~\cite{GolecBiernat:1999qd}.

\begin{figure*}[tb]
\centerline{
\includegraphics[width=4cm]{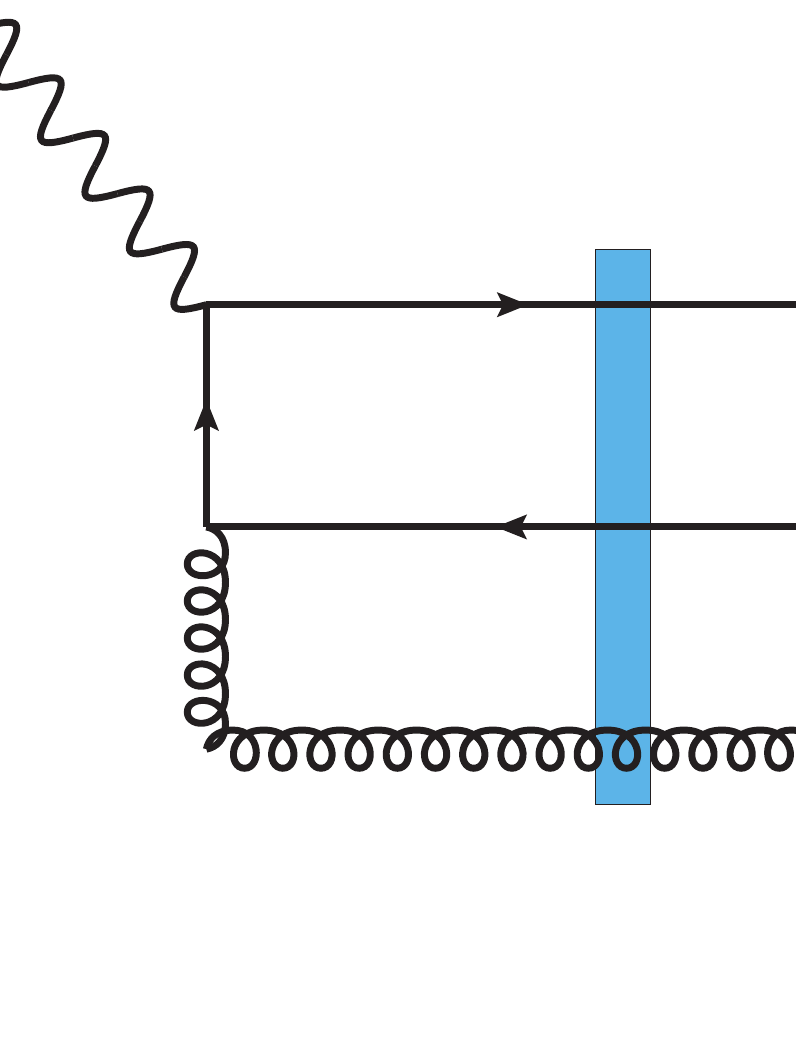}
\begin{tikzpicture}[overlay]
 \node[anchor=south east,colorkolme] at (-4cm,4.5cm) {$q^+$};
 \node[anchor=east,colorkolme] at (0.7cm,3.8cm) {$z_0 q^+$};
 \node[anchor=east,colorkolme] at (0.7cm,2.6cm) {$z_1 q^+$};
 \node[anchor=east,colorkolme] at (0.7cm,1.4cm) {$z_2 q^+$};
 \draw[line width=1pt,<->,coloryksi] (-2.9cm,3.7cm) -- (-2.9cm,2.7cm) ;
 \node[anchor=west,coloryksi] at (-2.9cm,3.3cm) {$\Pt_\qqb$};
 \draw[line width=1pt,<->,coloryksi] (-2.75cm,3.0cm) -- (-2.75cm,1.7cm) ;
 \node[anchor=west,coloryksi] at (-2.75cm,2.1cm) {$\Kt_\ggb$};
 \draw[line width=1pt,<->,coloryksi] (-0.7cm,3.7cm) -- (-0.7cm,2.7cm) ;
 \node[anchor=west,coloryksi] at (-0.7cm,3.3cm) {$\Pt_\qqb$};
 \draw[line width=1pt,<->,coloryksi] (-0.5cm,3.0cm) -- (-0.5cm,1.7cm) ;
 \node[anchor=west,coloryksi] at (-0.5cm,2.1cm) {$\widehat\Kt_\ggb$};
 \draw[line width=1pt,snake=brace,segment amplitude=5pt,colorkaksi] (0.6cm,4.0cm) -- (0.6cm,2.4cm) ;
 \node[anchor=west,colorkaksi] at (0.7cm,3.2cm) {$M^2_\qqb$}; 
 \draw[line width=1pt,snake=brace,segment amplitude=5pt,segment aspect=0.3,colorkaksi] (-2.1cm,4.0cm) -- (-2.1cm,1.2cm) ;
 \node[anchor=west,colorkaksi] at (-2.0cm,3.1cm) {$M^2_\qqbg$}; 
 \draw[line width=1pt,snake=brace,segment amplitude=5pt,colorkaksi] (1.4cm,4.0cm) -- (1.4cm,1.2cm) ;
 \node[anchor=west,colorkaksi] at (1.5cm,2.6cm) {$M^2_X$}; 
\end{tikzpicture}
\rule{5cm}{0pt}
\includegraphics[width=4.00cm]{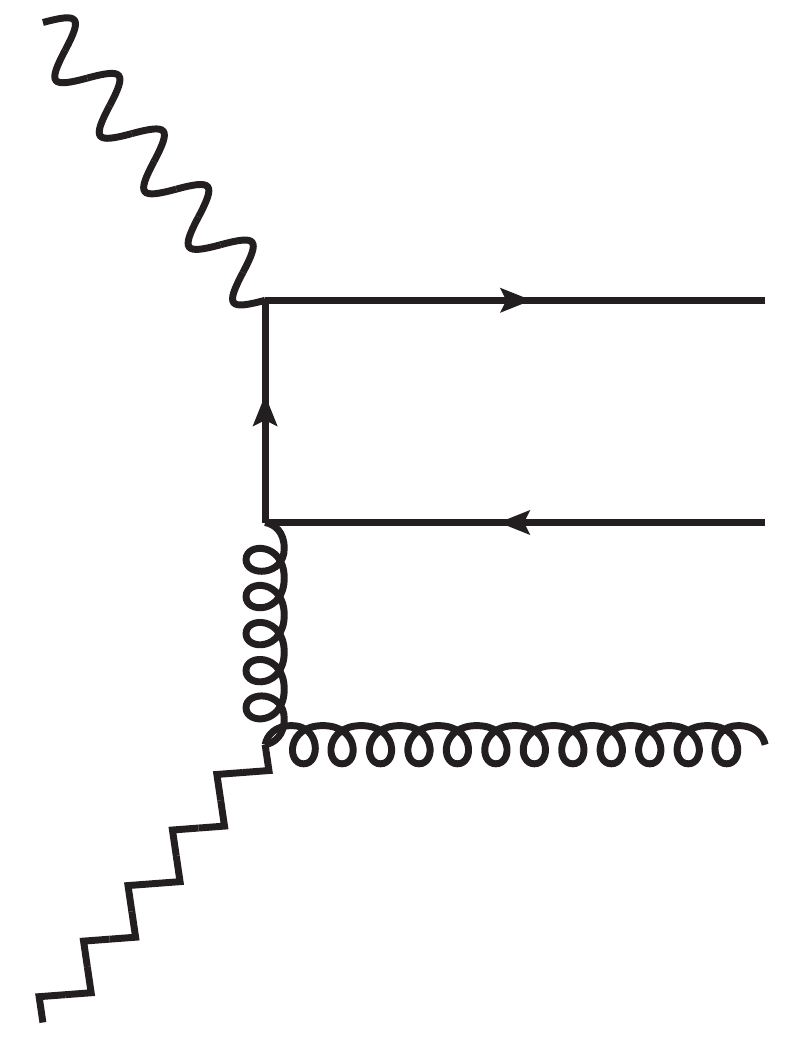}
\begin{tikzpicture}[overlay]
  \node[anchor=south east,colorkolme] at (-4cm,4.5cm) {$q$};
 \node[anchor=south east,colorkolme] at (-3.7cm,0.4cm) {$\xpom P^- = \frac{\xbj}{\beta} P^-$};
 \node[anchor=south east,colorkolme] at (-2.9cm,1.7cm) {$z \xpom P^- \equiv \frac{\xbj}{\xi} P^-$};
 \node[anchor=south east,colorkolme] at (-2.9cm,2.9cm) {$\xbj P^-$};
 \node[anchor=north east,colorkolme] at (-0.1cm,3.8cm) {$\xbj P^-$}; 
 \node[anchor=north east,colorkolme] at (-0.1cm,2.6cm) {$(1-\xi) z \xpom P^-$}; 
 \node[anchor=north east,colorkolme] at (-0.1cm,1.4cm) {$(1-z) \xpom P^-$}; 
 \draw[line width=1pt,snake=brace,segment amplitude=5pt,colorkaksi] (-0.1cm,4.0cm) -- (-0.1cm,2.4cm) ;
 \node[anchor=west,colorkaksi] at (0cm,3.2cm) {$M^2_\qqb$}; 
 \draw[line width=1pt,snake=brace,segment amplitude=5pt,colorkaksi] (0.7cm,4.0cm) -- (0.7cm,1.2cm) ;
 \node[anchor=west,colorkaksi] at (0.9cm,2.6cm) {$M^2_X$}; 
\end{tikzpicture}
}
\caption{
Kinematics for the  diffractive $q \Bar q g$-production diagram.  
Left:  dipole picture frame, where  the probe has a large $\qplus$ which is conserved in the interaction with the target. We show the plus and transverse momenta, with $\Pt_\qqb$ the relative transverse momentum of the $\qqb$ pair and $\Kt_\ggb$ of the gluon-effective gluon dipole. Right: infinite target momentum frame, where one tracks the minus component of the momentum, i.e. the target momentum fraction. 
The zigzag line refers to the pomeron emitted from the target, and is used to illustrate a generic diffractive interaction between the virtual photon and target.
The two scattering pictures are connected by the invariants of the scattering. In the GBW limit the $\qqb$ pair forms a hard system that is seen as pointlike by the target color field, thus $M_\qqb^2$ and $\Pt_\qqb$ are the same before and after the shockwave. In the aligned jet limit $z_2\ll z_1\ll z_0\approx 1$ we have  
$M_\qqb^2\approx \Pt^2_\qqb/z_1$, and from the collinear factorization picture on the on the right: $M_\qqb^2 = 2(\xbj/\xi)q^+P^-  + q^2 = (1/\xi-1)Q^2$. The invariant mass of the $\qqbg$ system  and $\Kt_\ggb$, on the other hand, are affected by the interaction with the shockwave. In the dipole picture, before the shockwave,  we have $M_\qqbg^2 \approx M_\qqb^2 + \Kt_\ggb^2/z_2$. The invariant mass after the shockwave is $M_X^2= (1/\beta-1)Q^2  \approx M_\qqb^2 + \widehat\Kt_\ggb^2/z_2 $.
}
\label{fig:kminus-flow}
\end{figure*}
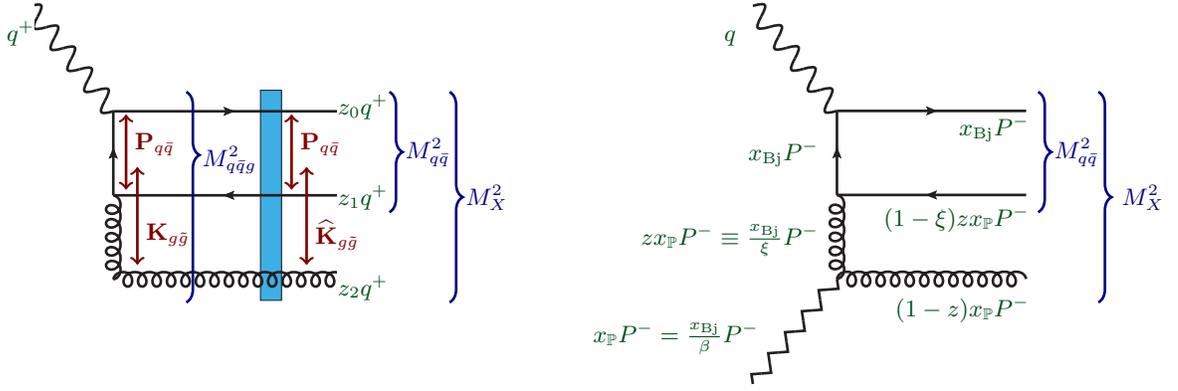

An essential feature of the large-$Q^2$ structure function is the manifestation of the
DGLAP splitting function for $g\to q \Bar{q}$ splitting. This is associated with the DGLAP evolution of the parton distributions of the pomeron, and is written in terms of the target minus momentum fractions $\beta$ and $z$ (recall that we work in a frame where the photon has a large plus momentum).
In the frame where the target has a large longitudinal momentum $\beta$ can be interpreted as the fraction of the pomeron momentum carried by the stuck parton, and it is also related to the invariant mass of the final state as $\beta = Q^2/(Q^2+M_X^2)$. The fraction $z$ is the fraction of the pomeron minus momentum transferred to the $\qqb$ system.
On the one hand, the presence of the splitting function in the large $Q^2$ limit is unavoidable in QCD. On the other hand, since we are using light cone perturbation theory to quantize the projectile photon, the light cone momentum fraction $\xi = \beta/z$ does not appear spontaneously in the calculation. These kinematical variables are illustrated in Fig.~\ref{fig:kminus-flow}.

The 3-particle phase space has been treated only approximately in the Wüsthoff result.
The result is written in terms of the size $r$ of  an ``effective adjoint dipole'' formed by the gluon and the quark-antiquark pair. Thus the interaction with the target color field appears in the form of an adjoint representation dipole cross section $\tilde \sigma_\text{dip}$. There is also an explicit $\log Q^2$ resulting from the integral over some of the ``internal'' kinematics of the $q\Bar{q}$ pair as we will demonstrate explicitly below.

The contribution corresponding to the DGLAP splitting function $[(1-\beta/z)^2+(\beta/z)^2]$ originates from the part of phase space where emissions are strongly ordered in transverse momenta. Since we want to obtain the DGLAP splitting function with fixed momentum fractions in $k^-$, this implies that the $k^+$ momentum fractions must also be strongly ordered. This is the Bjorken aligned jet~\cite{Bjorken:1973gc} configuration. Thus it is more convenient to work in momentum space to make the connection to the W\"usthoff result. 
To make this more specific, we choose the transverse momenta for the 3-particle $q\Bar{q}g$ final state (after the shockwave) as
\begin{align}
    \Pt_i & \coloneqq \pt_i - z_i \qt,
    \\
    \widehat \Pt_\qqb & \coloneqq \frac{z_0 \Pt_1 -z_1\Pt_0}{z_0+z_1},
    \\
    \widehat \Kt_\ggb & \coloneqq (z_0+z_1)\Pt_2 - z_2\left( \Pt_0+\Pt_1\right),
    \\
    \widehat \Deltat & \coloneqq -\Pt_0-\Pt_1-\Pt_2,
\end{align}
where $\pt_i$ are the transverse momenta of the final state partons, and the variable $\widehat \Pt_\qqb$ can be interpreted as the relative momentum between the quark and antiquark, whereas $\widehat \Kt_\ggb$ is the momentum of the gluon with respect to the quark-antiquark system, i.e. the ``effective gluon''. The total transverse momentum transfer in the scattering process is $\widehat \Deltat$. The corresponding conjugate variables in transverse coordinate space can be determined by writing  
\begin{equation}
\Pt_0 \cdot \xt_0 + \Pt_1\cdot \xt_1 + \Pt_2 \cdot \xt_2
\equiv
\widehat \Pt_\qqb \cdot \ut + \widehat \Kt_\ggb \cdot \rt + \widehat \Deltat \cdot \bt
\end{equation}
from which one obtains
\begin{align}
    \label{eq:fourier-coordinates}
    \ut & = \xt_1 - \xt_0 ,
    \\
    \rt & = \xt_2 - \frac{z_0\xt_0 + z_1\xt_1}{z_0+z_1} ,
    \\
    \bt & = z_0\xt_0 + z_1\xt_1 + z_2\xt_2.
\end{align}
Here $\ut$ is the size of the $q\Bar q$ dipole, $\rt$ is the distance from the gluon to the center-of-mass of the $q\Bar q$ system, i.e. the size of the effective gluonic dipole, and  $\bt$ is the center-of-mass.

In the derivation of the large $Q^2$ limit of our result for $F_T^D$, Eq.~\eqref{eq:ddis-FL-qqbarg-nlo}, our starting point is the expression~\eqref{eq:qqg_diffxs_Mx_t_copied} for the diffractive cross section. The calculation then proceeds roughly as follows. First we Fourier transform the LFWFs back into momentum space where we can apply the kinematic approximations related to the leading $\log Q^2$ limit. Then we write the virtual photon LFWF in momentum space using natural relative momenta from the perspective of the physical picture that emerges at the large-$Q^2$ limit discussed above. This enables us to apply the aligned jet limit (AJL) approximations at the LFWF-level, after which we proceed to integrate over surplus degrees of freedom at the cross section level.
In the aligned jet kinematics we have the strong transverse momentum ordering $Q^2\gg \Pt_\qqb^2 \gg \Kt_\ggb^2 \gg \Deltat^2$. The corresponding $k^+$ momentum fraction ordering is $z_2\ll z_1\ll z_0$ or, symmetrically, $z_2\ll z_0\ll z_1$. Since these two limits give the same contribution, we will just concentrate on the first one and in the end multiply the result by 2.

The transverse Fourier transform of the LFWF into momentum space is defined as~\cite{Beuf:2017bpd}
\begin{equation}
    \widetilde \psi^{\gamma^{\ast}_{\rm T}\rightarrow q\Bar{q}g}_{\nlo} 
    =
    \int \frac{\ud^2 \kt_0}{(2\pi)^2}
    \int \frac{\ud^2 \kt_1}{(2\pi)^2}
    \int \frac{\ud^2 \kt_2}{(2\pi)^2}
    (2\pi)^2 \delta^{(2)}(\kt_0 + \kt_1 + \kt_2 - \qt)
    e^{i(\kt_0 \cdot \xt_0 + \kt_1 \cdot \xt_1 + \kt_2 \cdot \xt_2)}
    \psi^{\gamma^{\ast}_{\rm T}\rightarrow q\Bar{q}g}_{\nlo}.
\end{equation}
Performing the Fourier transforms and rewriting the Fourier momenta in terms of the natural momenta using
\begin{eqnarray}
\kt_1 &=&-\frac{z_1}{z_0+z_1}\Kt_\ggb + \Pt_\qqb
\\
\kt_0 &=& -\frac{z_0}{z_0+z_1}\Kt_\ggb - \Pt_\qqb
\end{eqnarray}
the diffractive cross section from \eq\eqref{eq:qqg_diffxs_Mx_t_copied} becomes
\begin{align}
    \frac{\ud \sigma^{\text{D}}_{\lambda, \, q \Bar q g}}{\ud \mx^{2} }
    = 
    &
    \frac{\nc \cf}{4(2\pi)^2}
    \int_{-\infty}^0 \ud t
    \int_0^1 \frac{\ud z_0}{z_0}
	\int_0^1 \frac{\ud z_1}{z_1}
	\int_0^1 \frac{\ud z_2}{z_2} \delta(z_0+z_1+z_2-1)
    \int \frac{\ud^2 \widehat \Pt_\qqb}{(2 \pi)^2}
    \int \frac{\ud^2 \widehat \Kt_\ggb}{(2 \pi)^2}
    \int \frac{\ud^2 \widehat \Deltat}{(2 \pi)^2}
    \notag
    \\ \notag
    & \times
    \delta(\widehat \Deltat^2 - \abs{t})
    \delta \left( \frac{\widehat \Kt_\ggb^2}{z_2 (z_0+z_1)} + \frac{z_0 + z_1}{z_0 z_1} \widehat \Pt_\qqb^2 - \mx^2 \right)
    \\ \notag
    & \times
    \int_\ut \int_\rt \int_\bt \int_{\conj{\ut}} \int_{\conj{\rt}} \int_{\conj{\bt}} (2\pi)^6
    e^{i(\conj{\ut} - \ut) \cdot \widehat \Pt_\qqb}
    e^{i(\conj{\rt} - \rt) \cdot \widehat \Kt_\ggb}
    e^{i(\conj{\bt} - \bt) \cdot \widehat \Deltat}
    \\
    &
    \times
    \sum_{h_0, h_1, \lambda_2}
    \left(\widetilde{\psi}_{\gamma^{*}_\lambda \rightarrow q_{\Bar 0} \Bar q_{\Bar 1} g_{\Bar 2}} \right)^\dagger
    \left(\widetilde{\psi}_{\gamma^{*}_\lambda \rightarrow q_0 \Bar{q}_1 g_2} \right)
    \left[S_{\conj{\ut} \conj{\rt} \conj{\bt}}^{(3)\dagger} -1\right]
    \left[S_{\ut \rt \bt}^{(3)} -1\right],
    \label{eq:qqg_diffxs_Mx_t_momentum}
\end{align}
where the hatted quantities are the final state momenta. The squared virtual photon amplitude Fourier transformed to coordinate space reads
\begin{align}
\label{eq:gbwcoordmom}
    \left(\widetilde{\psi}_{\gamma^{*}_\lambda \rightarrow q_{\Bar 0} \Bar q_{\Bar 1} g_{\Bar 2}} \right)^\dagger
    \left(\widetilde{\psi}_{\gamma^{*}_\lambda \rightarrow q_0 \Bar{q}_1 g_2} \right)
    = & 
    \int \frac{\ud^2 \Bar \Pt_\qqb}{(2 \pi)^2}
    \int \frac{\ud^2 \Bar \Kt_\ggb}{(2 \pi)^2}
    \int \frac{\ud^2 \Bar \Deltat}{(2 \pi)^2}
    \left({\psi}_{\gamma^{*}_\lambda \rightarrow q_{\Bar 0} \Bar q_{\Bar 1} g_{\Bar 2}} \right)^\dagger
    e^{-i \Bar \ut \cdot \Bar \Pt_\qqb - i \Bar \rt \cdot \Bar \Kt_\ggb - i \Bar \bt \cdot \Bar \Deltat}
    (2\pi)^2 \delta^{(2)}(\Bar \Deltat)
    \notag
    \\
    & \times
    \int \frac{\ud^2 \Pt_\qqb}{(2 \pi)^2}
    \int \frac{\ud^2 \Kt_\ggb}{(2 \pi)^2}
    \int \frac{\ud^2 \Deltat}{(2 \pi)^2}
    \left({\psi}_{\gamma^{*}_\lambda \rightarrow q_0 \Bar{q}_1 g_2} \right)
    e^{i \ut \cdot \Pt_\qqb + i \rt \cdot \Kt_\ggb + i \bt \cdot \Deltat}
    (2\pi)^2 \delta^{(2)}(\Deltat).
\end{align}
Now we begin the application of the aligned jet limit approximations. First, we assume that the relative transverse momentum of the parent dipole $\norm{\Pt_\qqb}$ is much larger than that of either of the daughter dipoles, i.e. $\norm{\Pt_\qqb} \gg \norm{\Kt_\ggb}$. In position space this corresponds to a configuration where the gluon is emitted far away from the parent dipole, i.e. $\norm{\ut} \ll \norm{\rt}$. This makes the dipole amplitude independent of the parent dipole size, and  enables us to separate and perform the $\ut$-integrations in Eq.~\eqref{eq:qqg_diffxs_Mx_t_momentum}, which yields two delta-functions:
\begin{equation}
\label{eq:Pt-qqbar-deltas}
    \int \frac{\ud^2 \ut}{(2 \pi)^2} \int \frac{\ud^2 \Bar \ut}{(2 \pi)^2}
    e^{i \Bar{\ut} \cdot (\widehat \Pt_\qqb - \Bar \Pt_\qqb)}
    e^{-i {\ut} \cdot (\widehat \Pt_\qqb - \Pt_\qqb)}
    =
    \delta^{(2)} (\widehat \Pt_\qqb - \Bar \Pt_\qqb)
    \delta^{(2)} (\widehat \Pt_\qqb - \Pt_\qqb) .
\end{equation}
This enables us to perform the $\Pt_\qqb$ and $\Bar \Pt_\qqb$ integrations in Eq.~\eqref{eq:qqg_diffxs_Mx_t_momentum}.

To perform the $t$-integration of~\eqref{eq:qqg_diffxs_Mx_t_momentum}, we separate the closely related transverse momentum transfer integrals:
\begin{multline}
    \int_{-\infty}^0 \ud t
    \int \frac{\ud^2 \Bar \bt}{2\pi} \int \frac{\ud^2 \bt}{2\pi}
    \int \frac{\ud^2 \widehat \Deltat}{(2 \pi)^2}
    \int \frac{\ud^2 \Bar \Deltat}{(2 \pi)^2}
    \int \frac{\ud^2 \Deltat}{(2 \pi)^2}
    \delta \qty(\widehat \Delta^2 - \abs{t})
    (2\pi)^4
    \delta^{(2)} (\Bar \Deltat)
    \delta^{(2)} (\Deltat)
    e^{i \Bar{\bt} \cdot (\widehat \Deltat - \Bar \Deltat)}
    e^{-i {\bt} \cdot (\widehat \Deltat - \Deltat)}
    \\
    =
    \int_{-\infty}^0 \ud t
    \int \frac{\ud^2 \Bar \bt}{2\pi} \int \frac{\ud^2 \bt}{2\pi}
    \int \frac{\ud^2 \widehat \Deltat}{(2 \pi)^2}
    e^{i {\widehat \Deltat} \cdot (\Bar \bt - \bt)}
    \delta \qty(\widehat \Delta^2 - \abs{t})
    =
    \int \frac{\ud^2 \Bar \bt}{2\pi} \int \frac{\ud^2 \bt}{2\pi}
    \delta^{(2)} (\Bar \bt - \bt).
\end{multline}

An essential step tha we need to take before we proceed is to consider how to connect our results written in terms of the plus-momentum fractions to the minus momentum fractions in the Wüsthoff result \eqref{eq:wusthoffqqbarg}.  The essential idea here is to think in terms of invariant masses of different multiparton states. To connect the frames of reference, we need to approximate the invariant masses  in the aligned jet limit as:
\begin{eqnarray}
M_{\qqb}^2 &=& \frac{1-\xi}{\xi}Q^2 = \qty(1-z_2)^2 \frac{\Pt_\qqb^2}{z_0 z_1}  \approx  \frac{\Pt_\qqb^2}{z_1} ,
\\ 
M_{\qqb g}^2 +Q^2
& = &
Q^2 + \frac{z_0 + z_1}{z_0 z_1} \Pt_\qqb^2 + \frac{ \Kt_\ggb^2}{z_2(z_0+z_1)}
\approx Q^2 + \frac{\Pt_\qqb^2}{z_1}  + \frac{\Kt_\ggb^2}{z_2}
= \frac{Q^2}{\xi} + \frac{\Kt_\ggb^2}{z_2}
\end{eqnarray}
where $z\xpom = \xbj/\xi$ is the fraction of the target momentum carried by the $\qqb$ system.
Given the above relations, we can reparameterize the often occurring combination
\begin{equation}
    \Pt_\qqb^2 + z_1 Q^2 = z_1(M_\qqb^2 + Q^2) = z_1 Q^2/\xi.
\end{equation}
It is useful to write the $\qqb g$ energy denominator in terms of the diffractive state mass
\begin{equation}
\mx^2=\frac{1-\beta}{\beta}Q^2 
\approx \frac{\widehat\Pt_\qqb^2}{z_1} + \frac{\widehat\Kt_\ggb^2}{z_2}
,   
\end{equation}
where $\widehat\Kt_\ggb$ is the gluon relative momentum after the shockwave. Recall that $\Pt_\qqb \equiv \widehat\Pt_\qqb $ is conserved in the shockwave which enables us to leverage final state information about $\mx^2$ to simplify the $\gamma^\ast$ LFWF before the shockwave. 
We now want to eliminate $z_2$ using these variables, so that
\begin{equation}
    z_2 = \frac{\beta \widehat\Kt_\ggb^2}{Q^2(1-z)}.
\end{equation}
With this relation the three particle state invariant mass before the shockwave becomes
\begin{equation}
 M_{\qqb g}^2 = \frac{Q^2}{\beta \widehat\Kt_\ggb^2} \left[(z-\beta)\widehat\Kt_\ggb^2 + (1-z)\Kt_\ggb^2\right] ,
\end{equation}
and the ``outer'' (i.e. $\qqbg$ state) LC energy denominator in the NLO virtual photon LFWF will be
\begin{equation}
     M_{\qqb g}^2 + Q^2 = \frac{Q^2}{\beta \widehat\Kt_\ggb^2} \left[z \widehat\Kt_\ggb^2 + (1-z)\Kt_\ggb^2\right].
\end{equation}
Note the distinction that this is for the $\qqb g$ state before the shockwave. The momentum  $\widehat\Kt_\ggb^2$ after the shockwave is fixed by the final state kinematics. It, or the $\qqbg$ invariant mass,  is not the same before ($M_{\qqb g}^2$) and after ($\mx^2$) the shockwave. The momentum argument of the wavefunction before the shockwave $\Kt_\ggb$ will need to be Fourier-transformed into coordinate space, see \eq\nr{eq:gbwcoordmom}, in order to include the interaction with the target shockwave. The momenta $\Kt_\ggb$ and $\Bar \Kt_\ggb$ before the shockwave are separate in the DA and CCA, which have to be Fourier-transformed separately.

Next we move on to manipulate the virtual photon splitting light-front wavefunction in the aligned jet limit. The $\gamlam \to \qqbg$ LFWF in momentum space in the convention of Refs.~\cite{Beuf:2016wdz,Beuf:2017bpd} is
\begin{equation}
\label{eq:Tradfinal-GBC}
\begin{split}
\psi^{\gamma^{\ast}_{\rm T}\rightarrow q\Bar{q}g}_{\nlo}
=
4 
e e_f (gt^{a}_{\alpha\beta}) \sqrt{z_0 z_1}
&\left\{
    -\Bar \Sigma^{ijkl}_{\ref{diag:gwavef1}}\frac{(-\kt_1)^i\mt^{k}\epst^{j}_{\lambda}\epst^{\ast l}_{\sigma}}{\biggl [\kt_1^2 + \Bar{Q}^2_{\ref{diag:gwavef1}} \biggr ]\biggl [\mt^2 + \omega_{\ref{diag:gwavef1}} \left ( \kt_1^2 + \Bar{Q}^2_{\ref{diag:gwavef1}}\right )\biggr ]}
    \right.
    \\
&
    -\Bar \Sigma^{ijkl}_{\ref{diag:gwavef2}}\frac{\kt_0^i\lt^{k}\epst^{j}_{\lambda}\epst^{\ast l}_{\sigma}}{\biggl [\kt_0^2 + \Bar{Q}^2_{\ref{diag:gwavef2} }  \biggr ]\biggl [\lt^2 + \omega_{\ref{diag:gwavef2} } \left ( \kt_0^2 + \Bar{Q}^2_{\ref{diag:gwavef2} } \right )\biggr ]}
    \\
& 
    \left.
    -
    \Bar \Sigma^{ij}_{\ref{diag:gwavefinst1}} \frac{\epst^{\ast i}_{\sigma} \epst^{j}_{\lambda}}{\biggl [\mt^2 + \omega_{\ref{diag:gwavefinst1}}\left ( \kt_1^2 + \Bar{Q}^2_{\ref{diag:gwavefinst1}}\right ) \biggr ]}
    +
    \Bar \Sigma^{ij}_{\ref{diag:gwavefinst2}} \frac{\epst^{\ast i}_{\sigma} \epst^{j}_{\lambda}}{\biggl [\lt^2 + \omega_{\ref{diag:gwavefinst2}} \left ( \kt_0^2 + \Bar{Q}^2_{\ref{diag:gwavefinst2}}\right ) \biggr ]}
    \right\}
,
\end{split}
\end{equation}
where the momenta are defined as
\begin{eqnarray}
\lt & \coloneqq & \frac{z_1}{(z_0+z_1)(z_1+z_2)}\Kt_\ggb - \frac{z_2}{z_1+z_2}\Pt_\qqb,
\\
\mt & \coloneqq &  \frac{z_0}{(z_0+z_1)(z_0+z_2)}\Kt_\ggb + \frac{z_2}{z_0+z_2}\Pt_\qqb,
\end{eqnarray}
and the vertex factors are
\begin{align}
\Bar \Sigma^{ijkl}_{\ref{diag:gwavef1}}
    & =
    \frac{1}{4} \frac{1}{z_0 + z_2} 
    \biggl [\left ( 2 z_0 + z_2 \right ) \delta^{kl}  - i (2h_0) z_2 \epsilon ^{kl} \biggr ]
    \biggl [\left ( 2z_1 -1 \right )\delta^{ij}  - i (2h_0) \epsilon ^{ij} \biggr ],
\\
\Bar \Sigma^{ijkl}_{\ref{diag:gwavef2}}
    & =
    \frac{1}{4} \frac{1}{z_1 + z_2} 
    \biggl [\left (2 z_1 + z_2 \right ) \delta^{kl}  + i (2h_0) z_2 \epsilon ^{kl} \biggr ]
    \biggl [\left ( 2 z_0 -1 \right )\delta^{ij}  + i (2h_0) \epsilon ^{ij}\biggr ],
\\
\Bar \Sigma^{ij}_{\ref{diag:gwavefinst2}} 
    & = 
    \frac{1}{4}\frac{z_0z_2}{(z_0+z_2)^2}
    \biggl [\delta^{ij}  - i (2h_0) \epsilon^{ij}  \biggr ],
\\
\Bar \Sigma^{ij}_{\ref{diag:gwavefinst2}} 
    & = \frac{1}{4}\frac{z_1z_2}{(z_1+z_2)^2}
    \biggl [\delta^{ij}  + i (2h_0) \epsilon^{ij} \biggr ],
\end{align}
with
\begin{align}
\Bar{Q}^2_{\ref{diag:gwavef1}} & = z_1(z_0+z_2)Q^2, \quad\quad \omega_{\ref{diag:gwavef1}} = \frac{z_0z_2}{z_1(z_0+z_2)^2},
\\
\Bar{Q}^2_{\ref{diag:gwavef2}} & = z_0(z_2+z_1)Q^2, \quad\quad \omega_{\ref{diag:gwavef2} } = \frac{z_2z_1}{z_0(z_2+z_1)^2},
\\
\Bar{Q}^2_{\ref{diag:gwavefinst1}} & = z_1(z_0+z_2)Q^2,\quad\quad \omega_{\ref{diag:gwavefinst1}} = \frac{z_0z_2}{z_1(z_0+z_2)^2}, 
\\
\Bar{Q}^2_{\ref{diag:gwavefinst2}} & = z_0(z_2+z_1)Q^2, \quad\quad \omega_{\ref{diag:gwavefinst2}} \,\,= \frac{z_2z_1}{z_0(z_2+z_1)^2}.
\end{align}

In the aligned jet limit, these LC structures simplify significantly.  Let us now take the strongly ordered limit  $z_2\ll z_1\ll z_0$. Note that the leading term at $z_2=0$, $\Kt_\ggb \to 0$ cancels out, and thus one has to include subleading terms in the small-$\Kt_\ggb$ expansion to get the development right. We are left with:
\begin{eqnarray}
\kt_1 &\to& \Pt_\qqb \, ,
\\
\kt_0 &\to& -\Kt_\ggb - \Pt_\qqb \, ,
\\
\lt &\to& \Kt_\ggb -\frac{z_2}{z_1} \Pt_\qqb \, ,
\\
\mt &\to&  \Kt_\ggb \, ,
\end{eqnarray}
and
\begin{align}
\Bar \Sigma^{ijkl}_{\ref{diag:gwavef1}} & = -\frac{1}{2} \biggl [\delta^{ij} + i(2h_0)\epsilon^{ij} \biggr ]  \delta^{kl} \, ,
\\
\Bar \Sigma^{ijkl}_{\ref{diag:gwavef2}} & = \frac{1}{2} \biggl [\delta^{ij} + i(2h_0)\epsilon^{ij}\biggr ] \delta^{kl} \, ,
\\
\Bar \Sigma^{ij}_{\ref{diag:gwavefinst1}} & = \frac{1}{4} z_2 \biggl [\delta^{ij} - i(2h_0)\epsilon^{ij} \biggr ] \, ,
\\
\Bar \Sigma^{ij}_{\ref{diag:gwavefinst2}} & = \frac{1}{4} \frac{z_2}{z_1}\biggl [\delta^{ij} + i(2h_0)\epsilon^{ij} \biggr ] \, .
\end{align}
Additionally it will be useful to simplify the momentum dependence of the $\qqbg$ LFWF by using the three particle state invariant mass:
\begin{align}
    \mt^2 + \omega_{\ref{diag:gwavef1}} \left ( \kt_1^2 + \Bar{Q}^2_{\ref{diag:gwavef1}} \right )
    = 
    \frac{z_0 z_2}{z_0 + z_2} \left( \mxqqb^2 + Q^2 \right) \approx z_2 \left( \mxqqb^2 + Q^2 \right),
    \\
    \lt^2 + \omega_{\ref{diag:gwavef2}} \left ( \kt_0^2 + \Bar{Q}^2_{\ref{diag:gwavef2}} \right )
    =
    \frac{z_1 z_2}{z_1 + z_2} \left( \mxqqb^2 + Q^2 \right) \approx z_2 \left( \mxqqb^2 + Q^2 \right).
\end{align}
With these simplifications the normal emission part of the wavefunction~\eqref{eq:Tradfinal-GBC} becomes
\begin{multline}
\psi^{\gamma^{\ast}_{\rm T}\rightarrow q\Bar{q}g}_{\nlo}\bigg\vert_{\rm AJL}
=
4 e e_f (gt^{a}_{\alpha\beta})\sqrt{z_0z_1}
\frac{
    \epst^{j}_{\lambda} \epst^{\ast l}_{\sigma}
    \frac{1}{2} \left[\delta^{ij} + i(2h_0)\epsilon^{ij} \right] \delta^{kl}
    }{
    z_2
    \left( M_{\qqb g}^2 + Q^2 \right)
}
\\
\times
\left[
- \frac{\Pt_\qqb^i}{\Pt_\qqb^2 + z_1 Q^2}
\Kt_\ggb^k
+ 
\frac{(\Pt_\qqb+\Kt_\ggb)^i}{(\Pt_\qqb+ \Kt_\ggb)^2 + z_1 Q^2}
(\Kt_\ggb^k-\frac{z_2}{z_1}\Pt_\qqb^k)
\right] .
\end{multline}

In the kinematic regime $\Pt_\qqb^2 \gg \Kt_\ggb^2$ and we can first simplify the term in the square brackets as
\begin{multline}
    \Big[ \dots \Big] = \frac{-\Pt_\qqb^i((\Pt_\qqb+ \Kt_\ggb)^2 + z_1 Q^2)\Kt_\ggb^k
 + (\Pt_\qqb+\Kt_\ggb)^i(\Pt_\qqb^2 + z_1 Q^2)(\Kt_\ggb^k-\frac{z_2}{z_1}\Pt_\qqb^k)
}{((\Pt_\qqb+ \Kt_\ggb)^2 + z_1 Q^2)(\Pt_\qqb^2 + z_1 Q^2)}    
\\ \approx 
\frac{-2 \Pt_\qqb \cdot \Kt_\ggb \Pt_\qqb^i + (\Pt_\qqb^2 + z_1 Q^2)\Kt_\ggb^i}{(\Pt_\qqb^2 + z_1 Q^2)^2}    
\Kt_\ggb^k
- \frac{z_2}{z_1}\frac{\Pt_\qqb^i\Pt_\qqb^k}{\Pt_\qqb^2 + z_1 Q^2}    
+ \mathcal{O}(\Kt_\ggb^3),
\end{multline}
where the last term is kept since in this strong ordering limit we have $z_2/z_1 \sim \Kt_\ggb^2/\Pt_\qqb^2$, whereas the associated term proportional to $(z_2 / z_1) \Kt_\ggb^i \Pt_\qqb^k$ is discarded as a higher order term.
Out of the instantaneous emission terms, the contribution from diagram \ref{diag:gwavefinst2} is enhanced by a factor of $1/z_1$ w.r.t. diagram \ref{diag:gwavefinst1}, which means that in the aligned jet limit, we can neglect the contribution from the latter, and only the former contributes. Using the identity
\begin{equation}
    \frac{1}{2} \left[ \delta^{ij} + i (2h_0) \epsilon^{ij} \right] \epst^j_\lambda
    =
    \delta_{h_0, \lambda} \epst^i_\lambda
\end{equation}
we see that the quark (and consequently antiquark) helicity is completely fixed by the photon polarization. This is a common feature of LCPT vertices: the particle carrying all the longitudinal momentum in a splitting inherits the light front helicity of the parent~\cite{Lappi:2016oup}. 
We can now combine the leading instantaneous contribution with the regular emissions:
\begin{multline}
\psi^{\gamma^{\ast}_{\rm T}\rightarrow q\Bar{q}g}_{\nlo}\bigg\vert_{\rm AJL}
=
4 e e_f (gt^{a}_{\alpha\beta})\sqrt{z_0z_1}
\frac{
    \epst^{i}_{\lambda} \epst^{\ast j}_{\sigma}
    \delta_{h_0, \lambda}
    }{
    z_2
    \left( M_{\qqb g}^2 + Q^2 \right)
}
\\
\times
\left[
    \frac{-2 \Pt_\qqb \cdot \Kt_\ggb \Pt_\qqb^i + (\Pt_\qqb^2 + z_1 Q^2)\Kt_\ggb^i}{(\Pt_\qqb^2 + z_1 Q^2)^2} \Kt_\ggb^j
    - \frac{z_2}{z_1}\frac{\Pt_\qqb^i\Pt_\qqb^j}{\Pt_\qqb^2 + z_1 Q^2}
    + \half \frac{z_2}{z_1} \delta^{ij}
\right] 
\\
=
4 e e_f (gt^{a}_{\alpha\beta})\sqrt{z_0z_1}
\frac{
    \epst^{i}_{\lambda} \epst^{\ast j}_{\sigma}
    \delta_{h_0, \lambda}
    }{
    z_2
    \left( M_{\qqb g}^2 + Q^2 \right)
}
\frac{\xi}{z_1 Q^2}
\Bigg[
    -2 \frac{\xi}{z_1 Q^2} \left( \Pt_\qqb \cdot \Kt_\ggb \right) \Pt_\qqb^i \Kt_\ggb^j + \Kt_\ggb^i \Kt_\ggb^j
    - \xi \frac{z}{1-z} \frac{\widehat\Kt_\ggb^2}{z_1 Q^2} \Pt_\qqb^i \Pt_\qqb^j
    \\
    + \half \frac{z \widehat\Kt_\ggb^2}{1-z} \delta^{ij}
\Bigg] .
\end{multline}

Next we move on to calculate the squared amplitude. Here it is important to recall that  the momenta $\Kt_\ggb$ and $\Bar \Kt_\ggb$  in the DA and CCA, respectively, are separate, whereas $\Pt_\qqb \equiv \Bar \Pt_\qqb \equiv \widehat \Pt_\qqb$ due to Eq.~\eqref{eq:Pt-qqbar-deltas}.
We need the following algebra:
\begin{equation}
\left[
    - \xi  \frac{\widehat \Pt_\qqb^i \widehat \Pt_\qqb^j}{ z_1 Q^2} 
    + \frac{1}{2} \delta^{ij}
\right]
\left[
    - \xi  \frac{\widehat \Pt_\qqb^i \widehat \Pt_\qqb^j}{ z_1 Q^2} 
    + \frac{1}{2} \delta^{ij}
\right]
= \xi^2 \frac{\widehat \Pt_\qqb^4}{(z_1 Q^2)^2} + \xi \frac{\widehat \Pt_\qqb^2}{z_1 Q^22} + \frac{1}{2} = \frac{1}{2}\left(\xi^2 + (1-\xi)^2\right)
\end{equation}
and 
\begin{multline}
\left[
    -2 \frac{\xi}{z_1 Q^2} (\widehat \Pt_\qqb \cdot \Kt_\ggb) \widehat \Pt_\qqb^i \Kt_\ggb^j + \Kt_\ggb^i \Kt_\ggb^j
\right]
\left[
    - \xi  \frac{\widehat \Pt_\qqb^i \widehat \Pt_\qqb^j}{ z_1 Q^2} 
    + \frac{1}{2} \delta^{ij}
\right]
= 
2 \xi^2 \frac{\widehat \Pt_\qqb^2 \left(\widehat \Pt_\qqb \cdot \Kt_\ggb \right)^2}{(z_1 Q^2)^2}
\\
- 2\xi \frac{\left(\widehat \Pt_\qqb \cdot \Kt_\ggb \right)^2}{z_1 Q^2} +\frac{1}{2}\Kt_\ggb^2 .
\end{multline}
Now we remember that at the cross section level we are integrating over the phase space $\ud^2 \widehat \Pt_\qqb$ with $\widehat \Pt_\qqb^2 = z_1(1-\xi)Q^2/\xi$ fixed, i.e. over the angle of $\widehat \Pt_\qqb$. This means that we can replace 
\begin{equation}
\widehat \Pt_\qqb^i \widehat \Pt_\qqb^j \to \frac{1}{2} \widehat \Pt_\qqb^2 \delta_{ij} = \frac{1}{2} \frac{1-\xi}{\xi} z_1 Q^2 \delta_{ij} .
\end{equation}
Thus we get
\begin{equation}
2 \xi^2 \frac{\widehat \Pt_\qqb^2 \left(\widehat \Pt_\qqb \cdot \Kt_\ggb \right)^2}{(z_1 Q^2)^2} - 2 \xi \frac{\left(\widehat \Pt_\qqb \cdot \Kt_\ggb \right)^2}{z_1 Q^2} +\frac{1}{2}\Kt_\ggb^2
\to
\frac{1}{2}\left(\xi^2 + (1-\xi)^2\right) \Kt_\ggb^2.
\end{equation}
Analogously the other cross term yields $\frac{1}{2}\left(\xi^2 + (1-\xi)^2\right) \Bar \Kt_\ggb^2$.
Finally the last term gives
\begin{multline}
\left[
    -2 \frac{\xi}{z_1 Q^2} (\widehat \Pt_\qqb \cdot \Kt_\ggb) \widehat \Pt_\qqb^i \Kt_\ggb^j + \Kt_\ggb^i \Kt_\ggb^j
\right]   
\left[
    -2 \frac{\xi}{z_1 Q^2} (\widehat \Pt_\qqb \cdot \Bar \Kt_\ggb) \widehat \Pt_\qqb^i \Bar \Kt_\ggb^j + \Bar \Kt_\ggb^i \Bar \Kt_\ggb^j
\right]
\\
=
\left(\xi^2 + (1-\xi)^2\right) \left(\Kt_\ggb \cdot \Bar \Kt_\ggb \right)^2
\end{multline}
With these the full squared amplitude can be written as:
\begin{multline}
    \label{eq:ajl-psiTsquared}
    \half \sum_{h, \lambda, \sigma}
    \left( \psi^{\gamma^{\ast}_{\rm T}\rightarrow q\Bar{q}g}_{\nlo} \bigg\vert_{\rm AJL} \right)^\dagger
    \psi^{\gamma^{\ast}_{\rm T}\rightarrow q\Bar{q}g}_{\nlo} \bigg\vert_{\rm AJL}
    =
    8 
    e^2 e_f^2 g^2 \frac{z_0 z_1}{z_2^2}
    \frac{\left( \frac{\xi}{z_1 Q^2} \right)^2 \left( \xi^2 + (1-\xi)^2 \right)}{\left( \mxqqb^2 \qty(\Kt_\ggb) + Q^2 \right) \left( \mxqqb^2 \qty(\Bar \Kt_\ggb) + Q^2 \right)}
    \\
    \times
    \left[ \Kt_\ggb^i \Kt_\ggb^j + \half \frac{ z \widehat\Kt_\ggb^2}{1-z} \delta^{ij}\right]
    \left[ \Bar \Kt_\ggb^i \Bar \Kt_\ggb^j + \half \frac{ z \widehat\Kt_\ggb^2}{1-z} \delta^{ij}\right],
\end{multline}
where we note that after the aligned jet configuration approximations the squared amplitude took a form that factorizes into a DA-like and a CCA-like parts similar to what has been seen in literature~\cite{Hebecker:1997gp, Buchmuller:1998jv, Wusthoff:1997fz,GolecBiernat:1999qd}, even though this was not possible at earlier stages. To elaborate, we were only able to find the factorized form of the cross section with the ``effective $q \Bar q g$ wavefunction'' after squaring the amplitude and integrating over the angle of the relative momentum of the $\qqb$ pair. The same traceless rank-2 tensor structure is found in Ref.~\cite{Hebecker:1997gp}, also through a procedure of first squaring the amplitude, summing over internal degrees of freedom and then refactorizing the result. Appendix C of Ref.~\cite{Hebecker:1997gp} gives an interesting interpretation of the tensor structure of the $\gamma\to \ggb$ effective wavefunction in terms of polarization vectors in the projectile and target light cone gauges. 
We also note that the DGLAP splitting function $\left( \xi + ( 1-\xi)^2 \right)$ manifests at this stage of the calculation, i.e. it is an underlying feature of the $\gamma^*_{\rm T} \to q \Bar q g$ splitting function.

After this refactorization, the Fourier transforms over $\Kt_\ggb$ and $\Bar \Kt_\ggb$ now separate, and can be evaluated as:
\begin{align}
    \int \frac{\ud^2 \Kt}{(2\pi)^2} e^{i \Kt \cdot \rt} \frac{\Kt^i \Kt^j + \half \frac{z \widehat\Kt_\ggb^2}{1-z} \delta^{ij}}{\mxqqb^2(\Kt) + Q^2}
    = &
    \frac{\beta \widehat\Kt_\ggb^2}{(1-z)Q^2}
    \int \frac{\ud^2 \Kt}{(2\pi)^2} e^{i \Kt \cdot \rt} \frac{\Kt^i \Kt^j + \half \frac{z \widehat\Kt_\ggb^2}{1-z} \delta^{ij}}{\Kt^2 + \frac{z \widehat\Kt_\ggb^2}{1-z}}
    \\
    = &
    \frac{\beta \widehat\Kt_\ggb^2}{(1-z)Q^2}
    \left[
    \left( \frac{\rt^i \rt^j}{\rt^2} - \half \delta^{ij} \right)
    \left( \delta^{(2)}(\rt) - \frac{1}{2 \pi} \frac{z \widehat\Kt_\ggb^2}{1-z} \besk_2\left( \sqrt{ \frac{z \widehat\Kt_\ggb^2}{1-z} \rt^2} \right) \right)
    + \half \delta^{ij} \delta^{(2)}(\rt)
    \right]
    \notag
    \\
    \cong &
    \frac{1}{2 \pi}
    \frac{\beta \widehat\Kt_\ggb^2}{(1-z)Q^2}
    \frac{z \widehat\Kt_\ggb^2}{1-z} 
    \left( \half \delta^{ij} - \frac{\rt^i \rt^j}{\rt^2} \right)
    \besk_2\left( \sqrt{ \frac{z \widehat\Kt_\ggb^2}{1-z} \rt^2} \right),
\end{align}
where $\cong$ is used to imply effective equivalence since the contributions with $\rt \equiv 0$ vanish identically as the dipole amplitude vanishes at $\rt \equiv 0$. Thus after the transverse Fourier transforms we find for the squared virtual photon amplitude:
\begin{multline}
    \int \frac{\ud^2 \Kt_\ggb}{(2\pi)^2} e^{i \Kt_\ggb \cdot \rt}
    \int \frac{\ud^2 \Bar \Kt_\ggb}{(2\pi)^2} e^{-i \Bar \Kt_\ggb \cdot \rt}
    \half \sum_{h, \lambda, \sigma}
    \left( \psi^{\gamma^{\ast}_{\rm T}\rightarrow q\Bar{q}g}_{\nlo} \bigg\vert_{\rm AJL} \right)^\dagger
    \psi^{\gamma^{\ast}_{\rm T}\rightarrow q\Bar{q}g}_{\nlo} \bigg\vert_{\rm AJL}
    =
    2 \frac{(4 \pi)^2}{(2 \pi)^2} \aem \as e_f^2
    \frac{z_0}{z_1} \frac{\beta^4 \widehat\Kt_\ggb^8}{(1-z)^4 Q^8}
    \\
    \times
    \left[ \xi^2 + (1-\xi)^2 \right]
    \left( \delta^{ij} - 2 \frac{\rt^i \rt^j}{\rt^2} \right)
    \left( \delta^{ij} - 2 \frac{\Bar \rt^i \Bar \rt^j}{\Bar \rt^2} \right)
    \besk_2 \! \left( \sqrt{ \frac{z \widehat\Kt_\ggb^2}{1-z} \rt^2} \right)
    \besk_2 \! \left( \sqrt{ \frac{z \widehat\Kt_\ggb^2}{1-z} \Bar \rt^2} \right) .
\end{multline}

For the next stage we need the following identities which enforce the physical invariant masses at the cross section level:
\begin{equation}
    \int \frac{\ud^2 \widehat \Pt_\qqb}{(2 \pi)^2} \delta \left( \frac{\widehat \Pt_\qqb^2}{z_1} + \frac{\widehat\Kt_\ggb^2}{z_2} - \mx^2 \right)
    =
    \frac{z_1}{4 \pi},
\end{equation}
and
\begin{equation}
    \int_0^1 \frac{\ud z_2}{z_2^3} \delta \left( z - \qty(1 - \frac{\beta \widehat\Kt_\ggb^2}{Q^2 z_2}) \right)
    =
    \frac{(1-z) Q^4}{\beta^2 \widehat\Kt_\ggb^4}.
\end{equation}

We need to perform the change of variables $\left( z_2, \widehat \Kt_\ggb^2 \right) \mapsto \left( z, k^2 \right) $, where $k^2 \coloneqq \widehat\Kt_\ggb^2/(1-z)$ is the mean virtuality of the exchanged $t$-channel gluon in the two-gluon exchange model~\cite{Wusthoff:1997fz,GolecBiernat:1999qd}. This change is enacted by inserting the relations between these quantities as $\delta$-function integrals:
\begin{align}
    \int \frac{\ud^2 \widehat \Kt_\ggb}{(2 \pi)^2} \int \frac{\ud z_2}{z_2^3}
    = &
    \int \frac{\ud^2 \widehat \Kt_\ggb}{(2 \pi)^2} \int \frac{\ud z_2}{z_2^3}
    \int \ud k^2 \int \ud z \,
    \delta \left( z - \qty(1 - \frac{\beta \widehat\Kt_\ggb^2}{Q^2 z_2}) \right)
    \delta \left( k^2 - \frac{\widehat \Kt_\ggb^2}{1-z}\right)
    \notag
    \\
    = &
    \frac{1}{4 \pi} \int \ud k^2 \int \ud z \frac{Q^4}{\beta^2 k^4}.
\end{align}
After this the remaining longitudinal momentum fraction integrals separate and may be performed. Keeping in mind the assumption that $1 \gtrsim z_0 \gg z_1 \gg z_2$, we find:
\begin{align}
    \int_0^1 \ud z_0 \int_{z_{1,\mathrm{min}}}^{z_{1,\mathrm{max}}} \frac{\ud z_1}{z_1} \delta(z_0 + z_1 + z_2 -1)
    = &
    \int_{1-z_2-z_{1,\mathrm{max}}}^{1-z_2-z_{1,\mathrm{min}}} \frac{\ud z_0}{1-z_0-z_2} \nonumber 
    \\
    \label{eq:lnq2_gbw}
    = &
    \log \qty( \frac{z_{1,\mathrm{max}}}{z_{1,\mathrm{min}} } )
    \equiv
    \log \qty(  \frac{Q^2}{\beta k^2} ) \approx \log \qty(  \frac{Q^2}{k^2} ).
\end{align}
Here we have taken $ z_{1,\mathrm{max}}=1$ and $z_{1,\mathrm{min}} \equiv z_2$, reflecting the kinematics of the aligned jet limit. We are also taking $\beta$ to be of order one, and calculating in the leading large logarithmic limit in $Q^2$, thus the constant under the log is not under control at this point of the calculation.
We will additionally need the relation
\begin{equation}
    \int \frac{\ud^2 \rt}{2 \pi}
    \int \frac{\ud^2 \Bar \rt}{2 \pi}
    e^{i \kt_t \cdot (\rt - \Bar \rt)}
    \left( \delta^{ij} - 2 \frac{\rt^i \rt^j}{\rt^2} \right)
    \left( \delta^{ij} - 2 \frac{\Bar \rt^i \Bar \rt^j}{\Bar \rt^2} \right)
    =
    2
    \int \ud r \, r \besj_2 (k_t r) \int \ud \Bar r \, \Bar r \besj_2(k_t \Bar r)
\end{equation}
to simplify the tensor structure by computing the angular integrals.

Recalling that $\xi \coloneqq \frac{\beta}{z}$, we can finally collect our results and write the final result for the $\qqbg$ contribution to the transverse diffractive structure function using Eq.~\eqref{eq:structurefun_diffxs} at the large-$Q^2$ limit
\begin{multline}
    \label{eq:ddis-ft-aligned-jet-limit}
    \xpom F_{T, \,\qqbg}^{\textrm{D}(3), \,\mathrm{LL}(Q^2)}
    =
    2 \nc \cf \frac{\as \beta}{16 \pi^4} \sum_f e_f^2
    \int \ud^2 \bt
    \int_\beta^1 \ud z \left[ \qty( \frac{\beta}{z} )^2 + \qty( 1 - \frac{\beta}{z} )^2 \right]
    \int_0^{Q^2} \ud k^2\, k^4 \log \qty( \frac{z_{1,\mathrm{max}}}{\beta} \frac{Q^2}{k^2} )
    \\
    \times
    \int \ud r \int \ud \Bar r \,\, 
    r \besj_2 \qty(\sqrt{1-z} k r) \Bar r \besj_2 \qty(\sqrt{1-z} k \Bar r)
    \besk_2 \qty( \sqrt{z} k r)
    \besk_2 \qty( \sqrt{z} k \Bar r)
    \left[1-S_{\Bar \rt \bt} \right]^\dagger
    \left[1-S_{\rt \bt} \right]
    \\
    =
    \frac{\as \beta}{8 \pi^4} \sum_f e_f^2
    \int \ud^2 \bt
    \int_\beta^1 \ud z \left[ \qty( \frac{\beta}{z} )^2 + \qty( 1 - \frac{\beta}{z} )^2 \right]
    \int_0^{Q^2} \ud k^2\, k^4 \log \qty(  \frac{Q^2}{k^2} )
    \\
    \times
    \Bigg\lbrace
    \int \ud r \,\, 
    r \besj_2 \qty(\sqrt{1-z} k r)
    \besk_2 \qty( \sqrt{z} k r)
    \Big(2\left[1-S_{\rt \bt} \right] \Big)
    \Bigg\rbrace^2
\end{multline}
where the leading factor of $2$ accounts for the other momentum ordering where one has $z_1 \gg z_0 \gg z_2$. Explicitly writing out the color factor and taking into account the normalization of the dipole amplitude, we find exact agreement with the result of Ref.~\cite{Kowalski:2008sa} shown in Eq.~\eqref{eq:wusthoffqqbarg}. To reiterate, we were able to find exact agreement with the large-$Q^2$ limiting $\qqbg$-contribution to $F_T^D$ with a calculation beginning from the corresponding full
$\mathcal{O}\qty(\as)$-accuracy LCPT
result. This precise agreement is very reassuring, given that the Wüsthoff result was derived in a very different formalism more in the spirit of a perturbative two-gluon exchange~\cite{Wusthoff:1997fz, GolecBiernat:1999qd}.
Results similar to the Wüsthoff result  can be found in Refs.~\cite{Hebecker:1997gp,Buchmuller:1998jv}, where a semi-classical color field picture of the diffractive scattering is utilized.

In Eq.~\eqref{eq:ddis-ft-aligned-jet-limit} we can recognize the origins of some key features. The wavefunction of a gluon splitting to the effective adjoint dipole is a Bessel $\besk_2$ function and the wavefunction overlap of the effective dipole and a final state with invariant mass $\mx$ is a $\besj_2$.  This particular structure, in stead of the usual $\besk_1$, $\besj_1$ for a transverse photon, originates from the transverse tensor structure of the $\gamma\to \ggb$ wavefunction at the aligned jet limit. Furthermore, we see the DGLAP $g \to q \Bar q$ splitting function which emerged from the squared virtual photon wavefunction in momentum space at the aligned jet limit, corresponding to the first step in the DGLAP evolution of a diffractive quark parton distribution function.

The transverse structure function, Eq.~\eqref{eq:ddis-ft-aligned-jet-limit}, is proportional to  $\log Q^2$ and as such dominates at large $Q^2$. This logarithm 
originates from the aligned jet configuration part of the phase space when integrating over the kinematics of the $q\Bar q$ pair. This is explicitly visible in Eq.~\eqref{eq:lnq2_gbw} where the $\log Q^2/k^2$ contribution is obtained when integrating over the quark momentum fraction in the aligned jet limit. As the longitudinal photon wavefunction suppresses the aligned jet configurations where $z_0 \ll 1$ or $z_1 \ll 1$, the same log would not be present in the high-$Q^2$ limit of the longitudinal structure function.

\section{Discussion and conclusions}
\label{sec:conc}

In conclusion, we have here taken a significant step towards calculating diffractive structure functions at NLO accuracy in the color dipole picture applicable to the saturation regime of QCD. Our calculation includes the ``radiative'' part of the NLO correction, i.e. the $\qqbg$ component in the terminology used in earlier works. For the diffractive structure function one calculates the cross section for a fixed invariant mass, a much more inclusive final state than for jet production. Compared to dijet production, this observable thus does not require a jet definition. The subset of the NLO contribution calculated here should be finite by itself.

We have also checked that we can independently reproduce two earlier results for the diffractive structure functions appearing in the literature. In the limit of a soft gluon (i.e. large mass diffractive state), we reproduce an earlier result derived by many authors (including e.g. the derivation by Munier and Shoshi in a framework very similar to ours). More nontrivially, in the limit of a fixed $\beta =  Q^2/(M_X^2+Q^2)$ and large $Q^2$, we recover the earlier result used by Golec-Biernat and Wüsthoff (GBW) and in many other phenomenological studies. The original derivation of this result is perhaps not that clearly documented in the available literature. Certainly it is performed in a collinear factorization-type framework very different from ours. We have thus provided a completely independent rederivation of the large-$Q^2$ result in the dipole picture, fully agreeing with it\footnote{Apart from the treatment of the color factor of the adjoint dipole, which was noted already in Refs.~\cite{Marquet:2007nf,Kowalski:2008sa}.}. Specifically, our calculation shows how to obtain the ingredients of the large-$Q^2$ result: a DGLAP-type logarithm in $Q^2$, a splitting function $P_{g\to q\Bar{q}}$ and also the somewhat curious traceless rank-2 tensor ``photon to effective gluon dipole wavefunction'' (see \eq\nr{eq:ajl-psiTsquared}), from a dipole picture calculation. Thus we believe that the method of this calculation can be helpful in more general for matching the dipole picture with the collinear factorization limit. 

Experimentally, the diffractive structure function is a key part of the program in high energy DIS experiments, both at HERA and at the future EIC. Compared to e.g. diffractive dijets, it is a clean and well defined observable without requiring high transverse momentum or heavy quarks in the final state. This makes it possible to access smaller values of $\xbj$ at a finite collision energy than for dijet observables, and thus to achieve a better sensitivity to gluon saturation.  The nuclear modification of diffractive structure functions has already been identified as a key observable for gluon saturation at the EIC~\cite{Accardi:2012qut}.

Our results are presented in a form that can directly be applied to phenomenology. They generalize the large-$Q^2$ and large-$\mx$ results used in earlier phenomenological studies to a more precise kinematics.  Depending on assumptions on the impact parameter dependence of the dipole amplitude, various different simplifications are possible. However, our main result is completely general in this regard and can be applied to any impact parameter dependence. It will be interesting to evaluate the diffractive structure function numerically, both in order to compare to earlier limiting results, and to test dipole amplitude parametrizations against a new set of experimental data. On the theory side, we have in this paper also outlined the necessary steps to complete the NLO calculation of the diffractive structure function. Here there are many recent results that can be taken advantage of. The loop corrections to  the $\gamma \to \qqbg$ light cone wavefunction are known~\cite{Beuf:2016wdz,Beuf:2017bpd,Hanninen:2017ddy}. So is the procedure to factorize the large logarithms of $\xpom$ into the BK/JIMWLK evolution of the target~\cite{Ducloue:2017ftk,Hanninen:2021byo}, once the corresponding diagrams (with a gluon crossing the shockwave but not the cut) are calculated. The treatment of final state gluon exchanges poses interesting conceptual questions that are new in the context of LCPT.  While in many ways straightforward, these further calculations are sizeable enough projects that they are best left for future publications.

\section*{Acknowledgments}

We thank K. Golec-Biernat, A. van Hameren, C. Marquet, R. Paatelainen and  J. Penttala  for discussions. 
H.H, T.L and H.M are supported by the Academy of Finland, the Centre of Excellence in Quark Matter (project 346324) and projects 338263, 346567 and 321840.
G.B is supported in part by the National Science Centre (Poland) under the research grant no. 2020/38/E/ST2/00122 (SONATA BIS 10).
Y.M acknowledges financial support from Xunta de Galicia (Centro singular de investigaci\'on de Galicia accreditation 2019-2022); the ``Mar\'{\i}a de Maeztu'' Units of Excellence program MDM2016-0692 and the Spanish Research State Agency under project PID2020-119632GB-I00; European Union ERDF.
G.B and Y.M acknowledge financial support from MSCA RISE 823947 ”Heavy ion collisions: collectivity and precision in saturation physics” (HIEIC).
This work was also supported under the European Union’s Horizon 2020 research and innovation programme by the European Research Council (ERC, grant agreement No. ERC-2018-ADG-835105 YoctoLHC) and by the STRONG-2020 project (grant agreement No. 824093).
The content of this article does not reflect the official opinion of the European Union and responsibility for the information and views expressed therein lies entirely with the authors. 
    
\bibliographystyle{JHEP-2modlong.bst}
\bibliography{refs}

\end{document}

%% file: defs.tex
\newcommand{\Pt}{{\mathbf{P}}}

\newcommand{\rt}{{\mathbf{r}}}

\newcommand{\xt}{{\mathbf{x}}}
\newcommand{\cxt}{{\conj{\xt}}}
\newcommand{\bt}{{\mathbf{b}}}
\newcommand{\yt}{{\mathbf{y}}}
\newcommand{\Yt}{{\mathbf{Y}}}
\newcommand{\zt}{{\mathbf{z}}}
\newcommand{\pt}{{\mathbf{p}}}

\newcommand{\ut}{{\mathbf{u}}}
\newcommand{\qt}{{\mathbf{q}}}
\newcommand{\kt}{{\mathbf{k}}}

\newcommand{\mt}{{\mathbf{m}}}

\newcommand{\Kt}{\mathbf{K}}

\newcommand{\lt}{\mathbf{l}}

\newcommand{\epst}{\boldsymbol{\varepsilon}}

\newcommand{\Deltat}{{\boldsymbol{\Delta}}}

\newcommand{\pvec}{{\vec{p}}}

\newcommand{\besj}{{\mathrm{J}}}
\newcommand{\besk}{{\mathrm{K}}}

\newcommand{\vecp}[1]{\pvec_{#1}}

\newcommand{\nlo}{{\textnormal{NLO}}}

\newcommand{\ud}{\, \mathrm{d}}

\newcommand{\nc}{{N_\mathrm{c}}}
\newcommand{\Cf}{{C_\mathrm{F}}}

\newcommand{\half}{\frac{1}{2}}

\newcommand{\cf}{C_\mathrm{F}}

\newcommand{\nr}[1]{(\ref{#1})}

\newcommand{\gev}{\ \textrm{GeV}}

\newcommand{\as}{\alpha_{\mathrm{s}}}
\newcommand{\aem}{\alpha_{\mathrm{em}}}

\newcommand{\conj}[1]{\mathop{\overline{#1}}\nolimits}
\newcommand{\conz}{\conj{0}}
\newcommand{\cono}{\conj{1}}

\newcommand{\contrip}{\conj{0} \conj{1} \conj{2}}

\newcommand{\fig}{Fig.~}
\newcommand{\figs}{Figs.~}
\newcommand{\eq}{Eq.~}
\newcommand{\eqs}{Eqs.~}

\newcommand{\mx}{M_X}
\newcommand{\mxqqb}{M_{q \bar q g}}

\newcommand{\dsigmaadj}{{\frac{\ud \tilde{\sigma}_\textrm{dip}}{\ud^2 \bt}}}

\newcounter{diag}
\newcounter{subdiag}[diag]
\newcommand{\namediag}[1]{\refstepcounter{diag} \thediag \label{#1}}

\renewcommand{\thediag}{(\alph{diag})}

\newcommand{\qplus}{q^{+}}

\newcommand{\kplus}{k^{+}}

\newcommand{\pplus}{p^{+}}

\newcommand{\gamlam}{{\gamma^*_\lambda}}
\newcommand{\se}{{\Hat{S}_E}}

%% file: main.bbl
\providecommand{\href}[2]{#2}\begingroup\raggedright\begin{thebibliography}{100}

\bibitem{Accardi:2012qut}
A.~Accardi {\em et.~al.}, {\it {Electron Ion Collider: The Next QCD Frontier}:
  {Understanding the glue that binds us all}},
  \href{http://dx.doi.org/10.1140/epja/i2016-16268-9}{{\em Eur. Phys. J. A}
  {\bf 52} (2016)~no.~9 268} [\href{http://arXiv.org/abs/1212.1701}{{\tt
  arXiv:1212.1701 [nucl-ex]}}].

\bibitem{Aschenauer:2017jsk}
E.~C. Aschenauer, S.~Fazio, J.~H. Lee, H.~Mäntysaari, B.~S. Page, B.~Schenke,
  T.~Ullrich, R.~Venugopalan and P.~Zurita, {\it {The electron\textendash{}ion
  collider: assessing the energy dependence of key measurements}},
  \href{http://dx.doi.org/10.1088/1361-6633/aaf216}{{\em Rept. Prog. Phys.}
  {\bf 82} (2019)~no.~2 024301} [\href{http://arXiv.org/abs/1708.01527}{{\tt
  arXiv:1708.01527 [nucl-ex]}}].

\bibitem{AbdulKhalek:2021gbh}
R.~Abdul~Khalek {\em et.~al.}, {\it {Science Requirements and Detector Concepts
  for the Electron-Ion Collider: EIC Yellow Report}},
  \href{http://arXiv.org/abs/2103.05419}{{\tt arXiv:2103.05419
  [physics.ins-det]}}.

\bibitem{LHeCStudyGroup:2012zhm}
{\bf LHeC Study Group} collaboration, J.~L. Abelleira~Fernandez {\em et.~al.},
  {\it {A Large Hadron Electron Collider at CERN: Report on the Physics and
  Design Concepts for Machine and Detector}},
  \href{http://dx.doi.org/10.1088/0954-3899/39/7/075001}{{\em J. Phys. G} {\bf
  39} (2012) 075001} [\href{http://arXiv.org/abs/1206.2913}{{\tt
  arXiv:1206.2913 [physics.acc-ph]}}].

\bibitem{LHeC:2020van}
{\bf LHeC, FCC-he Study Group} collaboration, P.~Agostini {\em et.~al.}, {\it
  {The Large Hadron-Electron Collider at the HL-LHC}},
  \href{http://dx.doi.org/10.1088/1361-6471/abf3ba}{{\em J. Phys. G} {\bf 48}
  (2021)~no.~11 110501} [\href{http://arXiv.org/abs/2007.14491}{{\tt
  arXiv:2007.14491 [hep-ex]}}].

\bibitem{Iancu:2003xm}
E.~Iancu and R.~Venugopalan, {\it {The Color glass condensate and high-energy
  scattering in QCD}},
  \href{http://dx.doi.org/10.1142/9789812795533_0005}{{\em Quark gluon plasma
  3}  (2003) 249} [\href{http://arXiv.org/abs/hep-ph/0303204}{{\tt
  arXiv:hep-ph/0303204 [hep-ph]}}].

\bibitem{Weigert:2005us}
H.~Weigert, {\it {Evolution at small $x_{Bj}$: The Color glass condensate}},
  \href{http://dx.doi.org/10.1016/j.ppnp.2005.01.029}{{\em Prog. Part. Nucl.
  Phys.} {\bf 55} (2005) 461} [\href{http://arXiv.org/abs/hep-ph/0501087}{{\tt
  arXiv:hep-ph/0501087}}].

\bibitem{Gelis:2010nm}
F.~Gelis, E.~Iancu, J.~Jalilian-Marian and R.~Venugopalan, {\it {The Color
  Glass Condensate}},
  \href{http://dx.doi.org/10.1146/annurev.nucl.010909.083629}{{\em Ann. Rev.
  Nucl. Part. Sci.} {\bf 60} (2010) 463}
  [\href{http://arXiv.org/abs/1002.0333}{{\tt arXiv:1002.0333 [hep-ph]}}].

\bibitem{Nikolaev:1990ja}
N.~N. Nikolaev and B.~G. Zakharov, {\it {Color transparency and scaling
  properties of nuclear shadowing in deep inelastic scattering}},
  \href{http://dx.doi.org/10.1007/BF01483577}{{\em Z. Phys. C} {\bf 49} (1991)
  607}.

\bibitem{Nikolaev:1991et}
N.~Nikolaev and B.~G. Zakharov, {\it {Pomeron structure function and
  diffraction dissociation of virtual photons in perturbative QCD}},
  \href{http://dx.doi.org/10.1007/BF01597573}{{\em Z. Phys. C} {\bf 53} (1992)
  331}.

\bibitem{Mueller:1993rr}
A.~H. Mueller, {\it {Soft gluons in the infinite momentum wave function and the
  BFKL pomeron}},  \href{http://dx.doi.org/10.1016/0550-3213(94)90116-3}{{\em
  Nucl. Phys. B} {\bf 415} (1994) 373}.

\bibitem{Mueller:1994jq}
A.~H. Mueller and B.~Patel, {\it {Single and double BFKL pomeron exchange and a
  dipole picture of high-energy hard processes}},
  \href{http://dx.doi.org/10.1016/0550-3213(94)90284-4}{{\em Nucl. Phys. B}
  {\bf 425} (1994) 471} [\href{http://arXiv.org/abs/hep-ph/9403256}{{\tt
  arXiv:hep-ph/9403256}}].

\bibitem{Mueller:1994gb}
A.~H. Mueller, {\it {Unitarity and the BFKL pomeron}},
  \href{http://dx.doi.org/10.1016/0550-3213(94)00480-3}{{\em Nucl. Phys. B}
  {\bf 437} (1995) 107} [\href{http://arXiv.org/abs/hep-ph/9408245}{{\tt
  arXiv:hep-ph/9408245}}].

\bibitem{Balitsky:1995ub}
I.~Balitsky, {\it {Operator expansion for high-energy scattering}},
  \href{http://dx.doi.org/10.1016/0550-3213(95)00638-9}{{\em Nucl. Phys. B}
  {\bf 463} (1996) 99} [\href{http://arXiv.org/abs/hep-ph/9509348}{{\tt
  arXiv:hep-ph/9509348}}].

\bibitem{Kovchegov:1999ua}
Y.~V. Kovchegov, {\it {Unitarization of the BFKL pomeron on a nucleus}},
  \href{http://dx.doi.org/10.1103/PhysRevD.61.074018}{{\em Phys. Rev. D} {\bf
  61} (2000) 074018} [\href{http://arXiv.org/abs/hep-ph/9905214}{{\tt
  arXiv:hep-ph/9905214}}].

\bibitem{Kovchegov:1999yj}
Y.~V. Kovchegov, {\it {Small-$x$ $F_2$ structure function of a nucleus
  including multiple pomeron exchanges}},
  \href{http://dx.doi.org/10.1103/PhysRevD.60.034008}{{\em Phys. Rev. D} {\bf
  60} (1999) 034008} [\href{http://arXiv.org/abs/hep-ph/9901281}{{\tt
  arXiv:hep-ph/9901281}}].

\bibitem{Balitsky:2008zza}
I.~Balitsky and G.~A. Chirilli, {\it {Next-to-leading order evolution of color
  dipoles}},  \href{http://dx.doi.org/10.1103/PhysRevD.77.014019}{{\em Phys.
  Rev. D} {\bf 77} (2008) 014019} [\href{http://arXiv.org/abs/0710.4330}{{\tt
  arXiv:0710.4330 [hep-ph]}}].

\bibitem{Balitsky:2013fea}
I.~Balitsky and G.~A. Chirilli, {\it {Rapidity evolution of Wilson lines at the
  next-to-leading order}},
  \href{http://dx.doi.org/10.1103/PhysRevD.88.111501}{{\em Phys. Rev. D} {\bf
  88} (2013) 111501} [\href{http://arXiv.org/abs/1309.7644}{{\tt
  arXiv:1309.7644 [hep-ph]}}].

\bibitem{Kovner:2013ona}
A.~Kovner, M.~Lublinsky and Y.~Mulian, {\it {Jalilian-Marian, Iancu, McLerran,
  Weigert, Leonidov, Kovner evolution at next to leading order}},
  \href{http://dx.doi.org/10.1103/PhysRevD.89.061704}{{\em Phys. Rev. D} {\bf
  89} (2014)~no.~6 061704} [\href{http://arXiv.org/abs/1310.0378}{{\tt
  arXiv:1310.0378 [hep-ph]}}].

\bibitem{Balitsky:2014mca}
I.~Balitsky and A.~V. Grabovsky, {\it {NLO evolution of 3-quark Wilson loop
  operator}},  \href{http://dx.doi.org/10.1007/JHEP01(2015)009}{{\em JHEP} {\bf
  01} (2015) 009} [\href{http://arXiv.org/abs/1405.0443}{{\tt arXiv:1405.0443
  [hep-ph]}}].

\bibitem{Beuf:2014uia}
G.~Beuf, {\it {Improving the kinematics for low-$x$ QCD evolution equations in
  coordinate space}},  \href{http://dx.doi.org/10.1103/PhysRevD.89.074039}{{\em
  Phys. Rev. D} {\bf 89} (2014)~no.~7 074039}
  [\href{http://arXiv.org/abs/1401.0313}{{\tt arXiv:1401.0313 [hep-ph]}}].

\bibitem{Lappi:2015fma}
T.~Lappi and H.~M\"antysaari, {\it {Direct numerical solution of the coordinate
  space Balitsky-Kovchegov equation at next to leading order}},
  \href{http://dx.doi.org/10.1103/PhysRevD.91.074016}{{\em Phys. Rev. D} {\bf
  91} (2015)~no.~7 074016} [\href{http://arXiv.org/abs/1502.02400}{{\tt
  arXiv:1502.02400 [hep-ph]}}].

\bibitem{Iancu:2015vea}
E.~Iancu, J.~D. Madrigal, A.~H. Mueller, G.~Soyez and D.~N. Triantafyllopoulos,
  {\it {Resumming double logarithms in the QCD evolution of color dipoles}},
  \href{http://dx.doi.org/10.1016/j.physletb.2015.03.068}{{\em Phys. Lett. B}
  {\bf 744} (2015) 293} [\href{http://arXiv.org/abs/1502.05642}{{\tt
  arXiv:1502.05642 [hep-ph]}}].

\bibitem{Iancu:2015joa}
E.~Iancu, J.~D. Madrigal, A.~H. Mueller, G.~Soyez and D.~N. Triantafyllopoulos,
  {\it {Collinearly-improved BK evolution meets the HERA data}},
  \href{http://dx.doi.org/10.1016/j.physletb.2015.09.071}{{\em Phys. Lett. B}
  {\bf 750} (2015) 643} [\href{http://arXiv.org/abs/1507.03651}{{\tt
  arXiv:1507.03651 [hep-ph]}}].

\bibitem{Albacete:2015xza}
J.~L. Albacete, {\it {Resummation of double collinear logs in BK evolution
  versus HERA data}},
  \href{http://dx.doi.org/10.1016/j.nuclphysa.2016.07.008}{{\em Nucl. Phys. A}
  {\bf 957} (2017) 71} [\href{http://arXiv.org/abs/1507.07120}{{\tt
  arXiv:1507.07120 [hep-ph]}}].

\bibitem{Lappi:2016fmu}
T.~Lappi and H.~M\"antysaari, {\it {Next-to-leading order Balitsky-Kovchegov
  equation with resummation}},
  \href{http://dx.doi.org/10.1103/PhysRevD.93.094004}{{\em Phys. Rev. D} {\bf
  93} (2016)~no.~9 094004} [\href{http://arXiv.org/abs/1601.06598}{{\tt
  arXiv:1601.06598 [hep-ph]}}].

\bibitem{Lublinsky:2016meo}
M.~Lublinsky and Y.~Mulian, {\it {High Energy QCD at NLO: from light-cone wave
  function to JIMWLK evolution}},
  \href{http://dx.doi.org/10.1007/JHEP05(2017)097}{{\em JHEP} {\bf 05} (2017)
  097} [\href{http://arXiv.org/abs/1610.03453}{{\tt arXiv:1610.03453
  [hep-ph]}}].

\bibitem{Balitsky:2010ze}
I.~Balitsky and G.~A. Chirilli, {\it {Photon impact factor in the
  next-to-leading order}},
  \href{http://dx.doi.org/10.1103/PhysRevD.83.031502}{{\em Phys. Rev. D} {\bf
  83} (2011) 031502} [\href{http://arXiv.org/abs/1009.4729}{{\tt
  arXiv:1009.4729 [hep-ph]}}].

\bibitem{Balitsky:2012bs}
I.~Balitsky and G.~A. Chirilli, {\it {Photon impact factor and
  $k_T$-factorization for DIS in the next-to-leading order}},
  \href{http://dx.doi.org/10.1103/PhysRevD.87.014013}{{\em Phys. Rev. D} {\bf
  87} (2013)~no.~1 014013} [\href{http://arXiv.org/abs/1207.3844}{{\tt
  arXiv:1207.3844 [hep-ph]}}].

\bibitem{Beuf:2011xd}
G.~Beuf, {\it {NLO corrections for the dipole factorization of DIS structure
  functions at low x}},
  \href{http://dx.doi.org/10.1103/PhysRevD.85.034039}{{\em Phys. Rev. D} {\bf
  85} (2012) 034039} [\href{http://arXiv.org/abs/1112.4501}{{\tt
  arXiv:1112.4501 [hep-ph]}}].

\bibitem{Beuf:2016wdz}
G.~Beuf, {\it {Dipole factorization for DIS at NLO: Loop correction to the
  $\gamma^*_{T,L}\to q\overline q$ light-front wave functions}},
  \href{http://dx.doi.org/10.1103/PhysRevD.94.054016}{{\em Phys. Rev. D} {\bf
  94} (2016)~no.~5 054016} [\href{http://arXiv.org/abs/1606.00777}{{\tt
  arXiv:1606.00777 [hep-ph]}}].

\bibitem{Beuf:2017bpd}
G.~Beuf, {\it {Dipole factorization for DIS at NLO: Combining the $q\bar{q}$
  and $q\bar{q}g$ contributions}},
  \href{http://dx.doi.org/10.1103/PhysRevD.96.074033}{{\em Phys. Rev. D} {\bf
  96} (2017)~no.~7 074033} [\href{http://arXiv.org/abs/1708.06557}{{\tt
  arXiv:1708.06557 [hep-ph]}}].

\bibitem{Hanninen:2017ddy}
H.~H\"anninen, T.~Lappi and R.~Paatelainen, {\it {One-loop corrections to light
  cone wave functions: the dipole picture DIS cross section}},
  \href{http://dx.doi.org/10.1016/j.aop.2018.04.015}{{\em Annals Phys.} {\bf
  393} (2018) 358} [\href{http://arXiv.org/abs/1711.08207}{{\tt
  arXiv:1711.08207 [hep-ph]}}].

\bibitem{Beuf:2020dxl}
G.~Beuf, H.~H\"anninen, T.~Lappi and H.~M\"antysaari, {\it {Color Glass
  Condensate at next-to-leading order meets HERA data}},
  \href{http://dx.doi.org/10.1103/PhysRevD.102.074028}{{\em Phys. Rev. D} {\bf
  102} (2020) 074028} [\href{http://arXiv.org/abs/2007.01645}{{\tt
  arXiv:2007.01645 [hep-ph]}}].

\bibitem{Lappi:2020srm}
T.~Lappi, H.~M\"antysaari and A.~Ramnath, {\it {Next-to-leading order
  Balitsky-Kovchegov equation beyond large $N_c$}},
  \href{http://dx.doi.org/10.1103/PhysRevD.102.074027}{{\em Phys. Rev. D} {\bf
  102} (2020)~no.~7 074027} [\href{http://arXiv.org/abs/2007.00751}{{\tt
  arXiv:2007.00751 [hep-ph]}}].

\bibitem{Caucal:2021ent}
P.~Caucal, F.~Salazar and R.~Venugopalan, {\it {Dijet impact factor in DIS at
  next-to-leading order in the Color Glass Condensate}},
  \href{http://dx.doi.org/10.1007/JHEP11(2021)222}{{\em JHEP} {\bf 11} (2021)
  222} [\href{http://arXiv.org/abs/2108.06347}{{\tt arXiv:2108.06347
  [hep-ph]}}].

\bibitem{Taels:2022tza}
P.~Taels, T.~Altinoluk, G.~Beuf and C.~Marquet, {\it {Dijet photoproduction at
  low $x$ at next-to-leading order and its back-to-back limit}},
  \href{http://arXiv.org/abs/2204.11650}{{\tt arXiv:2204.11650 [hep-ph]}}.

\bibitem{Beuf:2021qqa}
G.~Beuf, T.~Lappi and R.~Paatelainen, {\it {Massive quarks in NLO dipole
  factorization for DIS: Longitudinal photon}},
  \href{http://dx.doi.org/10.1103/PhysRevD.104.056032}{{\em Phys. Rev. D} {\bf
  104} (2021)~no.~5 056032} [\href{http://arXiv.org/abs/2103.14549}{{\tt
  arXiv:2103.14549 [hep-ph]}}].

\bibitem{Beuf:2021srj}
G.~Beuf, T.~Lappi and R.~Paatelainen, {\it {Massive quarks at one loop in the
  dipole picture of Deep Inelastic Scattering}},
  \href{http://arXiv.org/abs/2112.03158}{{\tt arXiv:2112.03158 [hep-ph]}}.

\bibitem{Beuf:2022ndu}
G.~Beuf, T.~Lappi and R.~Paatelainen, {\it {Massive quarks in NLO dipole
  factorization for DIS: Transverse photon}},
  \href{http://arXiv.org/abs/2204.02486}{{\tt arXiv:2204.02486 [hep-ph]}}.

\bibitem{Golec-Biernat:1999qor}
K.~J. Golec-Biernat and M.~Wusthoff, {\it {Saturation in diffractive deep
  inelastic scattering}},
  \href{http://dx.doi.org/10.1103/PhysRevD.60.114023}{{\em Phys. Rev. D} {\bf
  60} (1999) 114023} [\href{http://arXiv.org/abs/hep-ph/9903358}{{\tt
  arXiv:hep-ph/9903358}}].

\bibitem{Kowalski:2007rw}
H.~Kowalski, T.~Lappi and R.~Venugopalan, {\it {Nuclear enhancement of
  universal dynamics of high parton densities}},
  \href{http://dx.doi.org/10.1103/PhysRevLett.100.022303}{{\em Phys. Rev.
  Lett.} {\bf 100} (2008) 022303} [\href{http://arXiv.org/abs/0705.3047}{{\tt
  arXiv:0705.3047 [hep-ph]}}].

\bibitem{Armesto:2019gxy}
N.~Armesto, P.~R. Newman, W.~S\l{}omi\'nski and A.~M. Sta\'sto, {\it {Inclusive
  diffraction in future electron-proton and electron-ion colliders}},
  \href{http://dx.doi.org/10.1103/PhysRevD.100.074022}{{\em Phys. Rev. D} {\bf
  100} (2019)~no.~7 074022} [\href{http://arXiv.org/abs/1901.09076}{{\tt
  arXiv:1901.09076 [hep-ph]}}].

\bibitem{Boussarie:2014lxa}
R.~Boussarie, A.~V. Grabovsky, L.~Szymanowski and S.~Wallon, {\it {Impact
  factor for high-energy two and three jets diffractive production}},
  \href{http://dx.doi.org/10.1007/JHEP09(2014)026}{{\em JHEP} {\bf 09} (2014)
  026} [\href{http://arXiv.org/abs/1405.7676}{{\tt arXiv:1405.7676 [hep-ph]}}].

\bibitem{Boussarie:2016ogo}
R.~Boussarie, A.~V. Grabovsky, L.~Szymanowski and S.~Wallon, {\it {On the one
  loop $ {\gamma}^{\left(\ast \right)}\to q\overline{q} $ impact factor and the
  exclusive diffractive cross sections for the production of two or three
  jets}},  \href{http://dx.doi.org/10.1007/JHEP11(2016)149}{{\em JHEP} {\bf 11}
  (2016) 149} [\href{http://arXiv.org/abs/1606.00419}{{\tt arXiv:1606.00419
  [hep-ph]}}].

\bibitem{Boussarie:2016bkq}
R.~Boussarie, A.~V. Grabovsky, D.~Y. Ivanov, L.~Szymanowski and S.~Wallon, {\it
  {Next-to-Leading Order Computation of Exclusive Diffractive Light Vector
  Meson Production in a Saturation Framework}},
  \href{http://dx.doi.org/10.1103/PhysRevLett.119.072002}{{\em Phys. Rev.
  Lett.} {\bf 119} (2017)~no.~7 072002}
  [\href{http://arXiv.org/abs/1612.08026}{{\tt arXiv:1612.08026 [hep-ph]}}].

\bibitem{Escobedo:2019bxn}
M.~A. Escobedo and T.~Lappi, {\it {Dipole picture and the nonrelativistic
  expansion}},  \href{http://dx.doi.org/10.1103/PhysRevD.101.034030}{{\em Phys.
  Rev. D} {\bf 101} (2020)~no.~3 034030}
  [\href{http://arXiv.org/abs/1911.01136}{{\tt arXiv:1911.01136 [hep-ph]}}].

\bibitem{Lappi:2020ufv}
T.~Lappi, H.~M\"antysaari and J.~Penttala, {\it {Relativistic corrections to
  the vector meson light front wave function}},
  \href{http://dx.doi.org/10.1103/PhysRevD.102.054020}{{\em Phys. Rev. D} {\bf
  102} (2020)~no.~5 054020} [\href{http://arXiv.org/abs/2006.02830}{{\tt
  arXiv:2006.02830 [hep-ph]}}].

\bibitem{Mantysaari:2021ryb}
H.~M\"antysaari and J.~Penttala, {\it {Exclusive heavy vector meson production
  at next-to-leading order in the dipole picture}},
  \href{http://dx.doi.org/10.1016/j.physletb.2021.136723}{{\em Phys. Lett. B}
  {\bf 823} (2021) 136723} [\href{http://arXiv.org/abs/2104.02349}{{\tt
  arXiv:2104.02349 [hep-ph]}}].

\bibitem{Mantysaari:2022bsp}
H.~M\"antysaari and J.~Penttala, {\it {Exclusive production of light vector
  mesons at next-to-leading order in the dipole picture}},
  \href{http://arXiv.org/abs/2203.16911}{{\tt arXiv:2203.16911 [hep-ph]}}.

\bibitem{Mantysaari:2022kdm}
H.~M\"antysaari and J.~Penttala, {\it {Complete calculation of exclusive heavy
  vector meson production at next-to-leading order in the dipole picture}},
  \href{http://arXiv.org/abs/2204.14031}{{\tt arXiv:2204.14031 [hep-ph]}}.

\bibitem{Iancu:2021rup}
E.~Iancu, A.~H. Mueller and D.~N. Triantafyllopoulos, {\it {Probing Parton
  Saturation and the Gluon Dipole via Diffractive Jet Production at the
  Electron-Ion Collider}},
  \href{http://dx.doi.org/10.1103/PhysRevLett.128.202001}{{\em Phys. Rev.
  Lett.} {\bf 128} (2022)~no.~20 202001}
  [\href{http://arXiv.org/abs/2112.06353}{{\tt arXiv:2112.06353 [hep-ph]}}].

\bibitem{Hatta:2022lzj}
Y.~Hatta, B.-W. Xiao and F.~Yuan, {\it {Semi-inclusive Diffractive Deep
  Inelastic Scattering at Small-$x$}},
  \href{http://arXiv.org/abs/2205.08060}{{\tt arXiv:2205.08060 [hep-ph]}}.

\bibitem{Lepage:1980fj}
G.~P. Lepage and S.~J. Brodsky, {\it {Exclusive Processes in Perturbative
  Quantum Chromodynamics}},
  \href{http://dx.doi.org/10.1103/PhysRevD.22.2157}{{\em Phys. Rev. D} {\bf 22}
  (1980) 2157}.

\bibitem{Wusthoff:1997fz}
M.~Wusthoff, {\it {Large rapidity gap events in deep inelastic scattering}},
  \href{http://dx.doi.org/10.1103/PhysRevD.56.4311}{{\em Phys. Rev. D} {\bf 56}
  (1997) 4311} [\href{http://arXiv.org/abs/hep-ph/9702201}{{\tt
  arXiv:hep-ph/9702201}}].

\bibitem{Golec-Biernat:1999qd}
K.~J. Golec-Biernat and M.~Wusthoff, {\it {Saturation in diffractive deep
  inelastic scattering}},
  \href{http://dx.doi.org/10.1103/PhysRevD.60.114023}{{\em Phys. Rev. D} {\bf
  60} (1999) 114023} [\href{http://arXiv.org/abs/hep-ph/9903358}{{\tt
  arXiv:hep-ph/9903358}}].

\bibitem{GolecBiernat:1999qd}
K.~J. Golec-Biernat and M.~Wusthoff, {\it {Saturation in diffractive deep
  inelastic scattering}},
  \href{http://dx.doi.org/10.1103/PhysRevD.60.114023}{{\em Phys. Rev. D} {\bf
  60} (1999) 114023} [\href{http://arXiv.org/abs/hep-ph/9903358}{{\tt
  arXiv:hep-ph/9903358}}].

\bibitem{Marquet:2007nf}
C.~Marquet, {\it {A Unified description of diffractive deep inelastic
  scattering with saturation}},
  \href{http://dx.doi.org/10.1103/PhysRevD.76.094017}{{\em Phys. Rev. D} {\bf
  76} (2007) 094017} [\href{http://arXiv.org/abs/0706.2682}{{\tt
  arXiv:0706.2682 [hep-ph]}}].

\bibitem{Kowalski:2008sa}
H.~Kowalski, T.~Lappi, C.~Marquet and R.~Venugopalan, {\it {Nuclear enhancement
  and suppression of diffractive structure functions at high energies}},
  \href{http://dx.doi.org/10.1103/PhysRevC.78.045201}{{\em Phys. Rev. C} {\bf
  78} (2008) 045201} [\href{http://arXiv.org/abs/0805.4071}{{\tt
  arXiv:0805.4071 [hep-ph]}}].

\bibitem{Kugeratski:2005ck}
M.~S. Kugeratski, V.~P. Goncalves and F.~S. Navarra, {\it {Saturation in
  diffractive deep inelastic eA scattering}},
  \href{http://dx.doi.org/10.1140/epjc/s2006-02517-7}{{\em Eur. Phys. J. C}
  {\bf 46} (2006) 413} [\href{http://arXiv.org/abs/hep-ph/0511224}{{\tt
  arXiv:hep-ph/0511224}}].

\bibitem{Bendova:2020hkp}
D.~Bendova, J.~Cepila, J.~G. Contreras, t.~V.~P. Gon\c{c}alves and M.~Matas,
  {\it {Diffractive deeply inelastic scattering in future electron-ion
  colliders}},  \href{http://dx.doi.org/10.1140/epjc/s10052-021-09006-x}{{\em
  Eur. Phys. J. C} {\bf 81} (2021)~no.~3 211}
  [\href{http://arXiv.org/abs/2009.14002}{{\tt arXiv:2009.14002 [hep-ph]}}].

\bibitem{Hanninen:2021byo}
H.~H\"anninen, {\em {Deep Inelastic Scattering in the Dipole Picture at
  Next-to-Leading Order}}.
\newblock PhD thesis, University of Jyväskylä, 2021.
\newblock \href{http://arXiv.org/abs/2112.08818}{{\tt arXiv:2112.08818
  [hep-ph]}}.

\bibitem{Munier:2003zb}
S.~Munier and A.~Shoshi, {\it {Diffractive photon dissociation in the
  saturation regime from the Good and Walker picture}},
  \href{http://dx.doi.org/10.1103/PhysRevD.69.074022}{{\em Phys. Rev. D} {\bf
  69} (2004) 074022} [\href{http://arXiv.org/abs/hep-ph/0312022}{{\tt
  arXiv:hep-ph/0312022}}].

\bibitem{Good:1960ba}
M.~L. Good and W.~D. Walker, {\it {Diffraction disssociation of beam
  particles}},  \href{http://dx.doi.org/10.1103/PhysRev.120.1857}{{\em Phys.
  Rev.} {\bf 120} (1960) 1857}.

\bibitem{Miettinen:1978jb}
H.~I. Miettinen and J.~Pumplin, {\it {Diffraction Scattering and the Parton
  Structure of Hadrons}},
  \href{http://dx.doi.org/10.1103/PhysRevD.18.1696}{{\em Phys. Rev.} {\bf D18}
  (1978) 1696}.

\bibitem{Caldwell:2009ke}
A.~Caldwell and H.~Kowalski, {\it {Investigating the gluonic structure of
  nuclei via $\mathrm{J}/\psi$ scattering}},
  \href{http://dx.doi.org/10.1103/PhysRevC.81.025203}{{\em Phys. Rev.} {\bf
  C81} (2010) 025203} [\href{http://arXiv.org/abs/0909.1254}{{\tt
  arXiv:0909.1254}}].

\bibitem{Mantysaari:2016ykx}
H.~M\"antysaari and B.~Schenke, {\it {Evidence of strong proton shape
  fluctuations from incoherent diffraction}},
  \href{http://dx.doi.org/10.1103/PhysRevLett.117.052301}{{\em Phys. Rev.
  Lett.} {\bf 117} (2016)~no.~5 052301}
  [\href{http://arXiv.org/abs/1603.04349}{{\tt arXiv:1603.04349 [hep-ph]}}].

\bibitem{Mantysaari:2020axf}
H.~Mäntysaari, {\it {Review of proton and nuclear shape fluctuations at high
  energy}},  \href{http://dx.doi.org/10.1088/1361-6633/aba347}{{\em Rept. Prog.
  Phys.} {\bf 83} (2020) 082201} [\href{http://arXiv.org/abs/2001.10705}{{\tt
  arXiv:2001.10705 [hep-ph]}}].

\bibitem{Aktas:2006hy}
{\bf H1} collaboration, A.~Aktas {\em et.~al.}, {\it {Measurement and QCD
  analysis of the diffractive deep-inelastic scattering cross-section at
  HERA}},  \href{http://dx.doi.org/10.1140/epjc/s10052-006-0035-3}{{\em Eur.
  Phys. J. C} {\bf 48} (2006) 715}
  [\href{http://arXiv.org/abs/hep-ex/0606004}{{\tt arXiv:hep-ex/0606004}}].

\bibitem{Aktas:2006hx}
{\bf H1} collaboration, A.~Aktas {\em et.~al.}, {\it {Diffractive
  deep-inelastic scattering with a leading proton at HERA}},
  \href{http://dx.doi.org/10.1140/epjc/s10052-006-0046-0}{{\em Eur. Phys. J. C}
  {\bf 48} (2006) 749} [\href{http://arXiv.org/abs/hep-ex/0606003}{{\tt
  arXiv:hep-ex/0606003}}].

\bibitem{Aktas:2006up}
{\bf H1} collaboration, A.~Aktas {\em et.~al.}, {\it {Diffractive open charm
  production in deep-inelastic scattering and photoproduction at HERA}},
  \href{http://dx.doi.org/10.1140/epjc/s10052-006-0206-2}{{\em Eur. Phys. J. C}
  {\bf 50} (2007) 1} [\href{http://arXiv.org/abs/hep-ex/0610076}{{\tt
  arXiv:hep-ex/0610076}}].

\bibitem{Aaron:2012hua}
{\bf H1, ZEUS} collaboration, F.~D. Aaron {\em et.~al.}, {\it {Combined
  inclusive diffractive cross sections measured with forward proton
  spectrometers in deep inelastic $ep$ scattering at HERA}},
  \href{http://dx.doi.org/10.1140/epjc/s10052-012-2175-y}{{\em Eur. Phys. J. C}
  {\bf 72} (2012) 2175} [\href{http://arXiv.org/abs/1207.4864}{{\tt
  arXiv:1207.4864 [hep-ex]}}].

\bibitem{Bjorken:1970ah}
J.~D. Bjorken, J.~B. Kogut and D.~E. Soper, {\it {Quantum Electrodynamics at
  Infinite Momentum: Scattering from an External Field}},
  \href{http://dx.doi.org/10.1103/PhysRevD.3.1382}{{\em Phys. Rev. D} {\bf 3}
  (1971) 1382}.

\bibitem{Kogut:1969xa}
J.~B. Kogut and D.~E. Soper, {\it {Quantum Electrodynamics in the Infinite
  Momentum Frame}},  \href{http://dx.doi.org/10.1103/PhysRevD.1.2901}{{\em
  Phys. Rev. D} {\bf 1} (1970) 2901}.

\bibitem{Iancu:2020mos}
E.~Iancu and Y.~Mulian, {\it {Forward dijets in proton-nucleus collisions at
  next-to-leading order: the real corrections}},
  \href{http://dx.doi.org/10.1007/JHEP03(2021)005}{{\em JHEP} {\bf 03} (2021)
  005} [\href{http://arXiv.org/abs/2009.11930}{{\tt arXiv:2009.11930
  [hep-ph]}}].

\bibitem{Chirilli:2012jd}
G.~A. Chirilli, B.-W. Xiao and F.~Yuan, {\it {Inclusive Hadron Productions in
  pA Collisions}},  \href{http://dx.doi.org/10.1103/PhysRevD.86.054005}{{\em
  Phys. Rev. D} {\bf 86} (2012) 054005}
  [\href{http://arXiv.org/abs/1203.6139}{{\tt arXiv:1203.6139 [hep-ph]}}].

\bibitem{Lappi:2016oup}
T.~Lappi and R.~Paatelainen, {\it {The one loop gluon emission light cone wave
  function}},  \href{http://dx.doi.org/10.1016/j.aop.2017.02.002}{{\em Annals
  Phys.} {\bf 379} (2017) 34} [\href{http://arXiv.org/abs/1611.00497}{{\tt
  arXiv:1611.00497 [hep-ph]}}].

\bibitem{Kovchegov:2012mbw}
Y.~V. Kovchegov and E.~Levin, {\em {Quantum chromodynamics at high energy}},
  vol.~33.
\newblock Cambridge University Press, 8, 2012.

\bibitem{Bartels:2000gt}
J.~Bartels, S.~Gieseke and C.~F. Qiao, {\it {The $\gamma^* \to q\bar q$ Reggeon
  vertex in next-to-leading order QCD}},
  \href{http://dx.doi.org/10.1103/PhysRevD.63.056014}{{\em Phys. Rev. D} {\bf
  63} (2001) 056014} [\href{http://arXiv.org/abs/hep-ph/0009102}{{\tt
  arXiv:hep-ph/0009102}}].
\newblock [Erratum: Phys.Rev.D 65, 079902 (2002)].

\bibitem{Bartels:2001mv}
J.~Bartels, S.~Gieseke and A.~Kyrieleis, {\it {The Process $\gamma^*_L + q \to
  q \bar q g + q$: Real corrections to the virtual photon impact factor}},
  \href{http://dx.doi.org/10.1103/PhysRevD.65.014006}{{\em Phys. Rev. D} {\bf
  65} (2002) 014006} [\href{http://arXiv.org/abs/hep-ph/0107152}{{\tt
  arXiv:hep-ph/0107152}}].

\bibitem{Bartels:2002uz}
J.~Bartels, D.~Colferai, S.~Gieseke and A.~Kyrieleis, {\it {NLO corrections to
  the photon impact factor: Combining real and virtual corrections}},
  \href{http://dx.doi.org/10.1103/PhysRevD.66.094017}{{\em Phys. Rev. D} {\bf
  66} (2002) 094017} [\href{http://arXiv.org/abs/hep-ph/0208130}{{\tt
  arXiv:hep-ph/0208130}}].

\bibitem{Bartels:2004bi}
J.~Bartels and A.~Kyrieleis, {\it {NLO corrections to the $\gamma^*$ impact
  factor: First numerical results for the real corrections to $\gamma^*_L$}},
  \href{http://dx.doi.org/10.1103/PhysRevD.70.114003}{{\em Phys. Rev. D} {\bf
  70} (2004) 114003} [\href{http://arXiv.org/abs/hep-ph/0407051}{{\tt
  arXiv:hep-ph/0407051}}].

\bibitem{Altinoluk:2015dpi}
T.~Altinoluk, N.~Armesto, G.~Beuf and A.~H. Rezaeian, {\it {Diffractive Dijet
  Production in Deep Inelastic Scattering and Photon-Hadron Collisions in the
  Color Glass Condensate}},
  \href{http://dx.doi.org/10.1016/j.physletb.2016.05.032}{{\em Phys. Lett. B}
  {\bf 758} (2016) 373} [\href{http://arXiv.org/abs/1511.07452}{{\tt
  arXiv:1511.07452 [hep-ph]}}].

\bibitem{Dominguez:2011wm}
F.~Dominguez, C.~Marquet, B.-W. Xiao and F.~Yuan, {\it {Universality of
  Unintegrated Gluon Distributions at small $x$}},
  \href{http://dx.doi.org/10.1103/PhysRevD.83.105005}{{\em Phys. Rev. D} {\bf
  83} (2011) 105005} [\href{http://arXiv.org/abs/1101.0715}{{\tt
  arXiv:1101.0715 [hep-ph]}}].

\bibitem{Jalilian-Marian:1996mkd}
J.~Jalilian-Marian, A.~Kovner, L.~D. McLerran and H.~Weigert, {\it {The
  Intrinsic glue distribution at very small x}},
  \href{http://dx.doi.org/10.1103/PhysRevD.55.5414}{{\em Phys. Rev. D} {\bf 55}
  (1997) 5414} [\href{http://arXiv.org/abs/hep-ph/9606337}{{\tt
  arXiv:hep-ph/9606337}}].

\bibitem{Jalilian-Marian:1997qno}
J.~Jalilian-Marian, A.~Kovner, A.~Leonidov and H.~Weigert, {\it {The BFKL
  equation from the Wilson renormalization group}},
  \href{http://dx.doi.org/10.1016/S0550-3213(97)00440-9}{{\em Nucl. Phys. B}
  {\bf 504} (1997) 415} [\href{http://arXiv.org/abs/hep-ph/9701284}{{\tt
  arXiv:hep-ph/9701284}}].

\bibitem{Jalilian-Marian:1997jhx}
J.~Jalilian-Marian, A.~Kovner, A.~Leonidov and H.~Weigert, {\it {The Wilson
  renormalization group for low x physics: Towards the high density regime}},
  \href{http://dx.doi.org/10.1103/PhysRevD.59.014014}{{\em Phys. Rev. D} {\bf
  59} (1998) 014014} [\href{http://arXiv.org/abs/hep-ph/9706377}{{\tt
  arXiv:hep-ph/9706377}}].

\bibitem{Iancu:2001md}
E.~Iancu and L.~D. McLerran, {\it {Saturation and universality in QCD at small
  x}},  \href{http://dx.doi.org/10.1016/S0370-2693(01)00526-3}{{\em Phys. Lett.
  B} {\bf 510} (2001) 145} [\href{http://arXiv.org/abs/hep-ph/0103032}{{\tt
  arXiv:hep-ph/0103032}}].

\bibitem{Ferreiro:2001qy}
E.~Ferreiro, E.~Iancu, A.~Leonidov and L.~McLerran, {\it {Nonlinear gluon
  evolution in the color glass condensate. 2.}},
  \href{http://dx.doi.org/10.1016/S0375-9474(01)01329-X}{{\em Nucl. Phys. A}
  {\bf 703} (2002) 489} [\href{http://arXiv.org/abs/hep-ph/0109115}{{\tt
  arXiv:hep-ph/0109115}}].

\bibitem{Iancu:2001ad}
E.~Iancu, A.~Leonidov and L.~D. McLerran, {\it {The Renormalization group
  equation for the color glass condensate}},
  \href{http://dx.doi.org/10.1016/S0370-2693(01)00524-X}{{\em Phys. Lett. B}
  {\bf 510} (2001) 133} [\href{http://arXiv.org/abs/hep-ph/0102009}{{\tt
  arXiv:hep-ph/0102009}}].

\bibitem{Iancu:2000hn}
E.~Iancu, A.~Leonidov and L.~D. McLerran, {\it {Nonlinear gluon evolution in
  the color glass condensate. 1.}},
  \href{http://dx.doi.org/10.1016/S0375-9474(01)00642-X}{{\em Nucl. Phys. A}
  {\bf 692} (2001) 583} [\href{http://arXiv.org/abs/hep-ph/0011241}{{\tt
  arXiv:hep-ph/0011241}}].

\bibitem{Mueller:2001uk}
A.~H. Mueller, {\it {A Simple derivation of the JIMWLK equation}},
  \href{http://dx.doi.org/10.1016/S0370-2693(01)01343-0}{{\em Phys. Lett. B}
  {\bf 523} (2001) 243} [\href{http://arXiv.org/abs/hep-ph/0110169}{{\tt
  arXiv:hep-ph/0110169}}].

\bibitem{Aaron:2009aa}
{\bf H1, ZEUS} collaboration, F.~D. Aaron {\em et.~al.}, {\it {Combined
  Measurement and QCD Analysis of the Inclusive e+- p Scattering Cross Sections
  at HERA}},  \href{http://dx.doi.org/10.1007/JHEP01(2010)109}{{\em JHEP} {\bf
  01} (2010) 109} [\href{http://arXiv.org/abs/0911.0884}{{\tt arXiv:0911.0884
  [hep-ex]}}].

\bibitem{Abramowicz:2015mha}
{\bf H1, ZEUS} collaboration, H.~Abramowicz {\em et.~al.}, {\it {Combination of
  measurements of inclusive deep inelastic ${e^{\pm }p}$ scattering cross
  sections and QCD analysis of HERA data}},
  \href{http://dx.doi.org/10.1140/epjc/s10052-015-3710-4}{{\em Eur. Phys. J. C}
  {\bf 75} (2015)~no.~12 580} [\href{http://arXiv.org/abs/1506.06042}{{\tt
  arXiv:1506.06042 [hep-ex]}}].

\bibitem{H1:2018flt}
{\bf H1, ZEUS} collaboration, H.~Abramowicz {\em et.~al.}, {\it {Combination
  and QCD analysis of charm and beauty production cross-section measurements in
  deep inelastic $ep$ scattering at HERA}},
  \href{http://dx.doi.org/10.1140/epjc/s10052-018-5848-3}{{\em Eur. Phys. J. C}
  {\bf 78} (2018)~no.~6 473} [\href{http://arXiv.org/abs/1804.01019}{{\tt
  arXiv:1804.01019 [hep-ex]}}].

\bibitem{H1:2012xnw}
{\bf H1, ZEUS} collaboration, H.~Abramowicz {\em et.~al.}, {\it {Combination
  and QCD Analysis of Charm Production Cross Section Measurements in
  Deep-Inelastic ep Scattering at HERA}},
  \href{http://dx.doi.org/10.1140/epjc/s10052-013-2311-3}{{\em Eur. Phys. J. C}
  {\bf 73} (2013)~no.~2 2311} [\href{http://arXiv.org/abs/1211.1182}{{\tt
  arXiv:1211.1182 [hep-ex]}}].

\bibitem{Lappi:2013zma}
T.~Lappi and H.~M\"antysaari, {\it {Single inclusive particle production at
  high energy from HERA data to proton-nucleus collisions}},
  \href{http://dx.doi.org/10.1103/PhysRevD.88.114020}{{\em Phys. Rev. D} {\bf
  88} (2013) 114020} [\href{http://arXiv.org/abs/1309.6963}{{\tt
  arXiv:1309.6963 [hep-ph]}}].

\bibitem{Albacete:2010sy}
J.~L. Albacete, N.~Armesto, J.~G. Milhano, P.~Quiroga-Arias and C.~A. Salgado,
  {\it {AAMQS: A non-linear QCD analysis of new HERA data at small-x including
  heavy quarks}},  \href{http://dx.doi.org/10.1140/epjc/s10052-011-1705-3}{{\em
  Eur. Phys. J. C} {\bf 71} (2011) 1705}
  [\href{http://arXiv.org/abs/1012.4408}{{\tt arXiv:1012.4408 [hep-ph]}}].

\bibitem{Ducloue:2019jmy}
B.~Duclou\'e, E.~Iancu, G.~Soyez and D.~N. Triantafyllopoulos, {\it {HERA data
  and collinearly-improved BK dynamics}},
  \href{http://dx.doi.org/10.1016/j.physletb.2020.135305}{{\em Phys. Lett. B}
  {\bf 803} (2020) 135305} [\href{http://arXiv.org/abs/1912.09196}{{\tt
  arXiv:1912.09196 [hep-ph]}}].

\bibitem{Mantysaari:2018zdd}
H.~M\"antysaari and B.~Schenke, {\it {Confronting impact parameter dependent
  JIMWLK evolution with HERA data}},
  \href{http://dx.doi.org/10.1103/PhysRevD.98.034013}{{\em Phys. Rev. D} {\bf
  98} (2018)~no.~3 034013} [\href{http://arXiv.org/abs/1806.06783}{{\tt
  arXiv:1806.06783 [hep-ph]}}].

\bibitem{Kowalski:2003hm}
H.~Kowalski and D.~Teaney, {\it {An Impact parameter dipole saturation model}},
   \href{http://dx.doi.org/10.1103/PhysRevD.68.114005}{{\em Phys. Rev. D} {\bf
  68} (2003) 114005} [\href{http://arXiv.org/abs/hep-ph/0304189}{{\tt
  arXiv:hep-ph/0304189}}].

\bibitem{Mantysaari:2018nng}
H.~M\"antysaari and P.~Zurita, {\it {In depth analysis of the combined HERA
  data in the dipole models with and without saturation}},
  \href{http://dx.doi.org/10.1103/PhysRevD.98.036002}{{\em Phys. Rev. D} {\bf
  98} (2018) 036002} [\href{http://arXiv.org/abs/1804.05311}{{\tt
  arXiv:1804.05311 [hep-ph]}}].

\bibitem{Rezaeian:2012ji}
A.~H. Rezaeian, M.~Siddikov, M.~Van~de Klundert and R.~Venugopalan, {\it
  {Analysis of combined HERA data in the Impact-Parameter dependent Saturation
  model}},  \href{http://dx.doi.org/10.1103/PhysRevD.87.034002}{{\em Phys. Rev.
  D} {\bf 87} (2013)~no.~3 034002} [\href{http://arXiv.org/abs/1212.2974}{{\tt
  arXiv:1212.2974 [hep-ph]}}].

\bibitem{Iancu:2018hwa}
E.~Iancu and Y.~Mulian, {\it {Forward trijet production in
  proton\textendash{}nucleus collisions}},
  \href{http://dx.doi.org/10.1016/j.nuclphysa.2019.02.003}{{\em Nucl. Phys. A}
  {\bf 985} (2019) 66} [\href{http://arXiv.org/abs/1809.05526}{{\tt
  arXiv:1809.05526 [hep-ph]}}].

\bibitem{GolecBiernat:2001mm}
K.~J. Golec-Biernat and M.~Wusthoff, {\it {Diffractive parton distributions
  from the saturation model}},
  \href{http://dx.doi.org/10.1007/s100520100661}{{\em Eur. Phys. J. C} {\bf 20}
  (2001) 313} [\href{http://arXiv.org/abs/hep-ph/0102093}{{\tt
  arXiv:hep-ph/0102093}}].

\bibitem{Bartels:2003yj}
J.~Bartels, K.~J. Golec-Biernat and K.~Peters, {\it {On the dipole picture in
  the nonforward direction}},  {\em Acta Phys. Polon. B} {\bf 34} (2003) 3051
  [\href{http://arXiv.org/abs/hep-ph/0301192}{{\tt arXiv:hep-ph/0301192}}].

\bibitem{Kowalski:2006hc}
H.~Kowalski, L.~Motyka and G.~Watt, {\it {Exclusive diffractive processes at
  HERA within the dipole picture}},
  \href{http://dx.doi.org/10.1103/PhysRevD.74.074016}{{\em Phys. Rev. D} {\bf
  74} (2006) 074016} [\href{http://arXiv.org/abs/hep-ph/0606272}{{\tt
  arXiv:hep-ph/0606272}}].

\bibitem{Hatta:2017cte}
Y.~Hatta, B.-W. Xiao and F.~Yuan, {\it {Gluon Tomography from Deeply Virtual
  Compton Scattering at Small-x}},
  \href{http://dx.doi.org/10.1103/PhysRevD.95.114026}{{\em Phys. Rev. D} {\bf
  95} (2017)~no.~11 114026} [\href{http://arXiv.org/abs/1703.02085}{{\tt
  arXiv:1703.02085 [hep-ph]}}].

\bibitem{Mantysaari:2020lhf}
H.~M\"antysaari, K.~Roy, F.~Salazar and B.~Schenke, {\it {Gluon imaging using
  azimuthal correlations in diffractive scattering at the Electron-Ion
  Collider}},  \href{http://dx.doi.org/10.1103/PhysRevD.103.094026}{{\em Phys.
  Rev. D} {\bf 103} (2021)~no.~9 094026}
  [\href{http://arXiv.org/abs/2011.02464}{{\tt arXiv:2011.02464 [hep-ph]}}].

\bibitem{Dumitru:2021tvw}
A.~Dumitru, H.~M\"antysaari and R.~Paatelainen, {\it {Color charge correlations
  in the proton at NLO: Beyond geometry based intuition}},
  \href{http://dx.doi.org/10.1016/j.physletb.2021.136560}{{\em Phys. Lett. B}
  {\bf 820} (2021) 136560} [\href{http://arXiv.org/abs/2103.11682}{{\tt
  arXiv:2103.11682 [hep-ph]}}].

\bibitem{Mantysaari:2019csc}
H.~M\"antysaari, N.~Mueller and B.~Schenke, {\it {Diffractive Dijet Production
  and Wigner Distributions from the Color Glass Condensate}},
  \href{http://dx.doi.org/10.1103/PhysRevD.99.074004}{{\em Phys. Rev. D} {\bf
  99} (2019)~no.~7 074004} [\href{http://arXiv.org/abs/1902.05087}{{\tt
  arXiv:1902.05087 [hep-ph]}}].

\bibitem{Salazar:2019ncp}
F.~Salazar and B.~Schenke, {\it {Diffractive dijet production in impact
  parameter dependent saturation models}},
  \href{http://dx.doi.org/10.1103/PhysRevD.100.034007}{{\em Phys. Rev. D} {\bf
  100} (2019)~no.~3 034007} [\href{http://arXiv.org/abs/1905.03763}{{\tt
  arXiv:1905.03763 [hep-ph]}}].

\bibitem{Iancu:2017fzn}
E.~Iancu and A.~H. Rezaeian, {\it {Elliptic flow from color-dipole orientation
  in pp and pA collisions}},
  \href{http://dx.doi.org/10.1103/PhysRevD.95.094003}{{\em Phys. Rev. D} {\bf
  95} (2017)~no.~9 094003} [\href{http://arXiv.org/abs/1702.03943}{{\tt
  arXiv:1702.03943 [hep-ph]}}].

\bibitem{Kovchegov:1999ji}
Y.~V. Kovchegov and E.~Levin, {\it {Diffractive dissociation including multiple
  pomeron exchanges in high parton density QCD}},
  \href{http://dx.doi.org/10.1016/S0550-3213(00)00125-5}{{\em Nucl. Phys. B}
  {\bf 577} (2000) 221} [\href{http://arXiv.org/abs/hep-ph/9911523}{{\tt
  arXiv:hep-ph/9911523}}].

\bibitem{Bartels:1999tn}
J.~Bartels, H.~Jung and M.~Wusthoff, {\it {Quark - anti-quark gluon jets in DIS
  diffractive dissociation}},
  \href{http://dx.doi.org/10.1007/s100520050618}{{\em Eur. Phys. J. C} {\bf 11}
  (1999) 111} [\href{http://arXiv.org/abs/hep-ph/9903265}{{\tt
  arXiv:hep-ph/9903265}}].

\bibitem{Kopeliovich:1999am}
B.~Z. Kopeliovich, A.~Schafer and A.~V. Tarasov, {\it {Nonperturbative effects
  in gluon radiation and photoproduction of quark pairs}},
  \href{http://dx.doi.org/10.1103/PhysRevD.62.054022}{{\em Phys. Rev. D} {\bf
  62} (2000) 054022} [\href{http://arXiv.org/abs/hep-ph/9908245}{{\tt
  arXiv:hep-ph/9908245}}].

\bibitem{Kovchegov:2001ni}
Y.~V. Kovchegov, {\it {Diffractive gluon production in proton nucleus
  collisions and in DIS}},
  \href{http://dx.doi.org/10.1103/PhysRevD.64.114016}{{\em Phys. Rev. D} {\bf
  64} (2001) 114016} [\href{http://arXiv.org/abs/hep-ph/0107256}{{\tt
  arXiv:hep-ph/0107256}}].
\newblock [Erratum: Phys.Rev.D 68, 039901 (2003)].

\bibitem{Golec-Biernat:2005prq}
K.~J. Golec-Biernat and C.~Marquet, {\it {Testing saturation with diffractive
  jet production in deep inelastic scattering}},
  \href{http://dx.doi.org/10.1103/PhysRevD.71.114005}{{\em Phys. Rev. D} {\bf
  71} (2005) 114005} [\href{http://arXiv.org/abs/hep-ph/0504214}{{\tt
  arXiv:hep-ph/0504214}}].

\bibitem{Bjorken:1973gc}
J.~D. Bjorken and J.~B. Kogut, {\it {Correspondence Arguments for High-Energy
  Collisions}},  \href{http://dx.doi.org/10.1103/PhysRevD.8.1341}{{\em Phys.
  Rev. D} {\bf 8} (1973) 1341}.

\bibitem{Hebecker:1997gp}
A.~Hebecker, {\it {Diffractive parton distributions in the semiclassical
  approach}},  \href{http://dx.doi.org/10.1016/S0550-3213(97)00512-9}{{\em
  Nucl. Phys. B} {\bf 505} (1997) 349}
  [\href{http://arXiv.org/abs/hep-ph/9702373}{{\tt arXiv:hep-ph/9702373}}].

\bibitem{Buchmuller:1998jv}
W.~Buchmuller, T.~Gehrmann and A.~Hebecker, {\it {Inclusive and diffractive
  structure functions at small x}},
  \href{http://dx.doi.org/10.1016/S0550-3213(98)00682-8}{{\em Nucl. Phys. B}
  {\bf 537} (1999) 477} [\href{http://arXiv.org/abs/hep-ph/9808454}{{\tt
  arXiv:hep-ph/9808454}}].

\bibitem{Ducloue:2017ftk}
B.~Duclou\'e, H.~H\"anninen, T.~Lappi and Y.~Zhu, {\it {Deep inelastic
  scattering in the dipole picture at next-to-leading order}},
  \href{http://dx.doi.org/10.1103/PhysRevD.96.094017}{{\em Phys. Rev. D} {\bf
  96} (2017)~no.~9 094017} [\href{http://arXiv.org/abs/1708.07328}{{\tt
  arXiv:1708.07328 [hep-ph]}}].

\end{thebibliography}\endgroup
